\documentclass[letterpaper]{article}
\usepackage{amsmath,longtable,graphicx}
\usepackage{amsfonts}
\usepackage{amssymb}
\begin{document}

\title{Orbital Parameters of Binary Radio Pulsars : Revealing Their Structure, Formation, Evolution and Dynamic History}
\author{Manjari Bagchi$^{1, *}$}
\date{\today}
\maketitle

\noindent{$^1$ Inter University Centre for Astronomy and Astrophysics, Pune 411 007, India \\  $*$ manjari.bagchi@gmail.com }

\begin{abstract}

Orbital parameters of binary radio pulsars reveal the history of the pulsars' formation and evolution including dynamic interactions with other objects. Advanced technology has enabled us to determine these orbital parameters accurately in most of the cases. Determination of post-Keplerian parameters of double neutron star binaries (especially of the double pulsar system PSR J0737-3039) provide clean tests of the general theory of relativity and in the future may lead us to constrain the dense matter Equations of State. For binary pulsars with main-sequence or white dwarf companions, knowledge about the values of the orbital parameters as well as of the spin periods and the masses of the pulsars and the companions might be useful to understand the evolutionary history of the systems. As accreting neutron star binaries lead to orbit circularization due to the tidal coupling during the accretion phase, their descendants $i.e.$ binary millisecond pulsars are expected to be in circular orbits. On the other hand, dense stellar environments inside globular clusters cause different types of interactions of single stars with pulsar binaries like ``fly-by", ``exchange", and ``merger". All these interactions impart high eccentricities to the pulsar binaries. So it is quite common to get eccentric millisecond pulsar binaries in globular clusters and we find that ``fly-by" causes intermediate values of eccentricities while ``exchange" or ``merger" causes high values of eccentricities. We also show that ``ionization" is not much effective in the present stage of globular clusters. Even in the absence of such kinds of stellar interactions, a millisecond pulsar can have an eccentric orbit as a result of Kozai resonance if the pulsar binary is a member of a hierarchical triple system. PSR J1903+0327 is the only one eccentric millisecond pulsar binary in the galactic disk where stellar interactions are negligible. The possibility of this system to be a member of a hierarchical triple system or past association of a globular cluster have been studied and found to be less likely.

\end{abstract}

\section{Introduction}
\label{sec:intro}

Radio pulsars are rotating, magnetized neutron stars (NSs) emitting radio beams from magnetic poles. Their masses are in the range of $1-2~M_{\odot}$, radii $\sim10$ km, densities $\sim10^{14}~{\rm g~cm^{3}}$ and surface values of the dipolar magnetic field are around $10^{8}-10^{14}$ gauss. Presently there are 1864 radio pulsars with their spin periods covering a wide range 1.39 ms $-$ 11.78 s \footnote{http://www.atnf.csiro.au/research/pulsar/psrcat/}. This number is increasing quite rapidly after the first serendipitous discovery in 1967 by Jocelyn Bell, a graduate student in Cambridge \cite{joc68}. 

Pulsars are believed to be born with spin period $\sim 0.1$ s which increases very slowly with time (at the rate of $10^{-20}-10^{-14}~{\rm s \, s^{-1}}$) as their rotational energy is emitted in the form of electromagnetic energy. Millisecond spin periods are explained as a result of spin-up process, which occurs when a neutron star has a normal star as its binary companion and accretes matter from the companion (at its red giant phase) causing transfer of angular momentum. Depending upon the evolution of the companion, the resultant millisecond pulsar can be either isolated or a member of a binary system. In this article we concentrate on binary pulsars and discuss how the knowledge of their orbital parameters help us to understand the structure, formation, evolution and dynamic history of pulsars. As an example, binary millisecond pulsars are expected to be in circular orbits as tidal coupling during the accretion phase lead to orbit circularization. But if there is no history of accretion phase, the orbit of the binary pulsar can be eccentric because of the asymmetric kick imparted to the pulsar during its birth in a supernova. A millisecond pulsar binary can also have an eccentric orbit if it experiences interactions with other near-by stars (which is possible in dense stellar environments like globular clusters) or if it is a member of a hierarchical triple. 

It is noteworthy to mention here that in addition to radio wavelengths, pulsars emit in x-rays and/or gamma rays too. But radio pulsars are the most useful astronomical objects firstly because the underlying neutron stars are more stable than in the case of X-ray pulsars which have episodically varying accretion torques; secondly high sensitivities of radio telescopes and data analysis software make it possible to determine their spin periods very accurately. Moreover, for radio pulsars in binary systems ($\sim 8\%$ of presently known radio pulsars are in binary systems), it is possible to determine the mass of the pulsar and its companion, size and shape of the orbit $etc$ by modelling Keplerian and post-Keplerian parameters. Such accurate determination of those parameters are difficult for X-ray pulsars at present. That is why we concentrate only on binary radio pulsars when we try to use the knowledge about their orbital parameters to understand the properties of pulsars. The first binary pulsar PSR B1913+16 was discovered in 1974 at Arecibo observatory by R. Hulse and J. H. Taylor \cite{hultal74} which is a 59 ms pulsar having orbital period 0.323 days and orbital eccentricities 0.671. Interestingly, its $\sim 1.06~M_{\odot}$ mass companion star is most probably another neutron star $i.e.$ the first discovered binary pulsar is also the first discovered double neutron star (candidate) system. Afterwords nine other double neutron star (DNS) systems have been discovered. Among these ten DNSs, seven systems are confirmed to be DNSs and the rest are most likely to be DNSs \cite{dunc08}. These DNSs are useful tools to test the validity of the general theory of relativity \cite{ingrid03} and even in principle to constrain the dense matter Equations of State \cite{lat06}. In Section \ref{sec:dns} we discuss how the study of DNSs can lead to a better understanding of the structure of neutron stars ($i.e$ the Equation of State), the physics of the pulsar magnetosphere and the formation scenario for the special case of the double pulsar system.

Most of the binary pulsars have either main sequence (MS) or white dwarf (WD) companions. Knowledge about the values of the orbital parameters as well as of the spin periods and the masses of these pulsars (and their companions) might be useful to understand the evolutionary history of these systems. The possible causes of eccentricities of millisecond pulsar binaries have been studied in this article. When these systems are in dense stellar environments as in globular clusters, the eccentricities can arise as a result of different types of interactions of single stars with pulsar binaries like ``fly-by", ``exchange" and ``merger" (Section \ref{sec:gc}). The eccentricity of the only eccentric millisecond pulsar PSR J$1903+0327$ in the galactic disk is difficult to explain as the hierarchical triple scenario faces difficulties to satisfy observed parameters of this system and the globular cluster origin also appears to be unlikely (section \ref{sec:triple}). All the findings of the present work have been summarized in section \ref{sec:summary}.

\section{Double Neutron Star Binaries}
\label{sec:dns}

After performing pulsar observations, one determines ``time of arrivals" (TOAs) of pulses at the telescope. Then various clock corrections are applied to the TOAs, followed by some other corrections $e.g.$ ``R\"omer delay", ``Shapiro delay" and ``Einstein delay" to calculate the TOAs with respect to the solar system barycenter. Then the correction for the frequency dependent delay (dispersion) by the interstellar medium is added. Finally, the pulse phase $\phi$ is fitted with time as a Taylor series expansion like $\phi=\phi_0+\nu(t-t_0)+\frac{1}{2}\dot{\nu}(t-t_0)^2$ where $\phi_0$  and $t_0$ are reference phase and time, $\nu$ and $\dot{\nu}$ are the spin frequency and its first time derivative. In case of a binary pulsar, one needs to model the orbital motion of the pulsar, and before doing that same type of delays for the binary system ($e.g.$ another set of ``R\"omer delay", ``Shapiro delay" and ``Einstein delay") should be taken into account. For a non-relativistic binary pulsar, five Keplerian parameters are required to describe the orbit, these are the orbital period ($P_{orb}$), projected semi-major axis of the pulsar orbit ($a_p~{\rm sin}~i$, where $i$ is the inclination of the orbit with respect to the sky plane), orbital eccentricity ($e$), longitude of periastron ($\omega$) and the epoch of the periastron passage ($T_0$). The semi major axes of the pulsar orbit and the companion orbit can be written as $a_p = a \, m_c / (m_p + m_c)$ and $a_c = a \, m_p / (m_p + m_c)$ respectively where $a$ is the semi major axis of the relative orbit coming in the Kepler's law $P^2_{orb} / 4 \pi^2 = a^3 /G(m_p+m_c)$. Here $m_p$ is the mass of the pulsar and $m_c$ is the mass of the companion. Pulsar timing analysis means first getting an initial fit for the orbital parameters by inspecting the variation of the pulse period (which is equal to the spin period $P_s$ of the pulsar) and then improving the values of the parameters through a phase connected timing solution. But for relativistic binaries ($i.e.$ when the companion of the pulsar is another neutron star or a white dwarf), five post-Keplerian (PK) parameters are needed, which are the relativistic advance of periastron (${\dot \omega}_{gr}$), redshift parameter ($\gamma$), Shapiro parameters (range ``$r$" and shape ``$s$") and the rate of change of orbital period ($\dot{P}_{orb}$) due to gravitational wave emission. Details of all the above mentioned parameters and how they are used in the timing models are described very nicely in  ``Handbook of Pulsar Astronomy" by D. R. Lorimer and M. Kramer \cite{lk05}. 

Post-Keplerian descriptions of the orbits are essential for ``double neutron star" (DNS) binaries which include PSR J0737-3039(A, B), PSR B1534+12, PSR J1756-2251, PSR J1811-173, PSR J1906+0746, PSR B1913+16, PSR B2127+11C (confirmed DNSs), PSR J1518+4904, PSR B1820-11, PSR J1829+2456 (candidate DNSs - orbital parameters hint DNS nature but companion mass informations are insufficient to confirm). Only one of the DNSs, namely PSR J0737-3039 is a double pulsar system $i.e.$ both the neutron stars emit radio pulses. This feature has made this system a unique astrophysical laboratory and we shall discuss about it in more details in subsections \ref{subsec:eos}, \ref{subsec:intAB} and \ref{subsec:form_double}. Moreover, there are three relativistic NS-WD systems - PSR J0751+1807, PSR J1757-5322, PSR J1141-6545 for which also post-Keplerian descriptions are unavoidable. Let us now try to understand how DNSs can help us in better understanding of the structure of neutron stars, physics of the pulsar magnetosphere and formation scenario (for the special case of the double pulsar system) of neutron stars.

\subsection{Constraining Dense Matter Equations of State}
\label{subsec:eos}

After the first proposal of neutron star as the ultimate fate of stars (of mass $\sim 5-60~M_{\odot}$) by Baade and Zwicky in 1934 \cite{bdzw34} and the immediate prediction of mass and radius by Oppenheimer and Volkoff in 1939 \cite {ov39}, many people worked to understand the nature of matter at such high densities, resulting a large number of Equations of States (EsoS) including the EsoS for strange quark matter. Different EsoS result different mass-radius relations - see Fig 2 of \cite{lm07} to realize how different the mass-radius curves can be. So the question arises, how to test the validity of those EsoS. Presently there are a number of efforts to constrain the EsoS through astronomical observations in X-rays. The usual approach is to determine the mass and the radius of the stars with the help of various observational features like gravitational redshifts (z) from spectral lines, cooling characteristics, kHz quasi-periodic oscillations (QPO) etc \cite{lm07, liqpo, ozel06, ozel08a, ozel08b, zhang07}. But these methods are not foolproof, $e.g.$ the value of z used in \"Ozel's \cite{ozel06} analysis of EXO 0748$-$676 could not be reproduced afterwords \cite{klahn06}. Moreover, to constrain EsoS from QPO observations, one needs to believe in a particular model of QPO which is again a subject of debate. An alternative method to constrain the dense matter EsoS might be the measurement of the moment of inertia of a neutron star in a relativistic binary, especially for a DNS system and to match this observationally determined value with the theoretically predicted one. The Equation of state (EoS) which gives the best agreement between observed and the theoretical values of the moment of inertia of the NS can be considered as the best description of NS structure. 

The moment of inertia of a neutron star of known mass can be theoretically calculated for any particular EoS either using Hartle's approximations \cite{kp99} or by performing exact calculations \cite{bbh05, bag10} for the rotating axisymmetric neutron star. It has been checked that the values of moment of inertia for a particular NS obtained by this two approaches do not differ more than 1\% for a star like PSR J0737-3039A \cite{bbh05, bag09, bag10}. We found the moment of inertia of PSR J0737-3039A to be in the range $0.67-0.74  \times 10^{45}{\rm g~ cm^2}$ for a set of EsoS having different stiffness \cite{bag09}. 

Let us now discuss about observational determination of the moment of a neutron star by accurate measurement of the periastron advance (${\dot \omega}_{gr}$) of the orbit of a relativistic binary pulsar (preferably a DNS) through timing analysis. The theoretical expression of ${\dot \omega}_{gr}$ including the effect of spin-orbit coupling is known to be as follow \cite{ds88}:
\begin{equation}
{ \dot \omega}_{gr}~=~\frac{3 \beta_0^2~ \Omega_{orb}}{1-e^2}\left[1+f_0 \beta_0^2 -(g_{s1} \beta_{s1}\beta_0+g_{s2} \beta_{s2}\beta_0) \right]
\label{eq:per_adv}
\end{equation}
where the first term inside the square bracket is the 1PN term, the second term is the 2PN term, and the third term represents the spin-orbit coupling (SO). The parameters used in the above expression are defined as :
\begin{equation}
\beta_0~=~\frac{(GM  \Omega_{orb})^{1/3}}{c}
\label{eq:beta0}
\end{equation}

\begin{equation}
\beta_{sa}~=~\frac{c I_a \omega_a}{G m_a^2}
\label{eq:betas}
\end{equation}

\begin{equation}
f_0~=~\frac{1}{1-e^2}\left( \frac{39}{4}x_1^2+\frac{27}{4}x_2^2+15 x_1 x_2 \right) -\left( \frac{13}{4}x_1^2+\frac{1}{4}x_2^2+\frac{13}{3} x_1 x_2 \right)
\label{eq:f0}
\end{equation}

\begin{eqnarray}
g_{sa}~=\nonumber~\frac{x_a \left(4 x_a+ 3x_{a+1}\right)}{6(1-e^2)^{1/2}sin^2 i}\left[ (3 ~sin^2 i-1)~{\it \bf   k~.~s_a}+cos~ i~ {\it \bf K_0~.~s_a} \right]   \\ 
\label{eq:gs}
\end{eqnarray}

where $G$ is the gravitational constant, $I_a$ is the moment of inertia of the $a^{th}$ body ($a~=~1,~2$), $\omega_a~=~2\pi/P_{s, a}$ is its angular velocity of rotation (here $a+1$ means modulo 2, $i.e.$ 2+1=1) and $ \Omega_{orb}~=~2\pi/P_{orb}$. $P_{s, a}$ is the spin period of the $a^{th}$ star and $P_{orb}$ is the orbital period of the binary. $x_1~=~m_1/M,~ x_2~=~m_2/M$ and $M~=~m_1+m_2$. $\bf k$ is the orbital angular momentum vector, $ \bf s_a$ is the spin vector, $\bf K_0$ is the line of sight vector perpendicular to the sky and $i$ is the inclination of the orbit with respect to the sky implying that the angle between $\bf K_0$ and $\bf k$ is also $i$. 

In the expression of ${\dot \omega}_{gr}$, the moment of inertia of both members of the DNS system are present (through $\beta_{s,1}$ and $\beta_{s,2}$ in Eqn. \ref{eq:per_adv}) which means that we have only one observable and two unknowns making Eqn. (\ref{eq:per_adv}) unsolvable. But fortunately, $\beta_{sa}$ is negligible for $P_{s, a}> 100$ ms. So if  $P_{s, 2}> 100$ ms, equation for periastron advance becomes 
\begin{equation}
{\dot \omega}_{gr}~=~\frac{3 \beta_0^2~ \Omega_{orb}}{1-e^2}\left[1+f_0 \beta_0^2 -(g_{s1} \beta_{s1}\beta_0) \right]
\label{eq:per_advnew}
\end{equation}
So DNS binaries having one fast neutron star ($P_{s}< 100$ ms) and another slow neutron star ($P_{s}> 100$ ms) are needed to extract the moment of inertia of the faster one by equating the observed ${\dot \omega}_{gr}$ with the prediction of Eqn. (\ref{eq:per_advnew}) using measured values of  $e$, $P_{orb}$, $i$, $m_1$, $m_2$ and $P_{s, 1}$. It is clear that the angle between $ \bf s_1$ and $\bf k$ plays a vital to in the expression of $g_{s1}$. If $\bf s_1 \parallel \bf k$, the angle between $\bf K_0$ and $ \bf s_1$ becomes $i$ and we get $g_{s1, \parallel}=\frac{x_1 \left(4x_1+3x_{2} \right)}{3 \left(1-e^2 \right)^{1/2}}$. But if this is not the case $i.e.$ the angle between $ \bf s_1$ and $ \bf k$ has a non-zero value, $g_{s1}$ will be maximum if $\bf s_1$ is parallel to the vector $\left[(3 ~sin^2 i-1)~{\it \bf   k~.~s_1}+cos~ i~ {\it \bf K_0~.~s_1} \right] $ giving $\mid g_{s1, max} \mid =\left[ 3+\frac{1}{sin^{2}i}\right]^{1/2}\left[\frac{x_1 \left(4x_1+3x_{2} \right)}{3 \left(1-e^2 \right)^{1/2}}\right]$, for other orientation of $ \bf s_1$, $g_{s1}$ will be between $+\mid g_{s1, max} \mid$ to $-\mid g_{s1, max} \mid$.  So the knowledge of the angle between $\bf k$ and $\bf s_1$ is needed for determination of the moment of inertia of the faster member of a DNS system.

At present, it seems that the use of Eqn. (\ref{eq:per_advnew}) to determine the moment of inertia of a neutron star is possible only for the double pulsar system PSR J0737-3039 \cite{ls05, bag09}, as for this system, timing analysis of the pulses from both the stars has already enabled us to determine $e$, $P_{orb}$, $m_1$, $m_2$, $P_{s, 1}$ and $P_{s, 2}$  \cite{burg03, lyn04}. The two pulsars which are named as PSR J0737-3039A and PSR J0737-3039B have masses and spin periods as $m_A=1.3381~M_{\odot}$, $m_B=1.2489~M_{\odot}$, $P_{s, A}=22.7$ ms and  $P_{s, B}=2.77$ s with orbital parameters as $e=0.0877$, $P_{orb}=0.102252$ days, ${\dot P}_{orb}=-1.252\times 10^{-12}$, $i=88.69^{\circ}$. So it is clear that we can easily neglect the spin-orbit term due to pulsar B. Although the angles between the vectors ${\it \bf k,~s_1}$ and ${\it \bf K_0,~s_1}$ are not known which are needed to be determined first to measure the moment of inertia of the pulsar A accurately. Just for the sake of completeness, we wish to mention that for PSR J0737-3039, the value of $\mid g_{sA, max} \mid$ is 1.2176 whereas the value of $ g_{sA, \parallel} $ is 0.60876. The accuracies of other relevant parameters are also needed to be improved.

In general, it is clear that the $(g_{s1} \beta_{s1}\beta_0)$ term in Eqn. (\ref{eq:per_advnew}) will have higher value for (i) smaller value of $P_{s,1}$, (ii) smaller value of $P_{orb}$ and (iii) higher value of $e$. Unfortunately, the eccentricity of PSR J0737-3039 is small in comparison to other DNSs. Future discovery of any double pulsar system satisfying these three conditions better than PSR J0737-3039 will be useful for the purpose of moment of inertia measurement (in addition to the necessary condition $P_{s,2}> 100$ ms).

\subsection{Interplay between Two Members of the Double Pulsar}
\label{subsec:intAB}

Among many interesting features observed from the double pulsar system PSR J0737-3039, one is the variation of the flux density of the slower pulsar B at its different orbital phases due to the distortion of B's magnetosphere by A's wind \cite{lyn04, lyut05a}. On the other hand, the flux density of pulsar A is constant for most of its orbit except $30$ s near its superior conjuction when the radio beam from the pulsar A is eclipsed by the magnetosphere of the pulsar B \cite{lyn04}. The eclipse is asymmetric, with the flux density decreasing more slowly during ingress than it increases during egress. The duration of the eclipse is very mildly dependent on the observing frequencies, with the eclipse lasting slightly longer at lower frequencies \cite{kaspi04}. Theoretical models explain these eclipse properties in the context of synchrotron absorption of A's radio emission by the plasma in the closed field line region of B's magnetosphere \cite{aron05, lyut05b, raf05}. By modeling the eclipse, the precession of B's spin axis about the orbital momentum vector has been found to match with the prediction of the general theory of relativity within the $13 \%$ measurement uncertainty \cite{bret07}. It is noteworthy to mention here that till date, the eclipse has been studied in the observing frequency range 427-1400 MHz. The study of PSR J0737-3039 (both A and B) at a much lower frequency of 326 MHz is being performed using Ooty Radio telescope (RAC-TIFR, Ooty, India). The details of eclipse characteristic at this low frequency will be useful to test the validity of the eclipse models and better understanding of the pulsar magnetosphere.

\subsection{Formation of the Double Pulsar}
\label{subsec:form_double}

Using observed parameters of the double pulsar system, different formation models were proposed \cite{dewi04, pod05, ps05}.  Initially, it was proposed that this system has originated from a close helium star plus a neutron star binary in which at the onset of the evolution the helium star had a mass in the range 4.0 $-$ 6.5 $M_{\odot}$ and an orbital period in the range 0.1 $-$ 0.2 day  \cite{dewi04}. In this model, a kick velocity in the range 70 $-$ 230 $ {\rm km ~s^{-1}} $ must have been imparted to the second neutron star at its birth. Afterwords an electron capture supernova scenario was proposed \cite{pod05}. At the same time, Piran \& Shaviv \cite{ps05} estimated that it is most likely that the progenitor of the pulsar B had a mass of $1.55 \pm 0.2~M_\odot$ with a low kick velocity 30 $ {\rm km ~s^{-1}} $. On the other hand, a study based on the population synthesis method \cite{will06} favored the standard formation scenario with reasonably high kick velocities (70-180 $ {\rm km ~s^{-1}} $). Later, it has been suggested  \cite{bag09} that the pulsar B is a strange star rather than a neutron star as the baryon loss is smaller for the first case supporting the models with small kick.    

\section{Binary Pulsars in Globular Clusters}
\label{sec:gc}

Globular clusters (GCs) are dense spherical collection of stars orbiting around the galactic
center containing a significant number of binary stars. The first radio pulsar discovered in a globular cluster is PSR B1821-24A in M28 \cite{lyn87} in 1987, 20 years after the discovery of first pulsar by Jocelyn Bell. This pulsar is at present an isolated pulsar but its very low spin period (3 millisecond) hints to a binary origin and spin-up due to accretion . How old and evolved stellar systems like GCs can contain apparently young, active objects like pulsars have led to suggestions, based on lifetime constraints from orbital decay by gravitational radiation \cite{pri91}, that some of these systems have been formed recently, or are forming even today. Interaction between binary stars and single stars in GCs is believed to be an important dynamical process as the resulting binary systems may provide a substantial source of energy for the host GC, as the binding energies of a few, very close binaries (like neutron star binaries) can approach that of a moderately massive GC \cite{spi87, hut03}. Observable parameters of the pulsars and their binary companions in GCs, such as spin, orbital period and eccentricity, projected radial position in the
cluster, companion mass and their distributions provide a tracer of the
past history of dynamical interactions of the binary NSs in individual GCs.
These parameters can even provide a valuable test-bed to examine the theoretical scenarios of
formation and evolution of recycled pulsars. 

\subsection{Archival Data of Binary Pulsars in Globular Clusters}
\label{sec:data}

Till today, a total of 140 pulsars in 26 GCs\footnote{Information on these
pulsars are found from P. Freire's webpage updated August 2008,
http://www.naic.edu/$\sim$pfreire/GCpsr.html.} have been discovered. Among them 74 are known as binaries (in 23 GCs), 59 are known as isolated and 7 have no published
timing/orbital solutions. Orbital parameters of one binary, namely PSR J2140$-$2310B in M30 are
not well determined, only lower limits of $P_{orb}$ and $e$ are
known as $P_{orb} > 0.8$ days and $e > 0.52$. So we exclude it
from our analysis and proceed with the rest 73 binaries. 

\begin{figure}[h]
{\includegraphics[width=0.9\textwidth,
height=0.45\textheight]{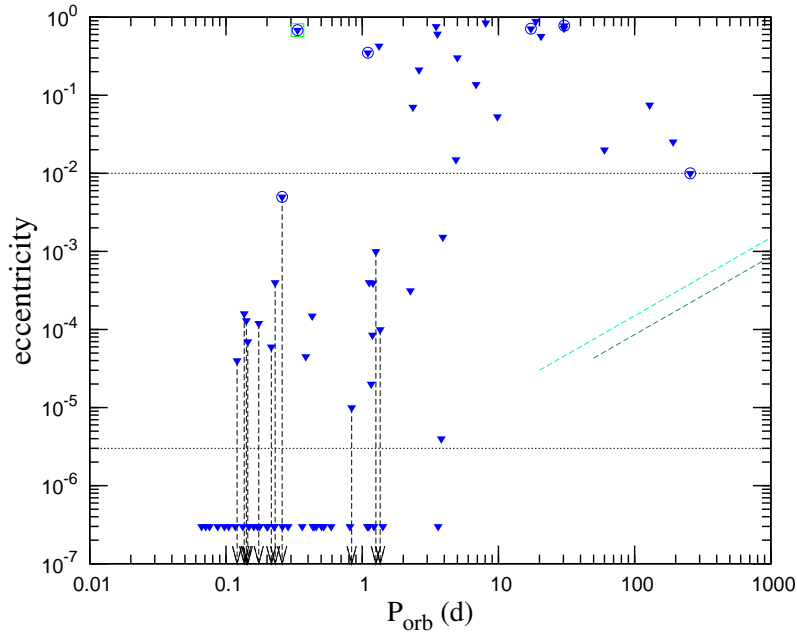}} 
\caption{ \footnotesize{Plot of eccentricity vs
orbital periods for binary radio pulsars in globular
clusters. The DNS system PSR B2127+11C ib M15 is enclosed by a green $\square$. Non-millisecond pulsars are enclosed by $\bigcirc$s. The dashed lines in the right side represents the eccentricity-orbital period relation as predicted by Phinney \cite{phi92} - the upper line is the original line drawn by Phinney \cite{phi92} and the lower line is using population II models. The black dotted lines parallel to the x-axis have been drawn to distinguish three groups of GC pulsars.}} \label{fig:porbecc_gc}
\end{figure}

In Fig \ref{fig:porbecc_gc}, we plot $e$ Vs $P_{orb}$ for these 73
binary pulsars in GCs with known orbital solutions. A logarithmic scale in
both eccentricity and orbital periods are chosen, as the enormous
range of both variables and the regions occupied by observed
pulsars are less obvious in linear scales. Downward arrows have been drawn for the systems for which 
only upper limits of eccentricities (see table \ref{tab:gcpsr_parms}) are known. Non-millisecond pulsars are enclosed by $\bigcirc$s and the DNS system PSR B2127$+$11C in M15 is enclosed by a green $\square$. In the present work, we define millisecond pulsars as the pulsars having $P_s < 30$ ms, we don't go for other definitions of millisecond pulsars based on both $P_s$ and $\dot{P_s}$ (see \cite{story07} for one such definition) as for GC binaries, the values of $\dot{P_s}$ are effected significantly by the cluster potential. 

Observed binaries in GCs can be categorized in three groups : (I) large
eccentricity pulsars ($e \geqslant 0.01$), 21 pulsars (22 if we
include M30B), (II) moderate eccentricity pulsars ($0.01 > e
\geqslant 3 \times 10^{-6}$), 20 pulsars and (III) small
eccentricity pulsars ($e \sim 0$), 32 pulsars. In the database,
orbital eccentricities of group (III) pulsars have been listed as
zero, and in this logarithmic plot we assign them an arbitrarily
small value of $e~=~3 \times 10^{-7}$. 

In addition we plot on the right of Fig \ref{fig:porbecc_gc}, dashed lines corresponding to the freeze-out eccentricity - orbital period relation predicted by Phinney \cite{phi92} on the
basis of the fluctuation dissipation theorem for convective eddies in
the erstwhile red giant envelopes surrounding the white-dwarf cores
which end up as companions of the neutron stars. This relation is obtained
from the following expression:
\begin{equation}
\frac{1}{2} \mu \Omega_{orb}^2 a ^2 <e^2>= 3.4 \times 10^{-5} (L^2
R_c^2 m_{env})^{1/3}
\end{equation} where $\Omega_{orb}~=~2\pi / P_{orb}$ is the orbital angular
rotation frequency, $a$ is the semi-major axis of the binary, $e$
is the orbital eccentricity, L is the luminosity of the companion,
$R_c$ is the radius of the companion, $m_{env}$ is the envelop
mass of the companion, $m_p$ is the pulsar mass, $m_c$ is the
companion mass and $\mu~=(m_p m_c)/(m_p+m_c)$ is the reduced mass
of the system. Phinney used $R_c=0.125~a$ when the mass transfer
ceases and $e$ freezes. Then the above equation reduces to
$<e^2>^{1/2}= C_1~(P_{orb}/100)$ where $C_1$ is a function of
the effective temperature ($T_{eff}$), $\mu$ and $m_{env}$. $P_{orb}$ is in days. Phinney calculated $C_1~ = ~1.5 \times 10^{-4}$ (see Eqn. 7.35 of \cite{phi92}) using average values of $m_{env}$ and $T_{eff}$ for population I stars ($z~=~0.04$) \cite{ref70} which gives the upper dashed green line in the right side of Fig \ref{fig:porbecc_gc}. However for globular clusters, population II stars are more appropriate. So we recalculated $C_1$ for population II red giant stars \cite{web83} and obtained $C_1=0.86 \times 10^{-4}$. This line is plotted as the lower dashed green line in Fig. \ref{fig:porbecc_gc}.

Earlier, it had been found that while the low mass X-ray binaries (LMXBs) are predominantly found in very dense GCs, the radio pulsars tend to be found more evenly distributed among low and high density clusters \cite{cam05}. We also don't find any obvious correlation of the number of binary radio pulsars with any GC parameters like $d$, $v_{10}$, $n_4$ or $v_{10}/n_4$ (where $n_4$ is the number density ($n$) of single stars in units of $10^4~ \rm{pc^{-3} }$ and $v_{10}$ is the velocity dispersion ($v$) in units of 10 ${\rm km~ s^{-1}}$ in GCs and $d$ is the distance of the GCs from sun in kpc see Table \ref{tb:gc_v_n}). But if we take only GCs with more than 3 observed binary pulsars, we find the number of binary pulsars to increase with $n_4$, but there are only five such GCs (Table \ref{tb:gc_v_n}).

\subsubsection{A comparison of binary pulsars in globular clusters and binary pulsars in the galactic disk}
\label{sec:data_gc_dick_comp}

There are total 83 binary pulsars in the galactic disk (and 1652 isolated pulsars) among which orbital period and/or eccentricity values are unknown for four binaries. So we exclude them from our study and we consider the rest 78 binary pulsars.  In the database, eccentricities of three binaries have assigned the value ``0" for which we put an arbitrary small value $3 \times 10^{-7}$ (as in the case of GC binary pulsars). 

\begin{figure}[h]
{\includegraphics[width=0.9\textwidth,
height=0.45\textheight]{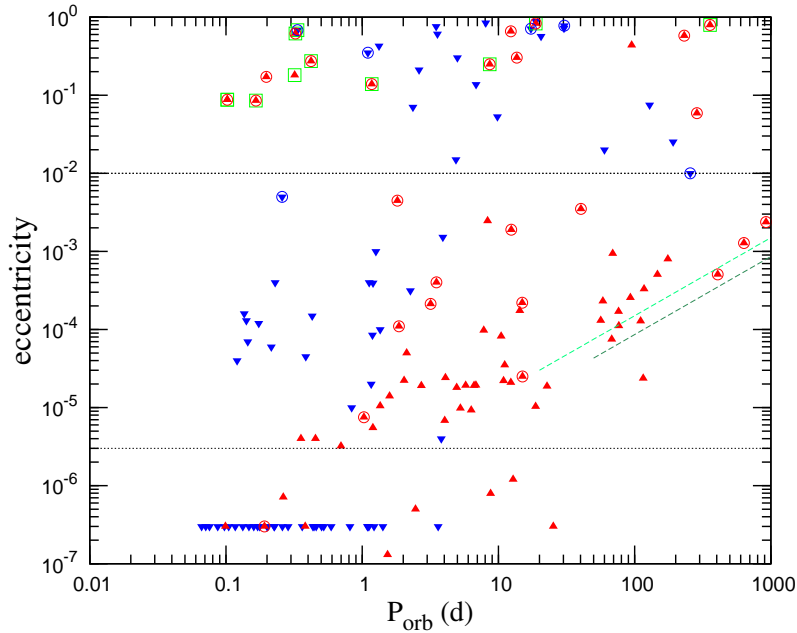}} 
\caption{\footnotesize{Plot of eccentricity vs
orbital periods for binary radio pulsars in and out globular
clusters. Blue $\blacktriangledown$s are for globular cluster
binaries and red $\blacktriangle$s are for disk binaries. Double
neutron star binaries are enclosed by green $\square$s. Non-millisecond pulsars are enclosed by $\bigcirc$s. Three disk binaries with $P_{orb} > 1000$ days have been excluded. The dashed lines in the right side represents the eccentricity-orbital period relation as predicted by Phinney \cite{phi92}
- the upper line is the original line drawn by Phinney \cite{phi92} and the lower line is using population II models. The black dotted lines parallel to the x-axis have been drawn to distinguish three groups of GC pulsars.}} \label{fig:porbecc_gc_gal}
\end{figure}

In Fig \ref{fig:porbecc_gc_gal}, we plot $e$ Vs $P_{orb}$ for all binary pulsars with known orbital solutions. Blue $\blacktriangledown$s are for GC binaries and red $\blacktriangle$s are for disk binaries. Three disk binaries with $P_{orb} > 1000$ days have not been plotted here as we have selected the range of $P_{orb}$ up to 1000 days in this plot. Here we take upper values of eccentricities as the actual values. DNSs are enclosed by green $\square$s. Non-millisecond pulsars are enclosed by $\bigcirc$s. It is believed that millisecond pulsars in binary systems formed from mass and angular momentum transfer due to Roche lobe overflow and the resultant tidal effects should appear in $e \sim 10^{-6} - 10^{-3}$ \cite{phi92}. Since many highly eccentric binary millisecond pulsars are found in GCs, this indicates that stellar interactions are important for inducing higher eccentricities which will be discussed later in this section. On the other hand, PSR J$1903+0327$ is the only one eccentric millisecond pulsar in the disk and there are lots of works in the recent past to explain the eccentricity of this system \cite{cha08, gop09, fre09}. We shall discuss about this system in somewhat details in section \ref{sec:triple}.

Observational selection effects may be influencing the
distribution seen in this figure for the two sets \cite{cam05},
the most important selection effect being operative towards the
left of the diagram, namely, it is more difficult to detect
millisecond pulsars in very short orbital period and highly
eccentric binaries; distance to the pulsars also is an important
selection effect, since at large distance only the brightest
pulsars can be observed. Nevertheless, it is clear that there is
a large abundance of pulsars with long orbital periods and
intermediate eccentricity near the eddy ``fluctuation dissipation"
lines among the galactic disk pulsars which are absent in
the GC pulsar population, where the important selection effects
are unlikely to play a major role.  The observed preponderance of DNSs in the disk with the very short
orbital periods and large eccentricities may be related to selection effects, but there is at least one DNS in GC - PSR J2127+1105C in M15 in the same phase-space. The disk DNSs are by and
large at small distances while M15 and other host GCs are usually far away.

\begin{figure}[h]
{\includegraphics[width=0.6\textwidth,
height=0.35\textheight, angle=-90]{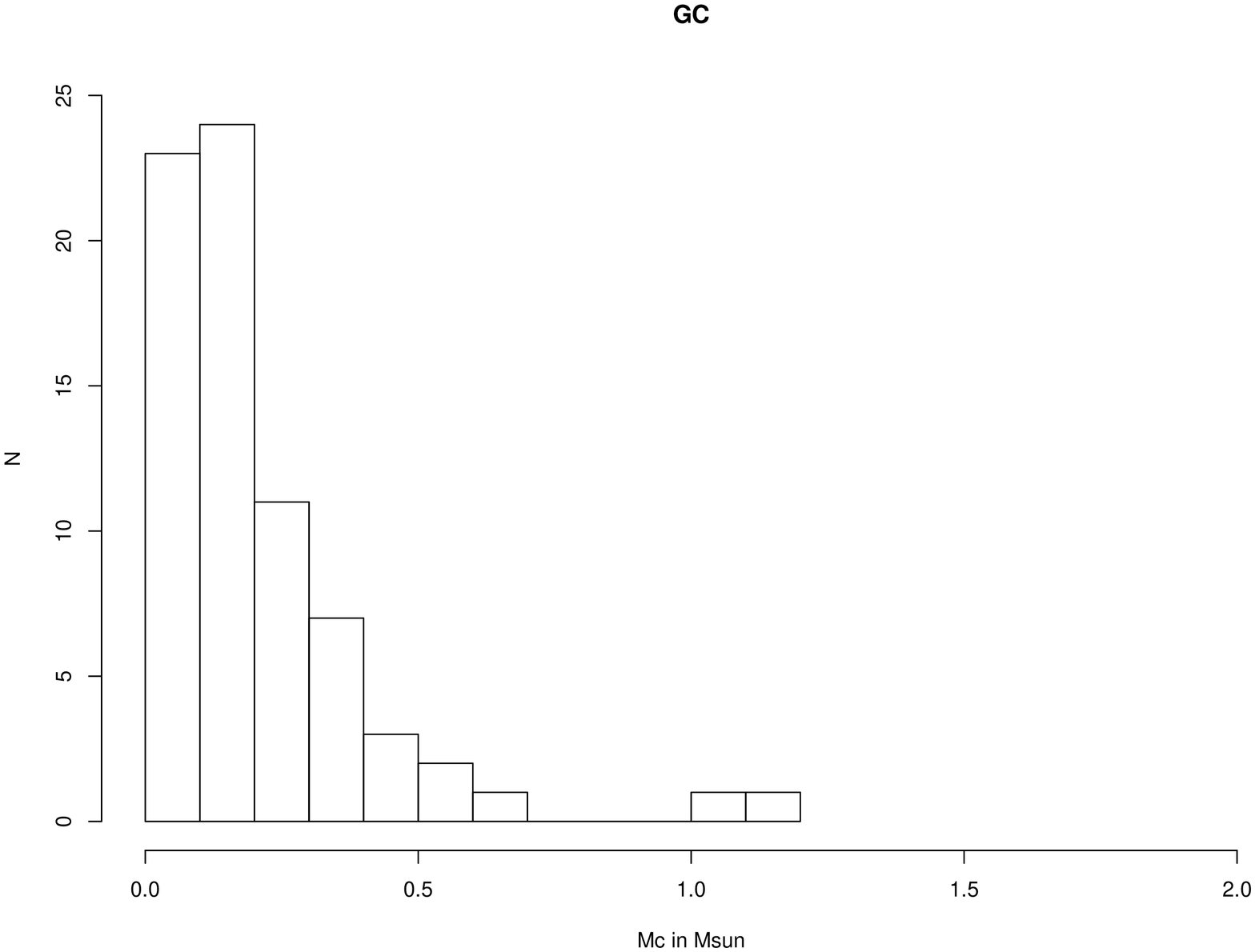}} 
{\includegraphics[width=0.6\textwidth,
height=0.35\textheight, angle=-90]{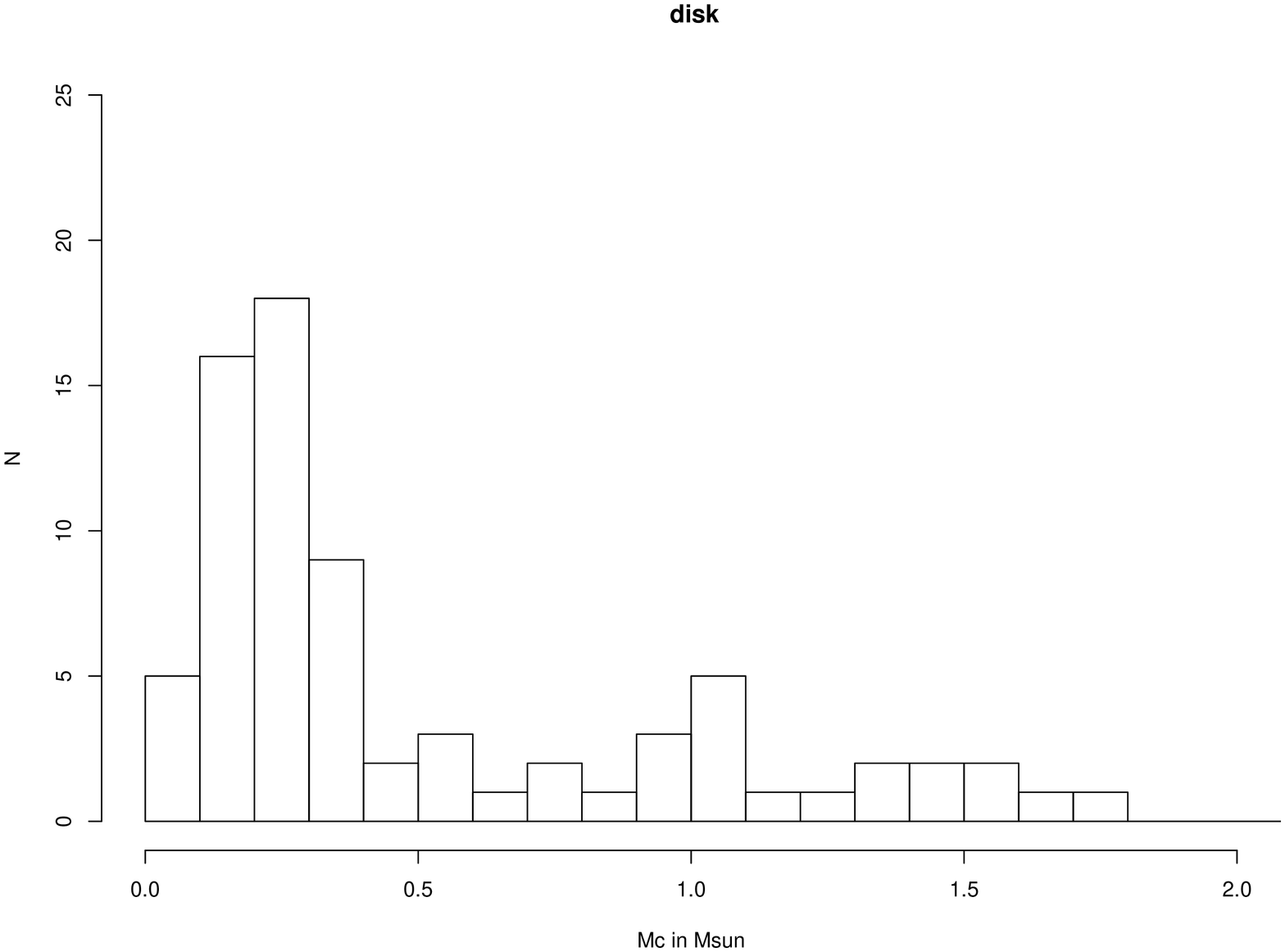}} 
{\includegraphics[width=0.6\textwidth,
height=0.35\textheight, angle=-90]{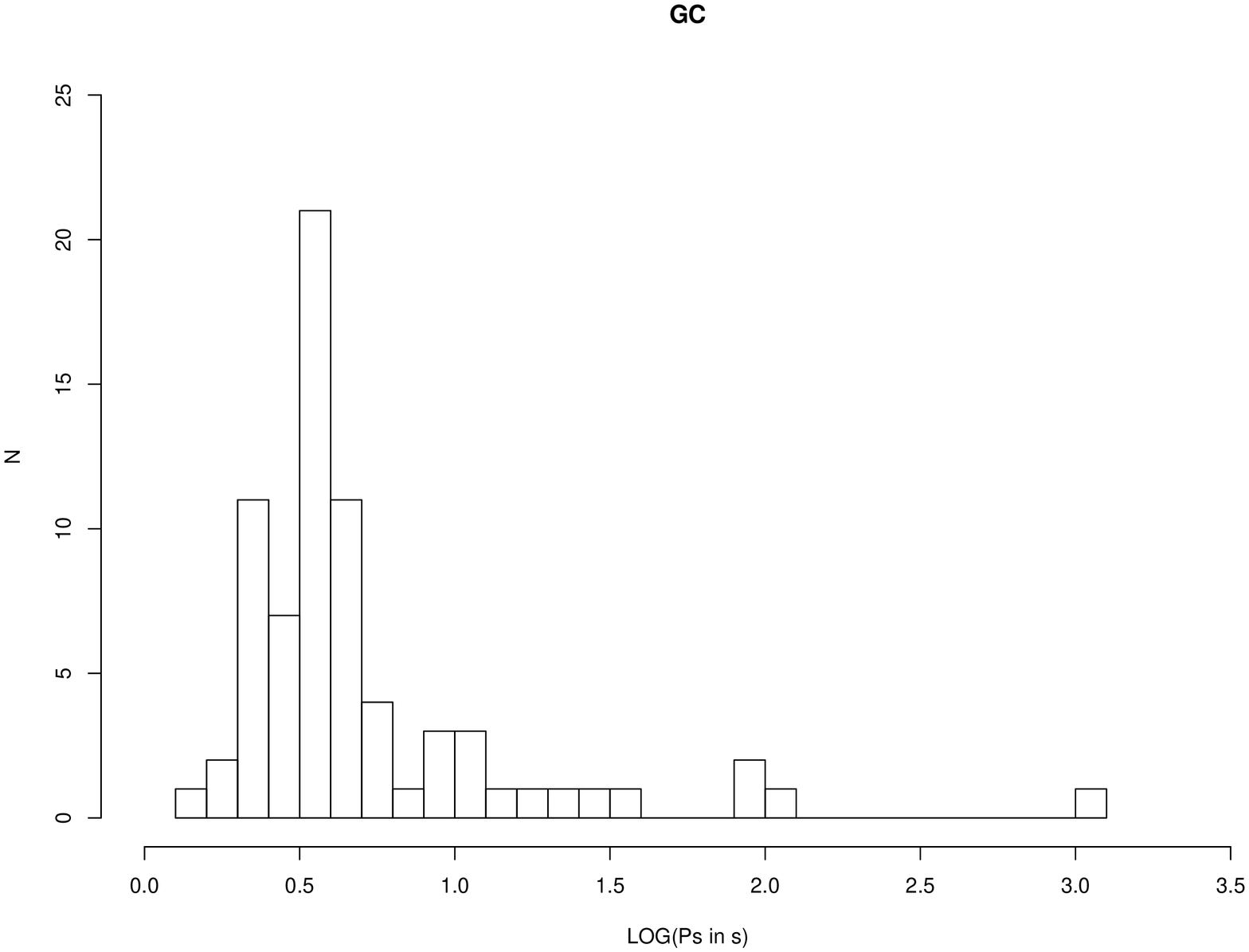}} 
{\includegraphics[width=0.6\textwidth,
height=0.35\textheight, angle=-90]{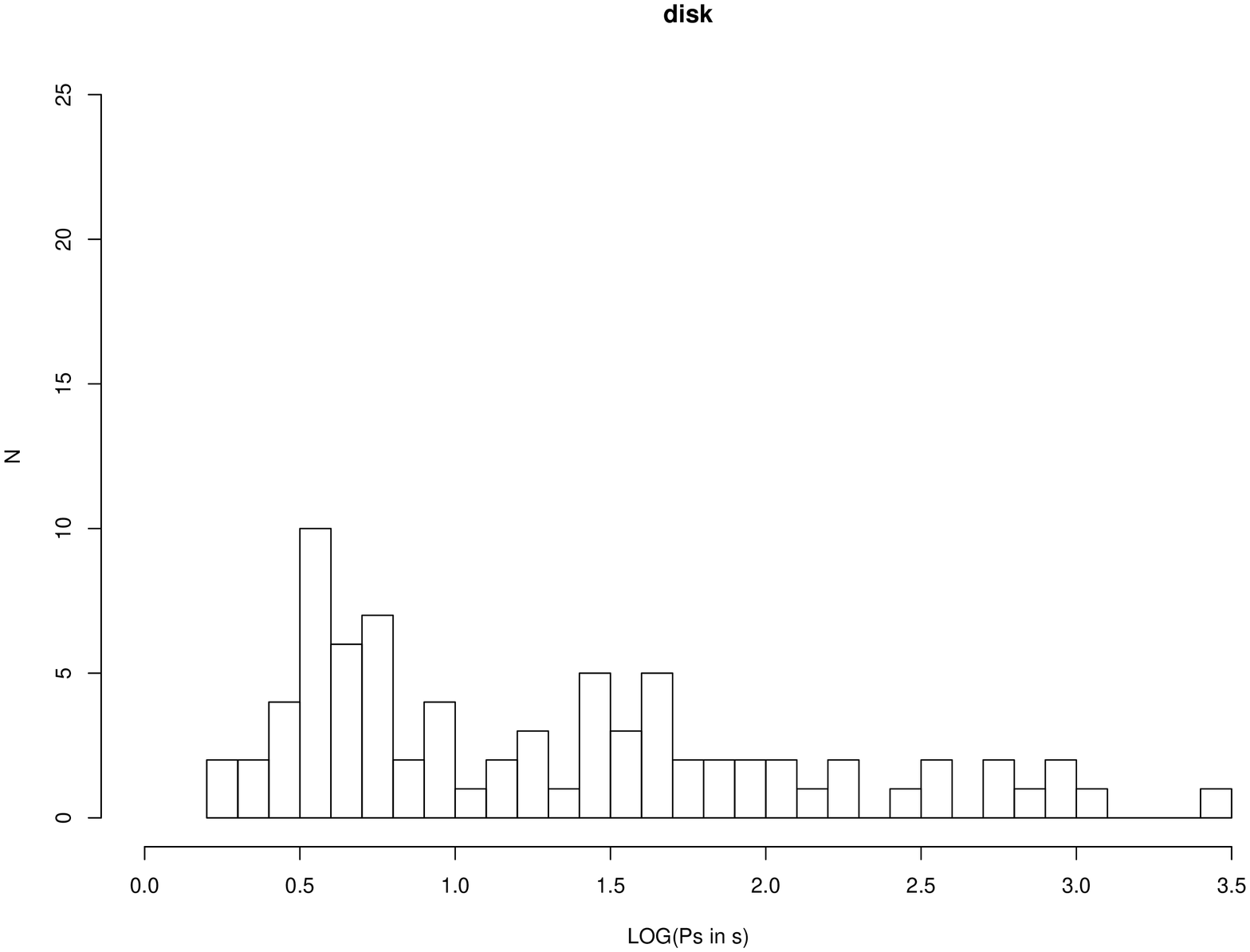}} 
\caption{\footnotesize{The distribution of companion masses for the binary radio pulsars (top) and the distribution of spin periods for the binary radio pulsars (bottom). Left panels are for the GC binaries and the right panels are for the disk binaries.}} \label{fig:mc_ps_hist}
\end{figure}

In the top panel of Fig. \ref{fig:mc_ps_hist} we compare the distribution of $m_c$s for the GC and the disk binaries and found that there are more low mass companions for GC binaries. In the bottom panel of Fig. \ref{fig:mc_ps_hist} we compare the distribution of $P_s$s for the GC and the disk binaries and found that in GCs, most are millisecond pulsars while for disk pulsars, $P_s$ are more uniformly distributed (not completely, more pulsars with lower $P_s$) over a wide range. 

\clearpage

\clearpage

\begin{figure}[h]
{\includegraphics[width=0.6\textwidth,
height=0.35\textheight]{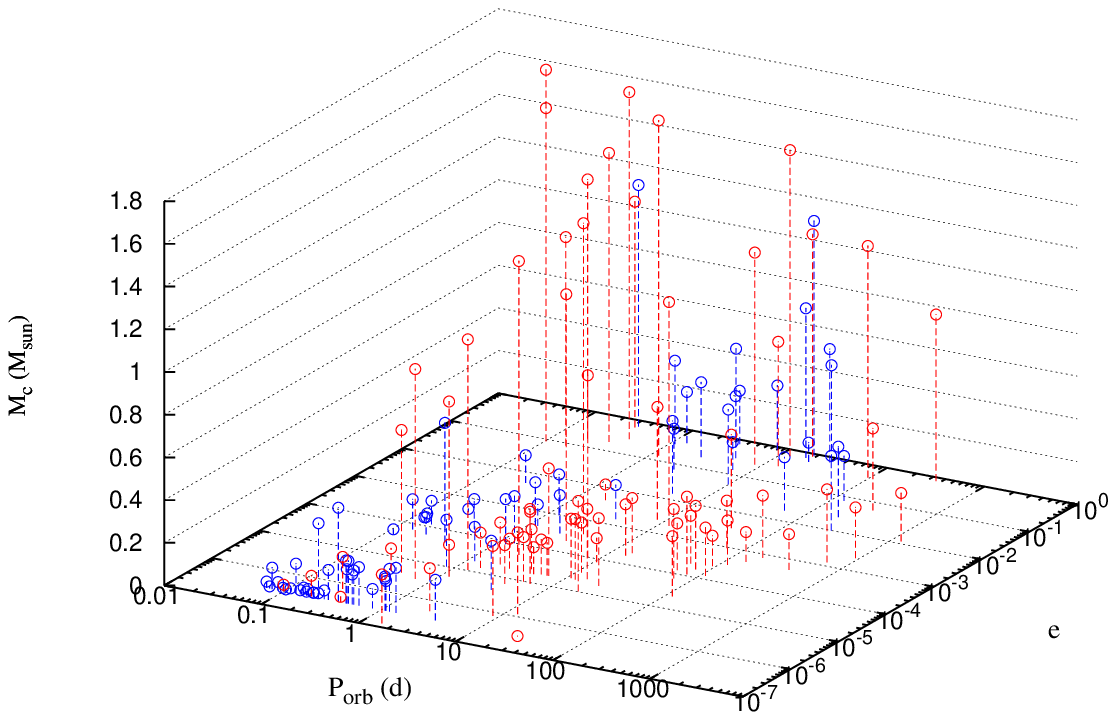}} 
{\includegraphics[width=0.6\textwidth,
height=0.35\textheight]{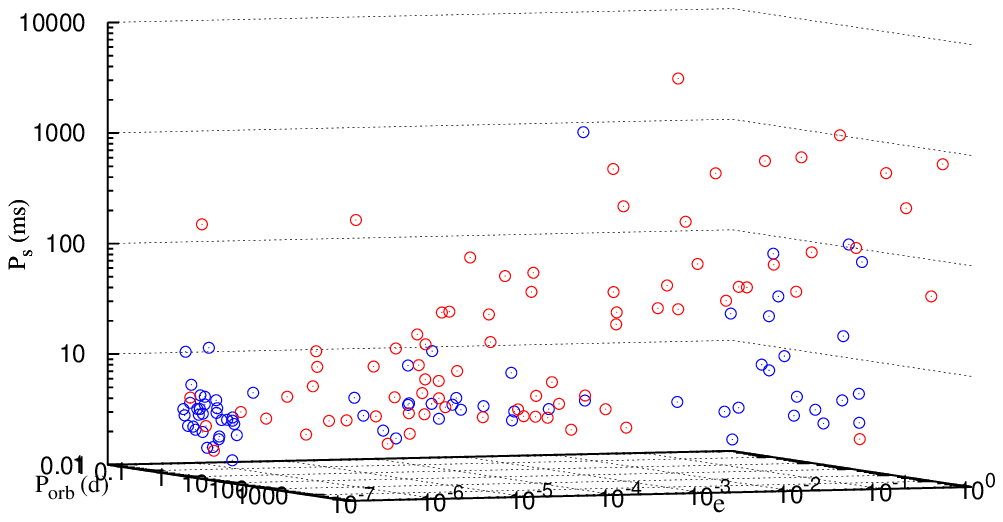}} 
\caption{\footnotesize{Left panel shows the  variation of companion masses with $P_{orb}$ and $e$, blue points are for GC binaries and red points are for disk binaries omitting three disk binaries with $m_c > 2.0~ M_{\odot}$(see table \ref{tab:diskpsr_parms}) for the ease of comparison. Right panel shows the variation of $P_{s}$ with $P_{orb}$ and $e$, blue points are for GC binaries and red points are for disk binaries. Note the presence of eccentric millisecond binaries in GCs.}} \label{fig:3dplots}
\end{figure}

In the left panel of Fig. \ref{fig:3dplots}, we plot $P_{orb}$, $e$ and $m_c$ for all binaries while in the right panel we plot $P_{orb}$, $e$ and $P_s$. In both cases, blue points represent GC binaries and red points represent disk binaries. Three disk binaries with $m_c > 2.0~ M_{\odot}$(see table \ref{tab:diskpsr_parms}) have been omitted from the plot in the left panel, even then we can see that companion masses of disk binaries are usually higher than those of GC binaries. Nearly circular binary pulsars have low mass companions.

It was proposed that orbital parameters of binary pulsars with white-dwarf companions satisfy the following relation \cite{rap95}:
\begin{equation}
P_{orb}\,(1-e)^{3/2}=0.374\, \left[ \frac{4950 \, m^{4.5}_{c}}{1+4 \, m^{4}_{c}}+0.5 \right]^{3/2}\,m^{-1/2}_{c}
\label{eq:rappaport}
\end{equation}

\begin{figure}[h]
{\includegraphics[width=0.9\textwidth,
height=0.5\textheight]{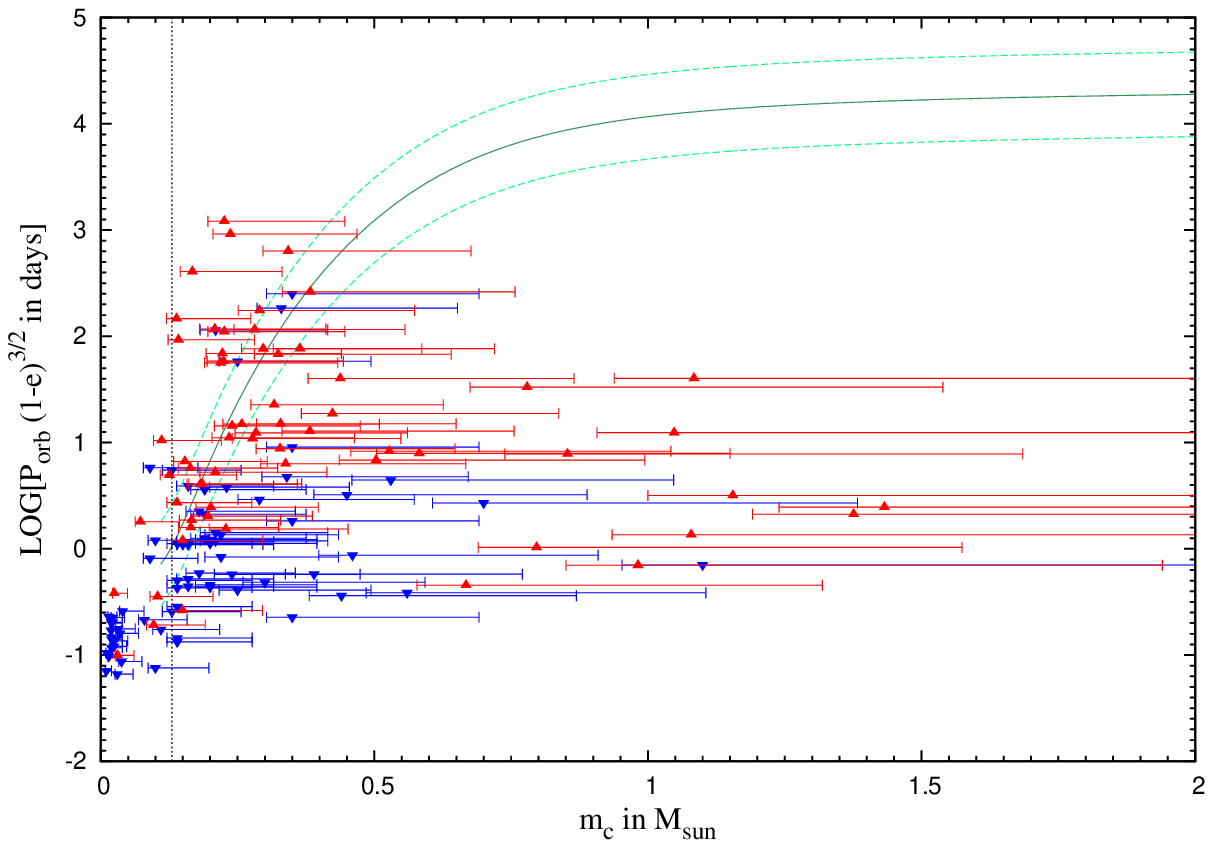}} 
{\includegraphics[width=0.9\textwidth,
height=0.5\textheight]{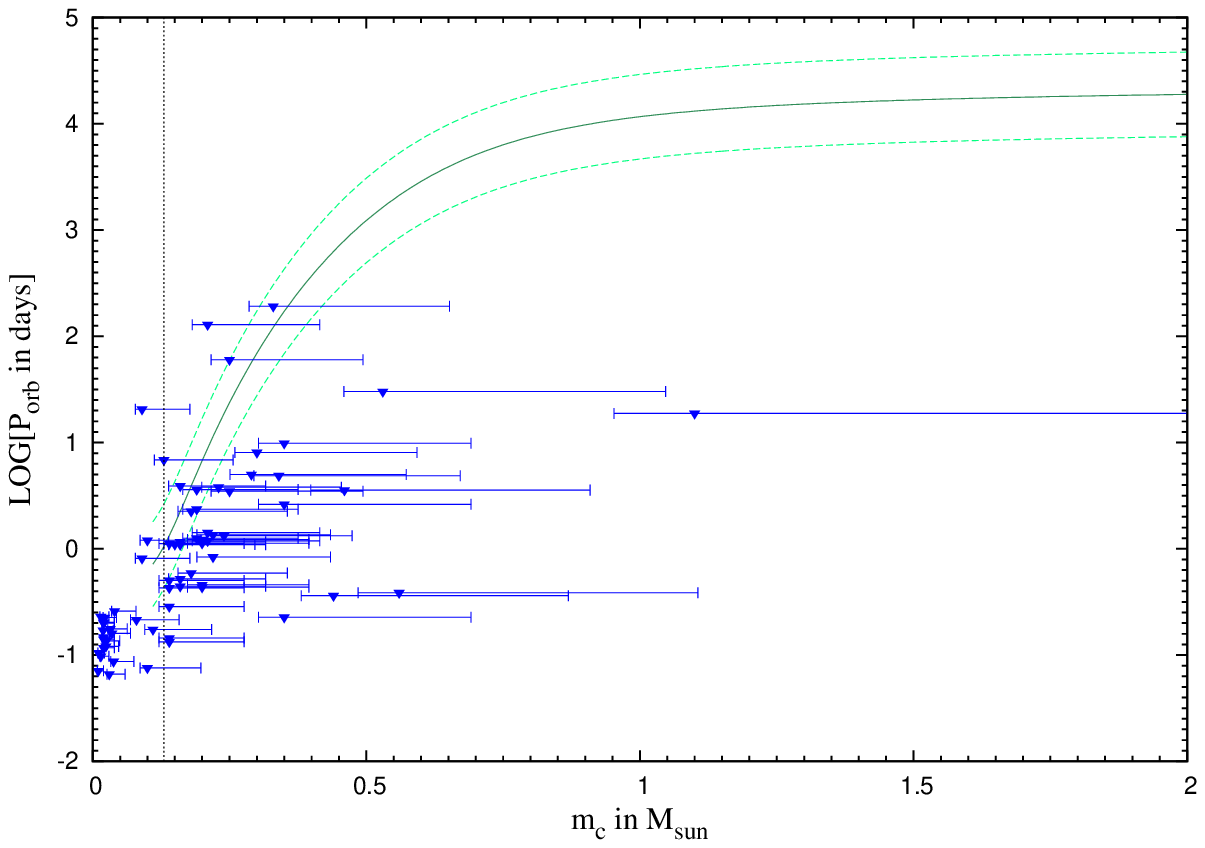}} 
\caption{\footnotesize{Upper Panel : Plot of $P_{orb}\,(1-e)^{3/2}$ against $m_c$ for binary pulsars in GCs (blue $\blacktriangledown$s) and in the disk (red $\blacktriangle$s). Lower Panel : Plot of $P_{orb}$ against $m_c$ for binary millisecond pulsars in GCs.
In both of the panels, DNSs and binaries with $m_c \geq 1.44~ M_{\odot}$ have been excluded. The points correspond to median values of $m_c$ taking $i=60^{\circ}$ while the upper and the lower limits correspond to $i=26^{\circ}$ and $i=90^{\circ}$ respectively. Theoretical prediction following Eqn. (\ref{eq:rappaport}) or Eqn. (\ref{eq:rappaport_zero}) are the middle lines. The upper and lower lines are obtained by multiplying the right hand side of the Eqns. (\ref{eq:rappaport}, \ref{eq:rappaport_zero}) by 2.5 (see \cite{rap95}). Note that the model used to derive the equations is not valid when $m_{c} < 0.13 ~ M_{\odot}$, though we have plotted the binaries with $m_{c} < 0.13 ~ M_{\odot}$ (located left of the vertical black dotted line) in both the panels. }} \label{fig:rappaport}
\end{figure}

In the upper panel of Fig. \ref{fig:rappaport}, we plot $P_{orb}\,(1-e)^{3/2}$ against $m_c$ for binary pulsars in GCs and also in the galactic disk. DNS binaries and the binaries with $m_c \geq 1.44~ M_{\odot}$ (Chandrasekhar limit) have been excluded. GC binaries are shown with blue $\blacktriangledown$s and disk binaries are shown with red $\blacktriangle$s. The points correspond to median values of $m_c$ taking $i=60^{\circ}$ while the upper and the lower limits of $m_c$s correspond to $i=26^{\circ}$ and $i=90^{\circ}$ respectively. Theoretical prediction following Eqn. (\ref{eq:rappaport}) is the middle solid green line. The upper and lower lines are obtained by multiplying the right hand side of Eqn. (\ref{fig:rappaport}) by 2.5 (see \cite{rap95}). Note that the model used to derive Eqn. (\ref{eq:rappaport}) is not valid when $m_{wd} < 0.13 ~ M_{\odot}$, though we have plotted the binaries with $m_{c} < 0.13 ~ M_{\odot}$ (located left of the vertical black dotted line) just to see how $P_{orb}\,(1-e)^{3/2}$ is related with $m_c$ for these systems. For recycled (millisecond) pulsars, where significant amount of mass had been transferred to the pulsars in past, one should expect lower values of $m_c$ and $e \sim 10^{-6} - 10^{-3}$. for these cases the relation becomes 
\begin{equation}
P_{orb} = 0.374\, \left[ \frac{4950 \, m^{4.5}_{c}}{1+4 \, m^{4}_{c}}+0.5 \right]^{3/2}\,m^{-1/2}_{c}
\label{eq:rappaport_zero}
\end{equation}

The deviation of GC binary pulsars from the theoretical prediction of Eqn. (\ref{eq:rappaport}) is well understood as their eccentricities have been effected by stellar interactions. So for the millisecond pulsar binaries in GCs (which naturally exclude the DNS in M15), we plot Eqn. (\ref{eq:rappaport_zero}) in the lower panel of Fig. (\ref{fig:rappaport}). Even then we found significant discrepancy between the theoretical prediction and the observed values. Moreover, for binary pulsars in the disk (upper panel of Fig. \ref{fig:rappaport}) where eccentricities and orbital periods should bear the clear signature of stellar evolutions, we also found disagreement between the model and the observation. 

We summarize the relevant properties of the host GCs and in Table (\ref{tb:gc_v_n}). 
The properties of binary pulsars in globular clusters and in the galactic disk are given in tables (\ref{tab:gcpsr_parms}) and (\ref{tab:diskpsr_parms}) respectively. Interested readers are requested to see original sources for better precision of the given parameters and for the values of other parameters.

Through a combination of two-body tidal captures and exchange interactions where an isolated neutron star replaces a normal star in a binary, numerous neutron star binaries are formed in GCs \cite{pod02}. This explains why the ratio of binary pulsars to isolated pulsars is so high in GCs in comparison to that in the galactic disk. Tidal capture leads to tight binaries whereas exchange usually leads to wider binaries (see Eqns. \ref{eq:afin_ex}). If the mass transfer in the tidally captured neutron star binary happen in a stable manner, it can spin up the neutron star to a millisecond spin period. Thus the observed preponderance of millisecond pulsars in GC binaries can be attributed to the tidal capture process (details will be discussed in the next subsection). We explain the eccentricities of millisecond pulsar binaries in GCs as a result of different types of interactions of single stars with pulsar binaries using present properties of the GCs. Some of the results can be found in our earlier papers \cite{bag09a, bag09b}. For the sake of simplicity, we don't go for a full population synthesis study of the evolution of the globular clusters including all kinds of stellar interactions $e.g.$ binary-binary, binary-single star and single star - single star interactions. We consider only binary-single star interactions faced by binary pulsars preceded by a discussion on the formation of binary pulsars through tidal capture.

\clearpage

\clearpage

\subsection{Orbital Parameter Changes Due to Stellar Interactions
and Gravitational Radiation}
\label{sec:types_int}

Globular clusters contain a large number of LMXBs
compared to the galactic disk; this  had led to the suggestion
\cite{fab75} that a binary is formed by the tidal capture of a
non-compact star by a neutron star in the dense stellar
environment of the GC cores. If mass and angular momentum transfer from the companion to the neutron star took place in a stable manner, this could lead to recycled pulsars in binaries or as single millisecond pulsars
\cite{alp82, rom87, ver87}. However large energies may be deposited in tides
after the tidal capture of a neutron star by a low mass MS star in a GC. The resultant structural readjustments of the star in response to the dissipation of the modes could be very
significant in stars with either convective or radiative damping
zones \cite{ray87, mcm87} and the companion star can undergo a size ``inflation"
due to its high tidal luminosity  which may exceed
that induced by nuclear reactions in the core, by a large factor. Efficiency of
viscous dissipation and orbit evolution is crucial to the
subsequent evolution of the system as viscosity regulates the
growth of oscillations and the degree to which the extended star is
bloated and shed. The evolution of the system could lead to
either a merger (leading to a Thorne Zytkow object) or a neutron
star surrounded by a massive disk comprising of the stellar
debris of its erstwhile companion or, in other cases, an LMXB or even a detached binary following envelope
ejection. Tidal oscillation energy can return to
the orbit (i.e. the tide orbit coupling is a dynamical effect)
thus affecting the orbit evolution or the extent of dissipative
heating in the less dense companion. The evolutionary transition
from initial tidal dynamics to any likely final, quiescent binary
system is thus regulated by viscosity. Tide orbit coupling can
even lead to chaotic evolution of the orbit in some cases
\cite{mar95} but in the presence of dissipation with non-liner
damping, the chaotic phase may last only for about 200 yrs after
which the binary may undergo a periodic circularization and after
about 5 Myrs finish circularizing \cite{mar96}. Of the stars that
do not directly lead to mergers, a roughly equal fraction of the
encounters lead to binaries that either become unbound as a
result of de-excitation or heating from other stars in the
vicinity or they are scattered into orbits with large pericenters
(compared to the size of the non-compact star) due to angular
momentum transfer from other stars \cite{koc92}. Thus ``intermediate pericenter" encounters can lead to widened orbit (even 100 days orbital period in an eccentric orbit) in tidal
encounters involving main sequence stars and neutron stars, which
in the standard scenario are attributed to encounters between
giants and neutron stars. The complexities of the dynamics of
tidal capture and subsequent evolution are manifold and attempts
to numerically simulate collisions of neutron stars with red
giants have been made with 3D Smoothed Particle Hydrodynamics
(SPH) codes \cite{ras91, dav91}. We note in this context that Podsiadlowski $et~al.$ \cite{pod02} have stated ``it is not only premature to rule out tidal capture as a formation scenario for LMXBs, but that LMXBs in globular clusters with well determined orbital periods
actually provide observational evidence in its favor". The
different formation channels of pulsars in binaries in GCs and
the birthrate problems of millisecond pulsar binaries vs LMXBs
were considered in \cite{ray90,kul90}.

Until the early eighties there was no evidence of a substantial population of primordial
binaries in any globular clusters, and it was even thought that GCs are significantly
deficient in binaries compared to a younger galactic population \cite{hut92}. Theoretical modeling of GCs was often started off as if all stars were singles. During the 1980s several observational techniques began to yield a rich population of binary stars in GCs \cite{bas08, yan94,pry89}. When the local binary fraction is substantial, the single star - binary interaction can exceed the encounter rate between single stars by a large factor \cite{sig93}. The existence of a significant population of primordial binary population
in GCs indicated that three body processes have to be accounted for in any dynamical
study of binaries involving compact stars. An encounter between an isolated star and a binary
may lead to a change of state of the latter, e.g.: i) the original binary may undergo a change of eccentricity and orbital period but otherwise remain intact -- a ``preservation" or a ``fly-by" process; ii) a member of the binary may be exchanged with the incoming field star, forming a new binary -- an ``exchange" process;
iii) two of the stars may collide and merge into a single object, and may or may not remain bound to the third star - a ``merger" process; iv) three of the stars may collide
and merge into a single object - a ``triple merger" process;
v) all three stars may become unbound - an ``ionization" process.
The ionization is always a ``prompt" process, whereas the others can be
either prompt or ``resonant" process. In resonant processes, the stars
form a temporarily bound triple system, with a negative total energy but
which decays into an escaping star and a binary, typically after 10-100
crossing times.

The value of the binding energy of the binary and the velocity of the incoming star determine the type of interaction the binary will encounter. The critical velocity $v_c$ of the incoming star, for which the total energy (kinetic plus potential) of the three body system
is zero, can be defined as
\begin{equation}
v_c^2=G \frac{m_1 m_2 (m_1+m_2+m_3)}{m_3(m_1+m_2)}\frac{1}{a_{in}}
\label{eq:vc}
\end{equation}
where $m_1$ and $m_2$ are the masses of the binary members, $m_3$ is the mass of the incoming star, $a_{in}$ is the semi-major axis of the binary and G is the gravitational constant.

In case of a binary-single star interaction, semi major axis of the final binary ($a_{fin}$) is related to the semi major axis of the initial binary ($a_{in}$) as follows \cite{sig93} :

\begin{equation}
a_{fin}=\frac{1}{1-\Delta}\frac{m_a m_b}{m_1 m_2} ~ a_{in}
\label{eq:afin_gen}
\end{equation}
where $m_1$ and $m_2$ are masses of the members of the initial
binary, $m_a$ and $m_b$ are masses of the members of the final
binary. $\Delta$ is the fractional change of binary binding
energy $i.e.$ $\Delta = \left(E_{in}-E_{fin}\right)/E_{in}$. The
binding energies of the initial and final binaries are
\begin{equation}
E_{in} = -G \frac{m_1 m_2}{2 a_{in}}, ~~~~ E_{fin} = -G \frac{m_a
m_b}{2 a_{fin}}
\end{equation}
For fly-by, $m_a = m_1$, $m_b = m_2$
giving
\begin{equation}
a_{fin}=\frac{1}{1-\Delta}~a_{in} \label{eq:afin_fly}
\end{equation}
For exchange, $m_a = m_1$, $m_b = m_3$
(star 2 is being replaced by star 3) giving

\begin{equation}
a_{fin}=\frac{1}{1-\Delta}\frac{m_1 m_3}{m_1 m_2} ~ a_{in}
=\frac{1}{1-\Delta}~\frac{m_3}{m_2}~ a_{in}\label{eq:afin_ex}
\end{equation}
For merger, $m_a = m_1$, $m_b = m_2 +
m_3 $ (star 2 is being merged with star 3) giving

\begin{equation}
a_{fin}=\frac{1}{1-\Delta}\frac{m_1 (m_2+m_3)}{m_1 m_2}~ a_{in}
=\frac{1}{1-\Delta}\frac{m_2+m_3}{m_2}~ a_{in}
\label{eq:afin_merg}
\end{equation}
The actual value of $\Delta$ is not known. Putting $\Delta~=~0$
simplifies equations (\ref{eq:afin_gen}), (\ref{eq:afin_fly}),
(\ref{eq:afin_ex}) and (\ref{eq:afin_merg}) giving $a_{fin} =
a_{in}$ for fly-by interactions. Similar expressions can be
derived for other cases $e.g.$ star 1 is being replaced by star 3
or star 1 being merged with star 3. But in the present study, we
always assume star 1 to be the neutron star, so these processes are irrelevant. Merger interactions which 
do not result in binary systems are also irrelevant for this work.

Interactions between binaries and single stars can enhance the
eccentricity of the binary orbit and this will be discussed in detail in the
subsequent sections. On the other hand, binary pulsars
emit gravitational waves which reduce both the sizes and eccentricities of the orbit
leading to mergers of the binary members on a timescale of
$t_{gw}$ which can be calculated as follows \cite{petmath63} :

\begin{equation}
t_{gw}= \frac{e}{\dot{e}_{gw}}
\label{eq:tgr_e}
\end{equation}
where
\begin{equation}
\dot{e}_{gw}=-\frac{304}{15}\frac{G^3 m_p m_c (m_p + m_c)}{c^5
a^4}g(e) \label{eq:dedtgr}
\end{equation}
and
\begin{equation}
g(e)=\left(1-e^2 \right)^{-5/2} e \left(1+\frac{121}{304}e^2
\right) \label{eq:ge}
\end{equation}
which is almost same as the expression of $t_{gw}$  calculated
from $\dot{a}_{gw}$.

\begin{equation}
t_{gw}=  \frac{a}{\dot{a}_{gw}}
\label{eq:tgr_a}
\end{equation}
\begin{equation}
\dot{a}_{gw}=-\frac{64}{5}\frac{G^3 m_p m_c (m_p + m_c)}{c^5
a^3}f(e) \label{eq:dadt}
\end{equation}
and
\begin{equation}
f(e)=\left(1-e^2 \right)^{-7/2} \left(1+\frac{73}{24}e^2 +
\frac{37}{96}e^4 \right) \label{eq:fe}
\end{equation}

In this work, we use eq. (\ref{eq:tgr_e}) to calculate $t_{gw}$.

\subsubsection{Fly-by interactions}
\label{sec:fly_by}

Figs. \ref{fig:porbecc_gc}, \ref{fig:porbecc_gc_gal} and Table \ref{tab:gcpsr_parms} show that most of the binary pulsars in GCs are millisecond pulsars (69 pulsars out of 73 have $P_s< 30$ ms, one has $P_s = 33.16$ ms and the other three having $P_s$ values as 80.34 ms, 111.61 ms and 1 s). Theoretically one expects that spun-up, millisecond pulsars in binary systems formed from mass and angular
momentum transfer due to Roche lobe overflow and the resultant
tidal effects should appear in low eccentricity orbits $e \sim
10^{-6} - 10^{-3}$ \cite{phi92}. Since many highly eccentric
binary millisecond pulsars are found in globular clusters (see the right panel of Fig \ref{fig:3dplots}), this
indicates that stellar interactions are important in GCs for inducing
higher eccentricities. Rasio and Heggie \cite{ras95a,heg96}
studied the change of orbital eccentricity ($\delta e = e_{fin} - e_{in}$ where $e_{in}$ is the initial eccentricity and $e_{fin}$ is the final eccentricity which is observable at present) of an initially
circular binary ($e_{in} = 0$) following a distant encounter with a third star
in a parabolic orbit. They used secular perturbation theory, i.e.
averaging over the orbital motion of the binary for sufficiently
large values of  the pericenter distance $r_p$ where the encounter is quasi-adiabatic and used
non secular perturbation theory for smaller values
 of $r_p$ where the encounter is non-adiabatic. In the first case
$\delta e$ varies as a power law with $r_p/a_{in}$ and in the second case
$\delta e$ varies exponentially with  $r_p/a_{in}$. The power law dominates
for $e_{fin}< 0.01$ wheras the exponential law dominates for $e_{fin} \gtrsim 0.01$. The timescales for these processes are \cite{ras95a} (where we write $e$ in place of $e_{fin}$ for the sake of simplicity) :

\begin{subequations}
\begin{eqnarray}
t_{fly}=4 \times 10^{11} n_4^{-1}v_{10}P_{orb}^{-2/3}e^{2/5}~ \rm {(e \lesssim 0.01) }\label{eq:tfly1}  \\
t_{fly}=2 \times 10^{11}
n_4^{-1}v_{10}P_{orb}^{-2/3}\left[-ln(e/4) \right]^{-2/3} ~\rm
{(e \gtrsim 0.01)}  \label{eq:tfly2}
\end{eqnarray}
\end{subequations}
where $P_{orb}$ is in days and $t_{fly}$ is in years. The values of $v_{10}/n_{4}$ are different for different GCs (table \ref{tb:gc_v_n}) and we grouped them into six different groups according
to their values of $v_{10}/n_{4}$ and calculated $t_{fly}$ for mean values
of $v_{10}/n_{4}$ for each group.  The unit of $v_{10}/n_{4}$ is $10^{-3}~{\rm km~s^{-1}~pc^{3}}$ which we will not mention explicitly anymore.

\clearpage

\begin{figure}[h]
{\includegraphics[width=0.5\textwidth,
height=0.3\textheight]{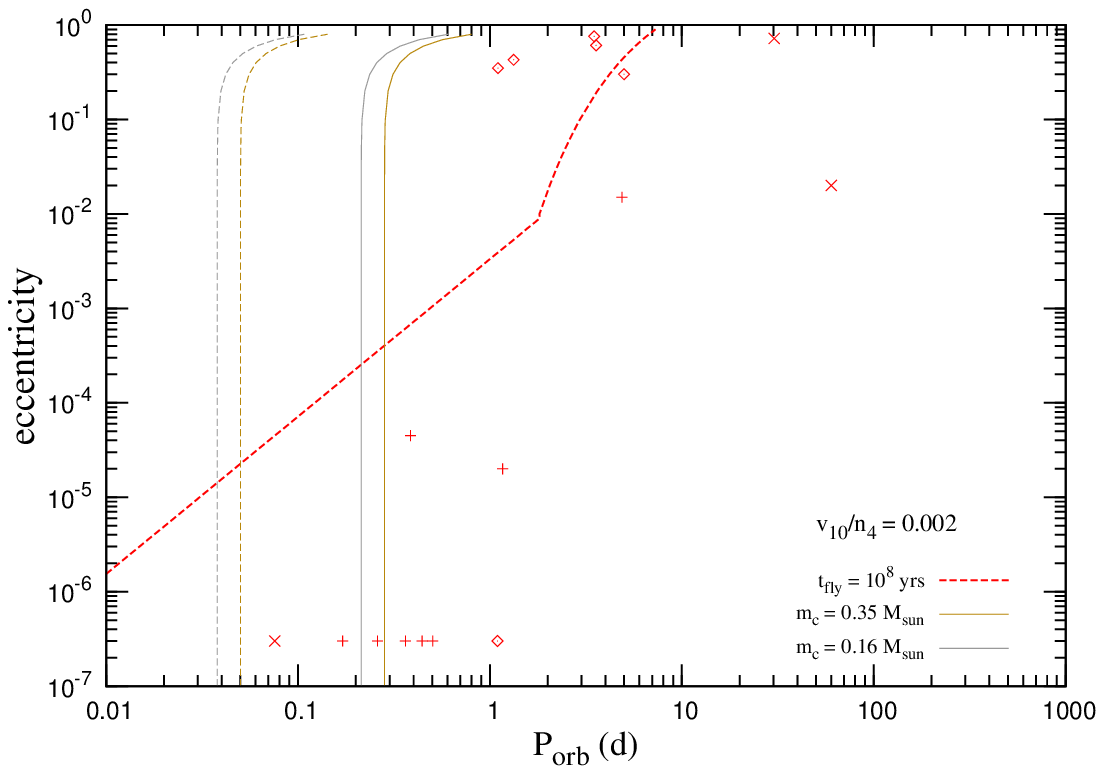}}
{\includegraphics[width=0.5\textwidth,
height=0.3\textheight]{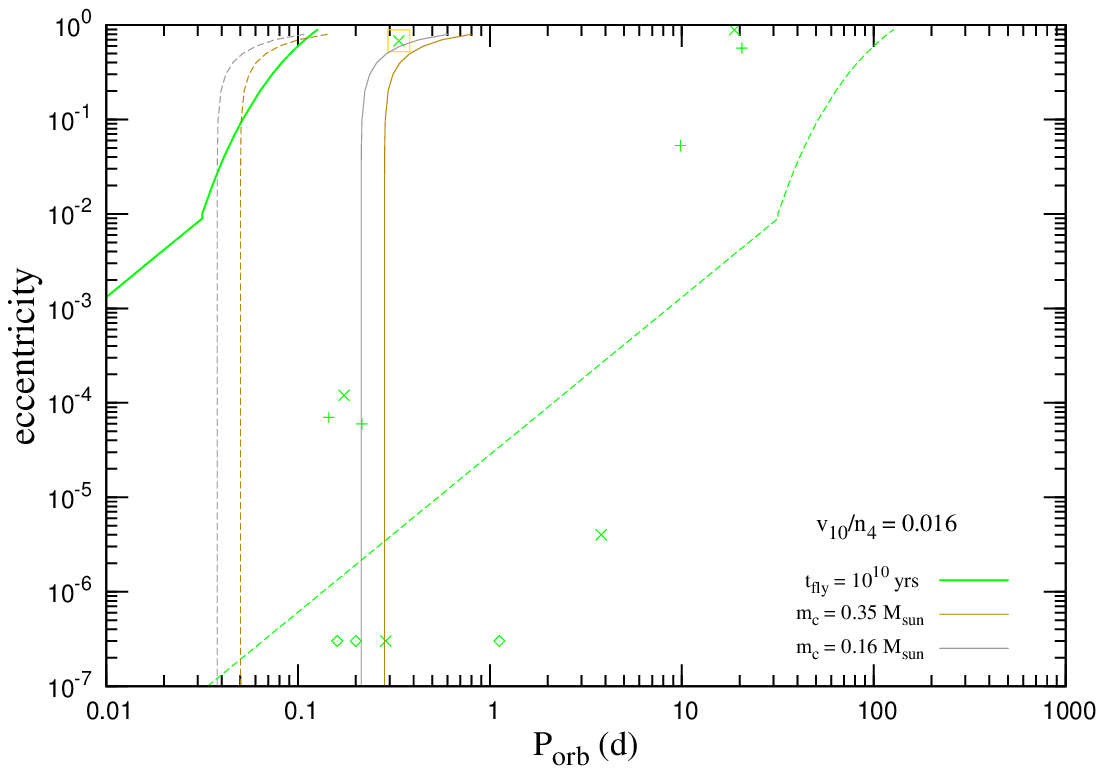}}
{\includegraphics[width=0.5\textwidth,
height=0.3\textheight]{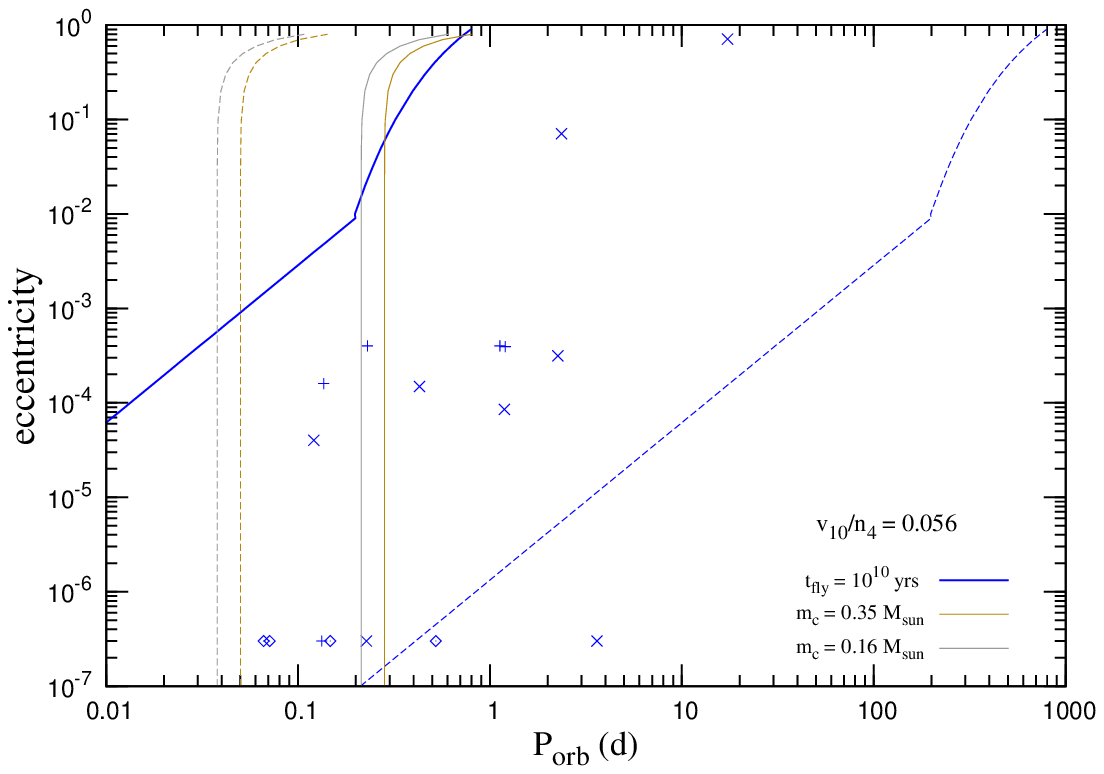}}
{\includegraphics[width=0.5\textwidth,
height=0.3\textheight]{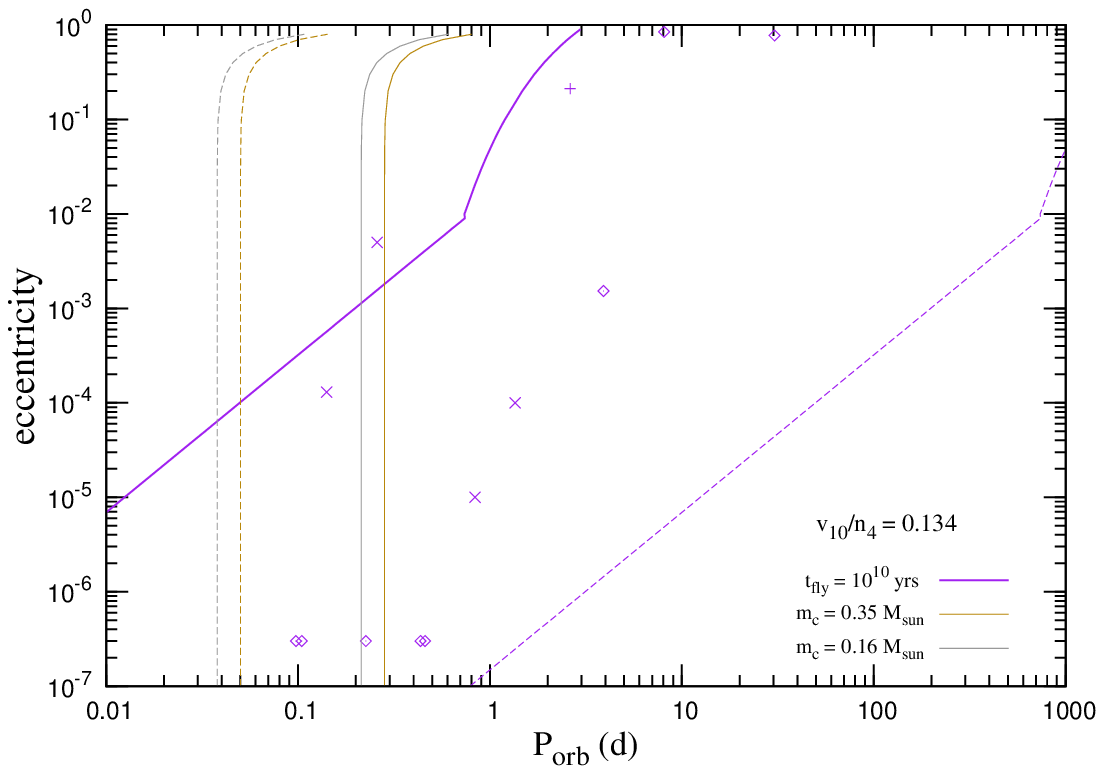}}
{\includegraphics[width=0.5\textwidth,
height=0.3\textheight]{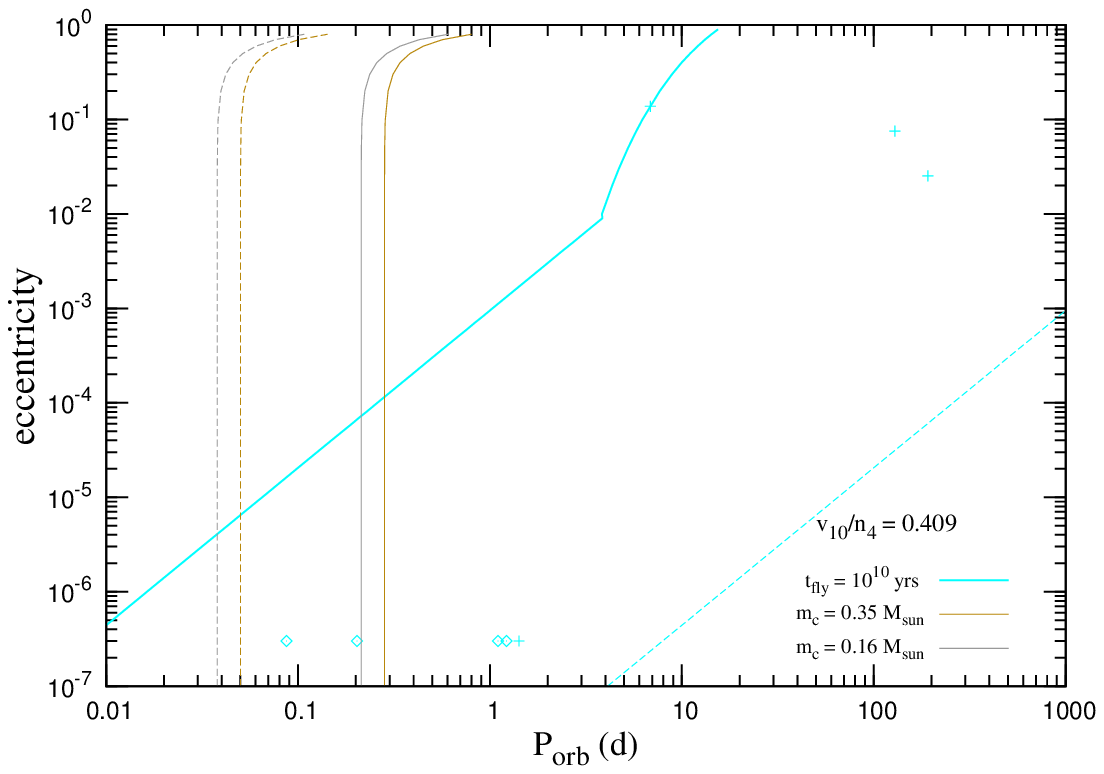}}
{\includegraphics[width=0.5\textwidth,
height=0.3\textheight]{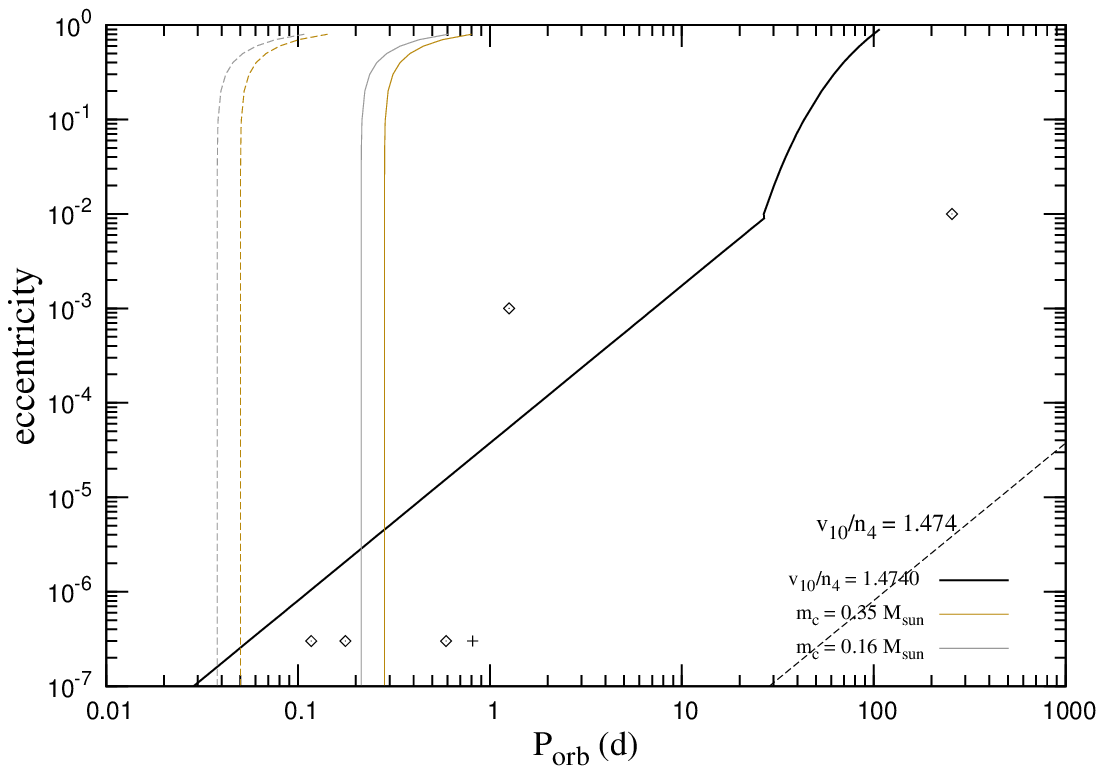}} \caption{\footnotesize{Cluster binary pulsars
in the $e-P_{orb}$ plane with contours of $t_{fly}=10^{10}$ yrs
(solid lines) and $t_{fly}=10^{8}$ yrs (dashed lines) for each
group. Contours of $t_{gw}=10^{10}$ yrs (solid lines) and
$t_{gw}=10^{8}$ yrs (dashed lines) for binaries with $m_p~=~1.4~
M_{\odot}$ and $m_c~=~0.35~ M_{\odot}$ or $0.16~ M_{\odot}$ are
also shown. Pulsars with projected positions inside the cluster
core are marked with $+$, those outside the cluster core are
marked with $\times$ and the pulsars with unknown position
offsets are marked with $\diamond$. Solid red line corresponding
to $t_{fly}=10^{10}$ years for $v_{10}/n_{4}= 0.0024$ is outside
the range plotted. Pulsars in a particular group are marked with the same color as of the $t_{fly}$ contours for that group. The upper limits of eccentricities are taken as the actual values of eccentricities.}}
\label{fig:psr_all_group_rasio_fly}
\end{figure}

\clearpage

Cluster binary pulsars in the $e-P_{orb}$ plane with contours
of $t_{fly}=10^{10}$ yrs (solid lines) and $t_{fly}=10^{8}$ yrs (dashed lines)
for each group are shown in Fig \ref{fig:psr_all_group_rasio_fly}.
Contours of $t_{gw}=10^{10}$ yrs (solid lines) and $t_{gw}=10^{8}$ yrs (dashed lines) for binaries with $m_p~=~1.4~ M_{\odot}$ and $m_c~=~0.35~ M_{\odot}$ or $0.16~ M_{\odot}$ are also shown. Pulsars with projected positions inside the cluster cores are marked with $+$, those outside the cluster cores are marked with $\times$ and the pulsars with unknown position offsets are marked with $\diamond$. Solid red line corresponding to $t_{fly}=10^{10}$ years for $v_{10}/n_{4}= 0.0024$ is outside the range of axes used in our plots (away from the upper left corner). Pulsars in a particular group are marked with the same color as of the $t_{fly}$ contours for that group. The upper limits of eccentricities are taken as the actual values of eccentricities. If a pulsar is located on the upper left half of the corresponding $t_{fly}=10^{10}$ years line $i.e.$ in the region where $t_{fly}>10^{10}$ years, then its eccentricity cannot be due to fly-by interactions. There are two such binaries,
one is PSR B1718-19\footnote{Note however that its cluster association
is sometimes doubted \cite{cam05}, as it has a large offset from the GC center.}
(in NGC 6342, in group 4) and the other is PSR B 1639+36B (in M13, in group 6).
But the first one is a normal pulsar with $P_{s} = 1$ s and as it is not spun up and may not
have been subject to a great deal of tidal forces, it can have escaped
circularization. On the other hand, the second pulsar is a millisecond pulsar with
$P_{s} = 3.528$ ms, so it should have been in a circular orbit if it has not
had the time to go through any significant kind of stellar interactions.
Therefore it is of interest to investigate if it is a result of an exchange or
merger interaction. Moreover, some of the eccentric pulsars which seem to be
explainable by fly-by interactions lie outside of globular cluster cores (where the stellar density is comparatively low and hence fly-by interactions are less efficient). For a few others positional offsets are not known. Even if a pulsar appears to be inside the cluster core in the projected image, it can still be outside the cluster core in the three dimensional space. GC models show that values of
both  $v_{10}$ and $n_4$ fall outwards from the cluster center and as the fall of
$n_4$ is much more rapid, the value of $v_{10}/n_4$ is higher outside the
cluster core. As we have already seen (different panels in Fig.
\ref{fig:psr_all_group_rasio_fly}) that an increase in the value of
$v_{10}/n_4$ shifts the contour of $t_{fly}~=~10^{10}$ years rightwards, this
will make the pulsar fall in the region where $t_{fly}~>~10^{10}$ years excluding the possibility of fly-by interaction as the cause of its eccentricity. In those cases also, we need to think about exchange and/or merger interactions. 

We have seen that there are many zero eccentricity pulsars which
lie in the region where $t_{fly}~<~10^{10}$ years. So it is of interest
to understand why they still appear in circular orbits despite the possibility
of fly-by interactions. We will discuss about this point in subsection \ref{sec:zero}.

\subsubsection{Exchange and merger interactions}
\label{sec:starlab}

Since exchange and merger interactions have not been amenable to
analytic treatment one needs to use numerical techniques. We performed numerical scattering experiments using the STARLAB \footnote{www.ids.ias.edu/$\sim$starlab/} task ``sigma3" which gives the scaled cross section $X$ for different types of interactions, $e.g.$ fly-by, exchange, merger and ionization (both resonant and non-resonant). The inputs given are : velocity ($v$) of the incoming star
(with mass and radius $m_3, R_3$) and the masses, the radii, the semi major axis
of the initial binary ($m_1, R_1;~ m_2, R_2;~a_{in}$) as well as the trial density which determines the number of scattering experiments to be done. $X$ is defined as :
\begin{equation}
X~=~\frac{\sigma}{\pi~a_{in}^2}\left(\frac{v}{v_c} \right)^2
\label{eq:X}
\end{equation}
where $\sigma$ is the cross section and $v_c$ is the critical velocity as
defined earlier (eq. \ref{eq:vc}). So from the reported values of $X$, one can calculate $\sigma$ and then the interaction time scale $t_{int}$ as $t_{int} = 1 / n \sigma v$ where $n$ is the number density of single stars. We took different sets of stellar parameters ($m_1, R_1;~ m_2, R_2$) to understand the effect of each of them and for each set of stellar parameters, we varied $a_{in}$ over a wide range in such a way that  the initial orbital periods ($P_{orb,~in}$, calculated using Kepler's law) are in the range of $0.01 - 1000$ days. 

For each set of parameters ($m_1, R_1;~ m_2, R_2;~a_{in}$), we take the maximum trial density ($n_{trial}$) as 5000 which is uniformly distributed in the impact parameter ($\rho$) over the range
$ 0 \leq \rho \leq \rho_0$ where $\rho_0$ simply corresponds to a periastron
separation of $2a$. The impact parameter range is then systematically expanded
to cover successive annuli of outer radii $\rho_i = 2^{i/2} \rho_0$ with
$n_{trial}$ trials each until no interesting interaction takes place in the
outermost zone \cite{mcm96}. Thus the total number of trials become
significantly large. As an example, for the parameters $m_1~= 1.4~M_{\odot},
R_1~=10 ~{\rm km };~ m_2~= 0.16~M_{\odot}, R_2~= 0.16~R_{\odot};~ m_3~=0.33~M_{\odot},
R_3~=~R_{\odot};~a_{in}~=~0.2 ~{\rm AU}; ~v~= 11.76~{\rm  km~ s^{-1}}$
we had a sample total of 15570 scatterings. Out of these 10960 were fly-bys, 3722 exchanges, 982 two mergers and 6 three-mergers.

We performed simulations over widely different values of $v$ $e.g.$ 11.76 ${\rm km~ s^{-1}}$, 13.15 ${\rm km~ s^{-1}}$, 7.79 ${\rm km~ s^{-1}}$ and 3.29 ${\rm km~ s^{-1}}$ chosen in such a way that almost the entire range of $v_{10}/n_{4}$ ($0.0024, 0.062, 0.131, 0.934$ respectively) has been covered. The variation in $v$ is also significant. But we mainly concentrate on the runs for $v=11.76$ ${\rm km~ s^{-1}}$ as in Ter 5, the host of the largest number of known binary radio pulsars. 

\begin{figure}[h]
{\includegraphics[width=0.5\textwidth,
height=0.3\textheight]{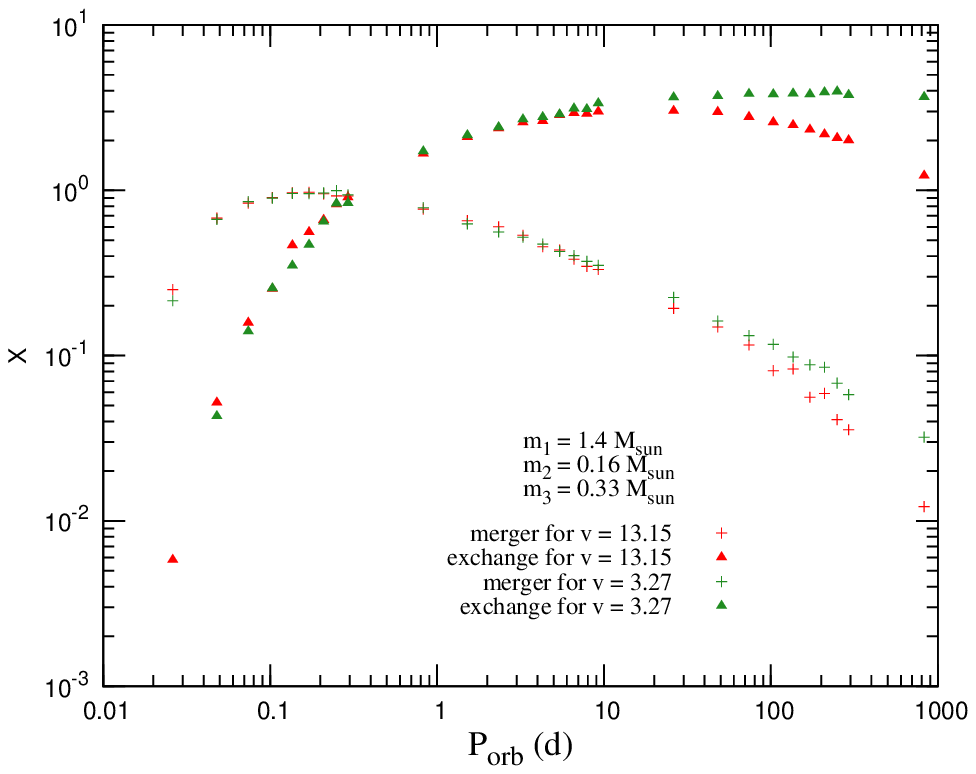}}
{\includegraphics[width=0.5\textwidth,
height=0.3\textheight]{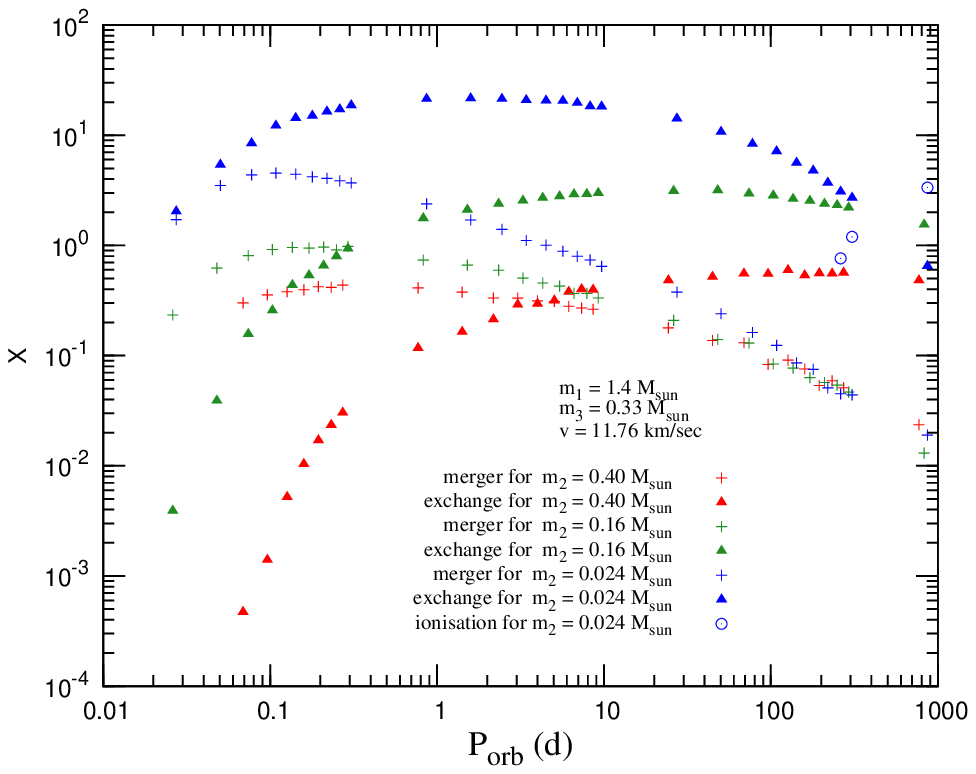}}
{\includegraphics[width=0.5\textwidth,
height=0.3\textheight]{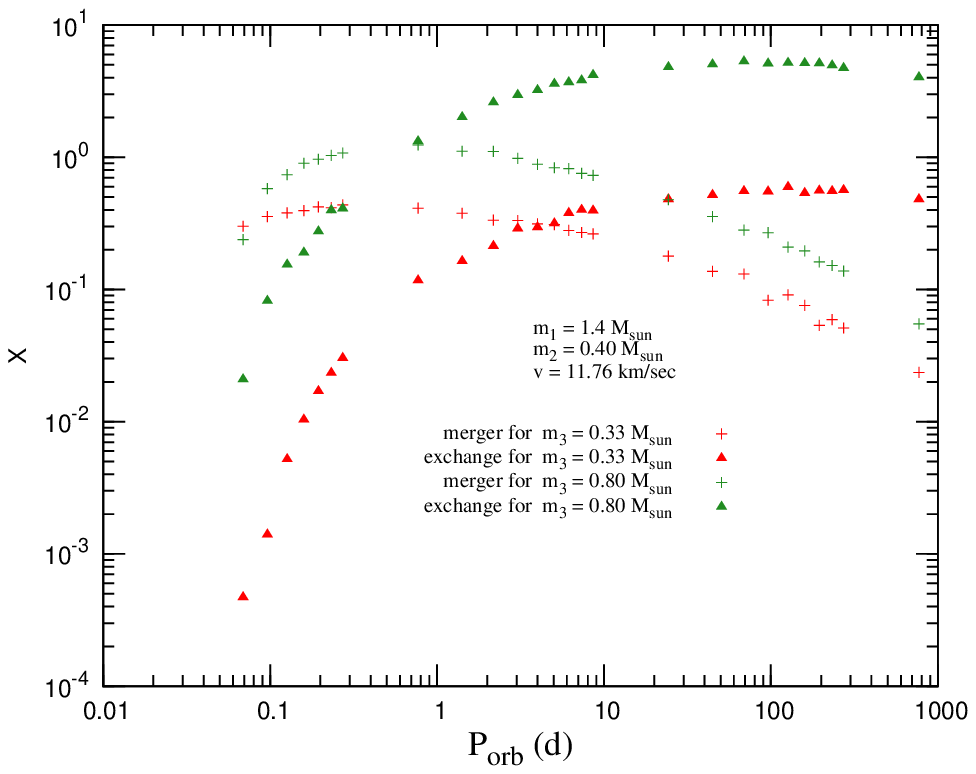}}
{\includegraphics[width=0.5\textwidth,
height=0.3\textheight]{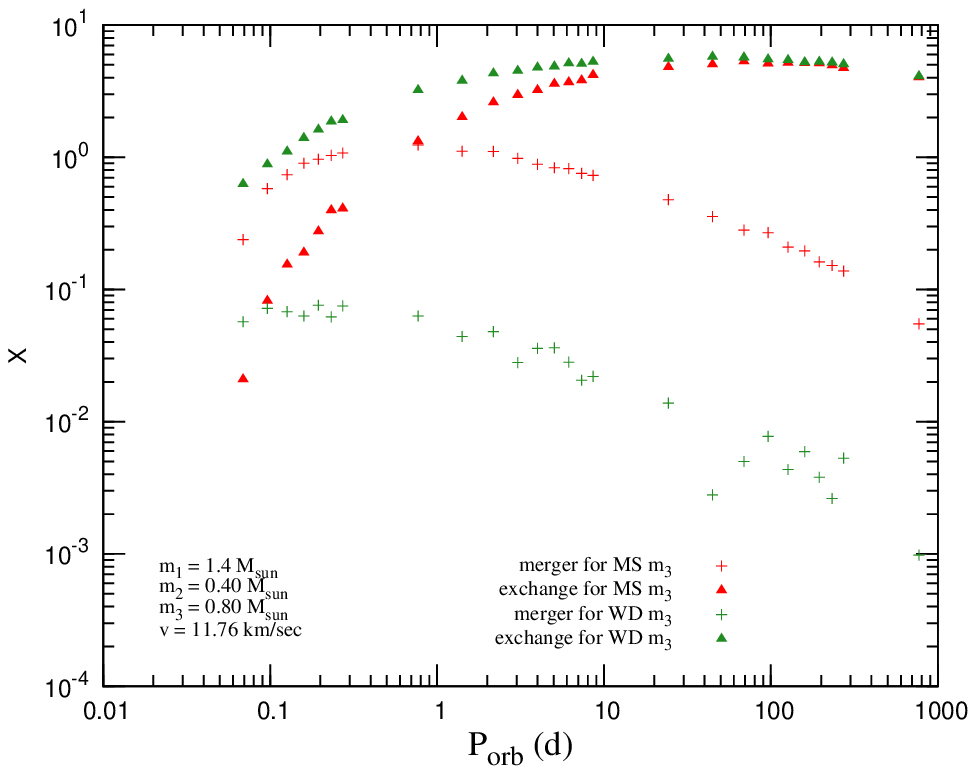}} \caption{\footnotesize{Variation of $X$ for
exchange ($+$), merger ($\blacktriangle$) and ionization
($\circ$, whenever significant) with $P_{orb,~in}$. (i) The first
figure is for $m_1=1.4~M_{\odot},~m_2=0.16~M_{\odot},
~m_3=0.33~M_{\odot}$ and two values of $v$ $e.g$ 13.16 ${\rm km~ s^{-1}}$
(red) and 3.27 ${\rm km~ s^{-1}}$ (green).  (ii) The second figure is for
$m_1=1.4~M_{\odot},~m_3=0.33~M_{\odot},$ $v=11.76$ ${\rm km~ s^{-1}}$ and
different values of $m_2$ $e.g$ $0.40~M_{\odot}$ (red),
$0.16~M_{\odot}$ (green) and $0.024~M_{\odot}$ (blue). (iii) The
third figure is for $m_1=1.4~M_{\odot},~m_2=0.16~M_{\odot},$
$v=11.76$ ${\rm km~ s^{-1}}$ and different values of $m_3$ $e.g.$
$0.33~M_{\odot}$ (red), $0.80~M_{\odot}$ (green).  (iv) The fourth
figure is for $m_1=1.4~M_{\odot},~m_2=0.16~M_{\odot},
~m_3=0.80~M_{\odot}$, $v=11.76$ ${\rm km~ s^{-1}}$ and  the $3^{rd}$ star is
either a MS (red) or a WD (green).}} \label{fig:X_porb}
\end{figure}

In Fig. (\ref{fig:X_porb}) we plot the variation of $X$ for exchange ($+$), merger ($\blacktriangle$) and ionization ($\circ$, whenever significant) with $P_{orb,~in}$ and check the
dependencies on different parameters - (i) In the first figure,
we set $m_1~=~1.4~M_{\odot},~m_2~=~0.16~M_{\odot}$,
$m_3~=~0.33~M_{\odot}$, and take two values of $v$ $e.g.$ 13.16
${\rm km~ s^{-1}}$ (red) and 3.27 ${\rm km~ s^{-1}}$ (green) which are the highest and
the lowest values of $v$ among our choice (see Table \ref{tb:gc_v_n}).  (ii) In the second figure, we set $m_1~=~1.4~M_{\odot},~m_3~=~0.33~M_{\odot},$ $v~=11.76~{\rm km~ s^{-1}}$,
and take different values of $m_2$ $e.g.$ $0.40~M_{\odot}$ (red),
$0.16~M_{\odot}$ (green) and $0.024~M_{\odot}$ (blue). 
Ionization starts for $m_2=~0.024~M_{\odot}$ at
$P_{orb,~in}\approxeq 200$ days.  (iii) In the third figure, we
set $m_1~=~1.4~M_{\odot},~m_2~=~0.16~M_{\odot},$ $v~=11.76$ ${\rm km~ s^{-1}}$, and take different values of $m_3$ $e.g.$ $0.33~M_{\odot}$
(red), $0.80~M_{\odot}$ (green).  (iv) In the fourth
figure, we set $m_1~=~1.4~M_{\odot},~m_2~=~0.16~M_{\odot},
~m_3~=~0.80~M_{\odot}$, $v~=~11.76 ~{\rm km~ s^{-1}}$ and take the $3^{rd}$
star to be either an MS (red) or a WD (green).  The results can be summarized in a tabular form :

\vskip 0.5 cm
\begin{tabular}{cc}
\hline \hline
change of variable &  result \\ \hline
decrease in the value of $v$ & $X_{exchange}$  and $X_{merger}$  \\ 
 & increase slightly at higher $P_{orb}$ \\  \hline
decrease in the value of $m_2$ &  $X_{exchange}$  and $X_{merger}$  \\
 & increase significantly at lower $P_{orb}$.  \\ 
 & Ionization is significant at very low   \\  
 &   values of $m_2$ and higher values of  $P_{orb}$ \\ \hline
increase in the value of $m_3$ &  $X_{exchange}$  and $X_{merger}$ \\
 &   increases significantly.  \\  \hline
nature of the incoming star  &  $X_{exchange}$ is slightly higher at lower $P_{orb}$ when the \\
(of mass $m_3$) &  incoming star is a WD rather than a MS. \\
 &  $X_{merger}$ is significantly lower when the \\ 
&  incoming star is a WD rather than a MS. \\
\hline \hline
\end{tabular}
\vskip 0.5 cm

\begin{figure}[h]
{\includegraphics[width=0.5\textwidth,
height=0.35\textheight]{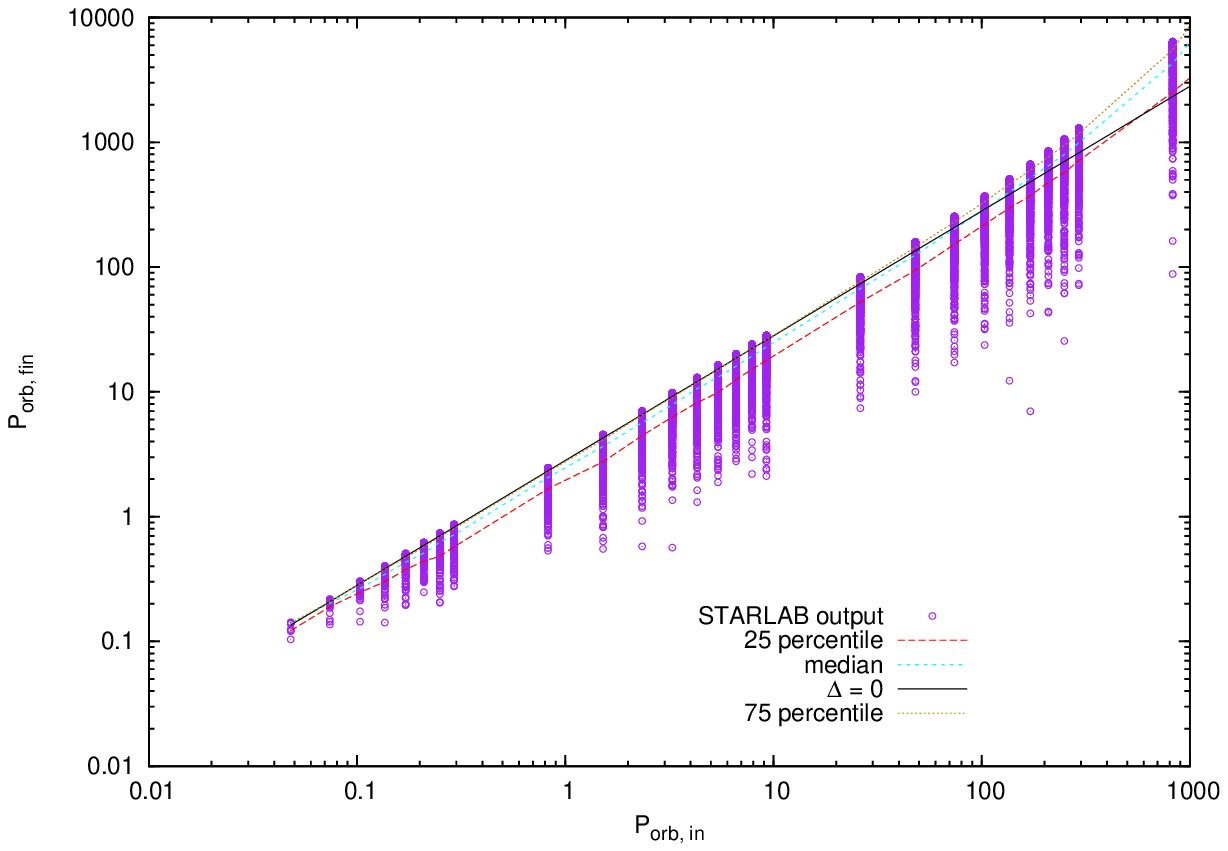}}
{\includegraphics[width=0.5\textwidth,
height=0.35\textheight]{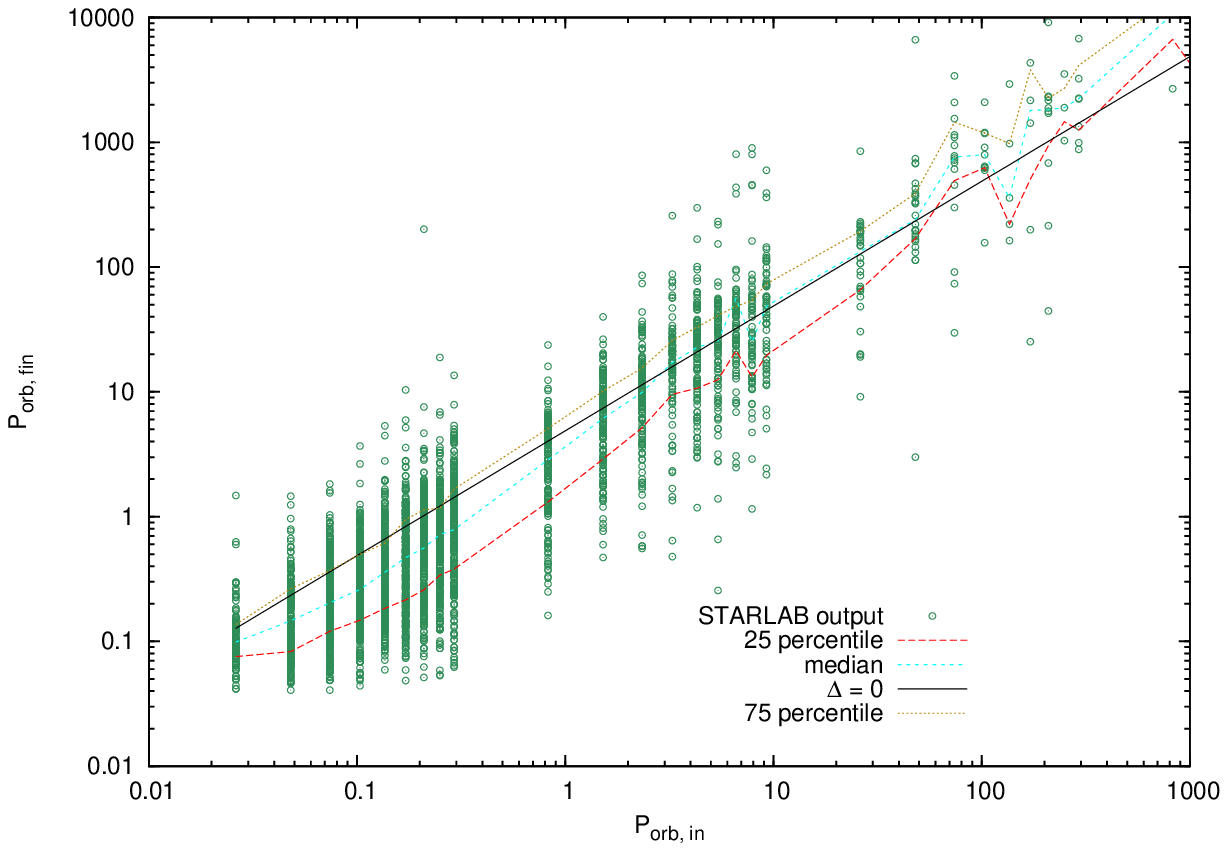}} \caption{\footnotesize{$P_{orb,~in}$ Vs
$P_{orb,~fin}$ plots. Right hand side (purple points) diagram is
for exchanges and the left hand side diagram (green points) is
for mergers. The stellar parameters are as follows :
$m_1=1.4~M_{\odot}$ (NS), $m_3=0.33~M_{\odot}$ (MS),
$m_2=~0.16~M_{\odot}$ (MS) and $v=11.76~{\rm km~ s^{-1}}$.}}
\label{fig:pin_pfin_stat}
\end{figure}

Eqn. (\ref{eq:X}) reveals that $\sigma \propto X a_{in}^{2} v_{c}^{2} / v^{2}$ for any fixed set of stellar parameters ($m_1, m_2, m_3$) which becomes $\sigma \propto X a_{in} / v^{2}$ after using the expression of $v_c$ as in Eqn. (\ref{eq:vc}). So the interaction timescale $t_{int} = 1 / n \sigma v$ can be written as  $t_{int} \propto v / ( n X a_{in} )$. As it is clear from the top-left panel of Fig. (\ref{fig:X_porb}), that for a fixed set of $m_1, m_2, m_3$, the variation of $X$ is negligible for any given $a_{in}$. So we can write $t_{int} \propto v /  n $. Remember, the same relation has been used for $t_{fly}$ in Eqns. (\ref{eq:tfly1}, \ref{eq:tfly2}).

STARLAB also gives the properties of final states, $i.e.$ eccentricities and semi major axes of
the final binaries for each set of inputs ($m_1, R_1;~ m_2, R_2;~ m_3, R_3;~a_{in}$). So for each
value of $P_{orb,~in}$, we get a number of values of $P_{orb,~fin}$ and $e_{fin}$.
For each value of $P_{orb,~in}$, we calculated $25$ percentile,
median and  $75$ percentile values of $P_{orb,~fin}$ as well as the values from
analytic expressions given in Eqns. (\ref{eq:afin_ex}) and (\ref{eq:afin_merg})
putting $\Delta=0$. As all these values are very close to each other,
we use the $P_{orb,~in}-P_{orb,~fin}$ relation corresponding to $\Delta~=~0$
throughout (eq. \ref{eq:afin_gen}). As an example, in Fig. \ref{fig:pin_pfin_stat}, we have plotted
$P_{orb,~in}$ Vs $P_{orb,~fin}$ as reported by STARLAB (scatter plots)
for both merger and exchange with $m_1~=~1.4~M_{\odot},~m_2~=~0.16~M_{\odot}$, $m_3~=~0.33~M_{\odot}$, $v~=~11.76~{\rm km~ s^{-1}}$. The line for 25 percentile,
median, 75 percentile along with the analytic relation of $P_{orb,~in}-P_{orb,~fin}$ with $\Delta~=~0$ (eq. \ref{eq:afin_gen}) are shown in the same plot.

\begin{figure}
{\includegraphics[width=0.5\textwidth,
height=0.3\textheight]{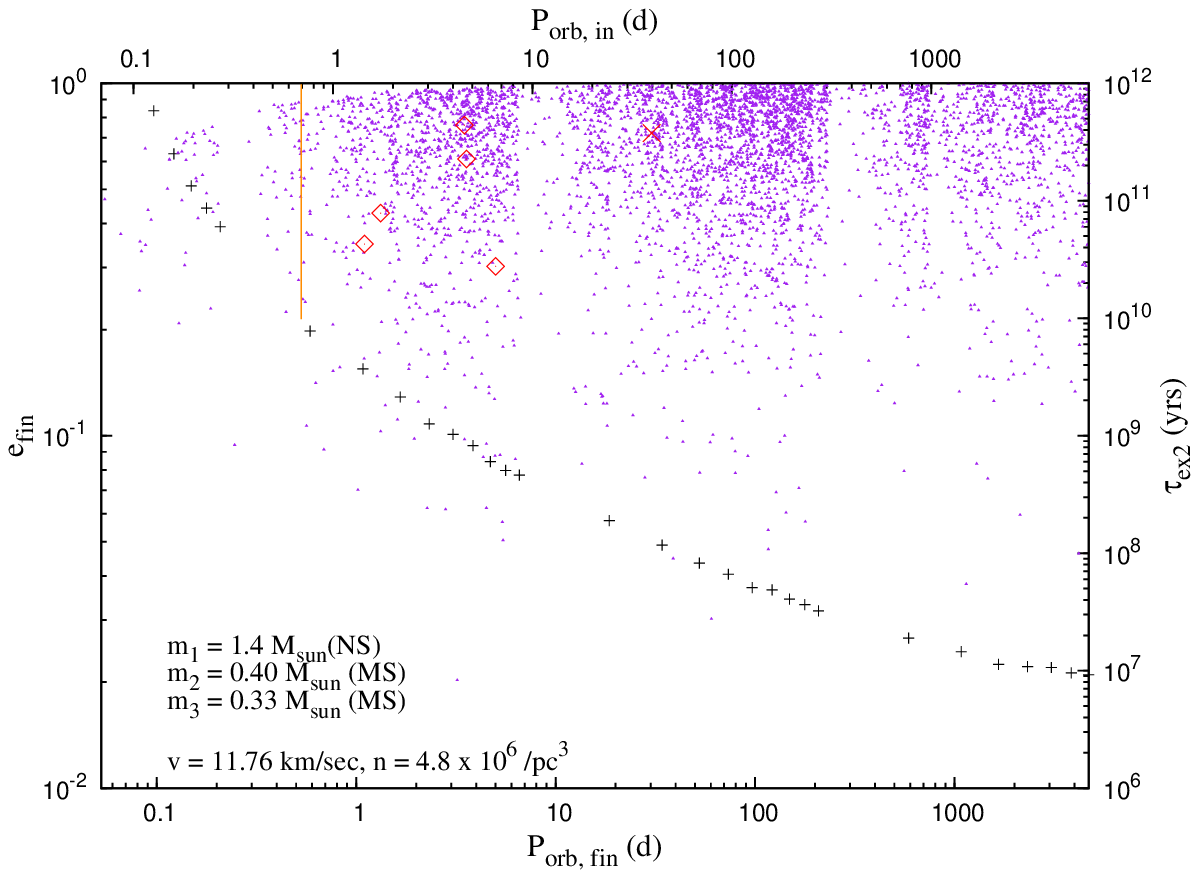}}
{\includegraphics[width=0.5\textwidth,
height=0.3\textheight]{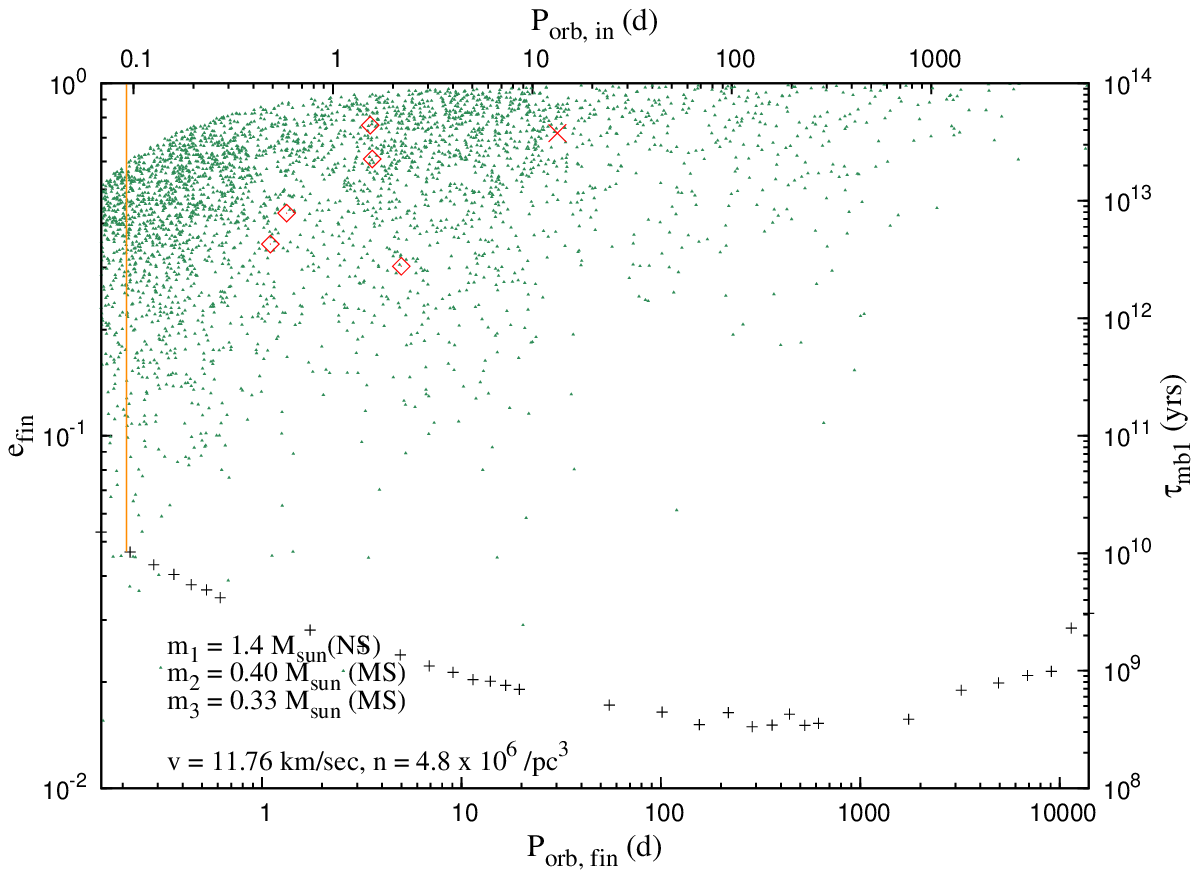}}
{\includegraphics[width=0.5\textwidth,
height=0.3\textheight]{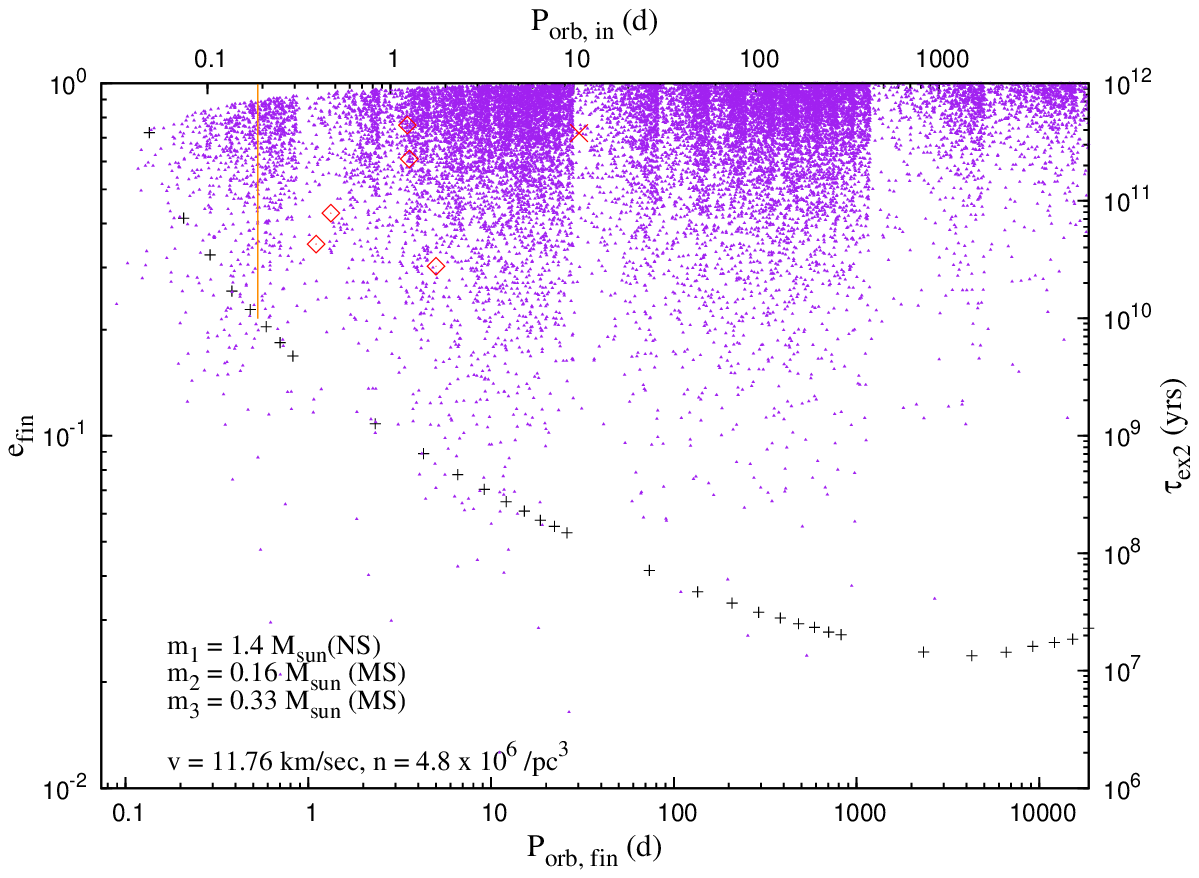}}
{\includegraphics[width=0.5\textwidth,
height=0.3\textheight]{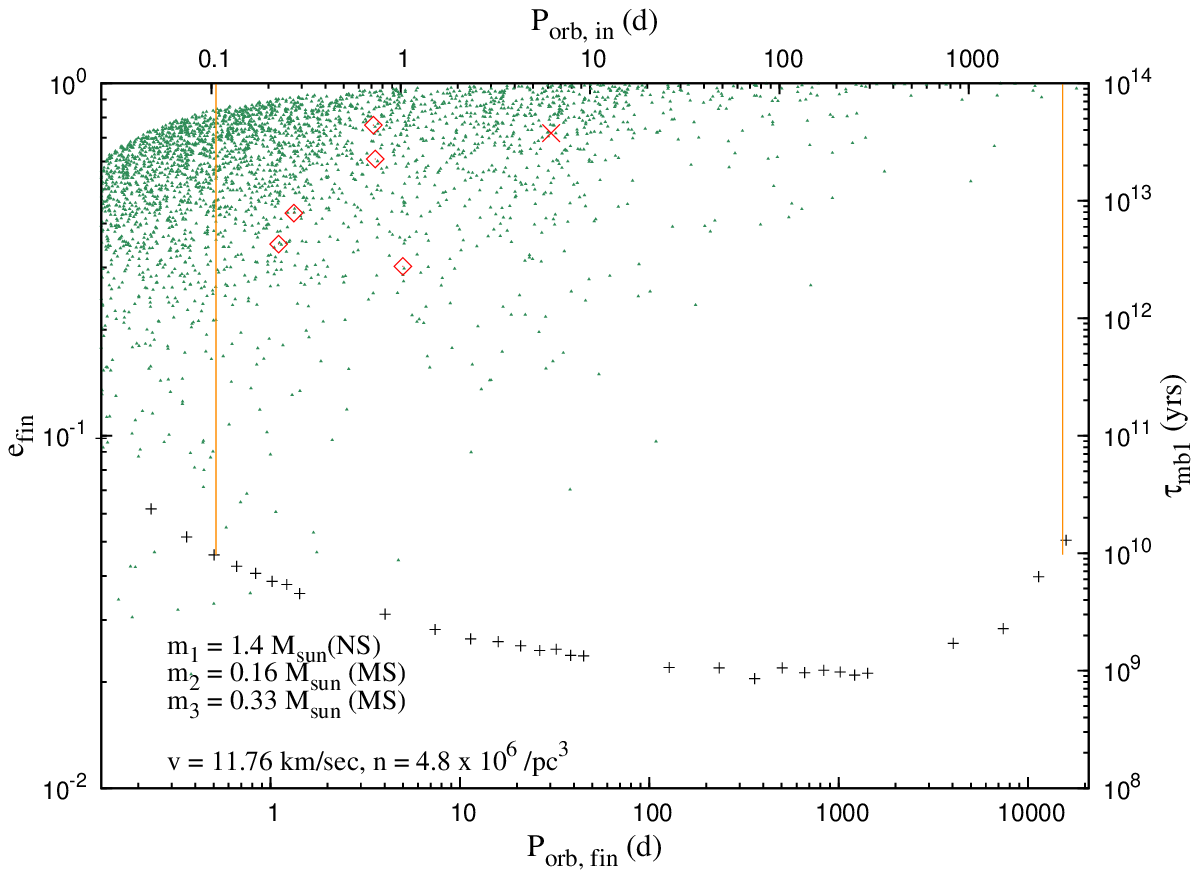}}
{\includegraphics[width=0.5\textwidth,
height=0.3\textheight]{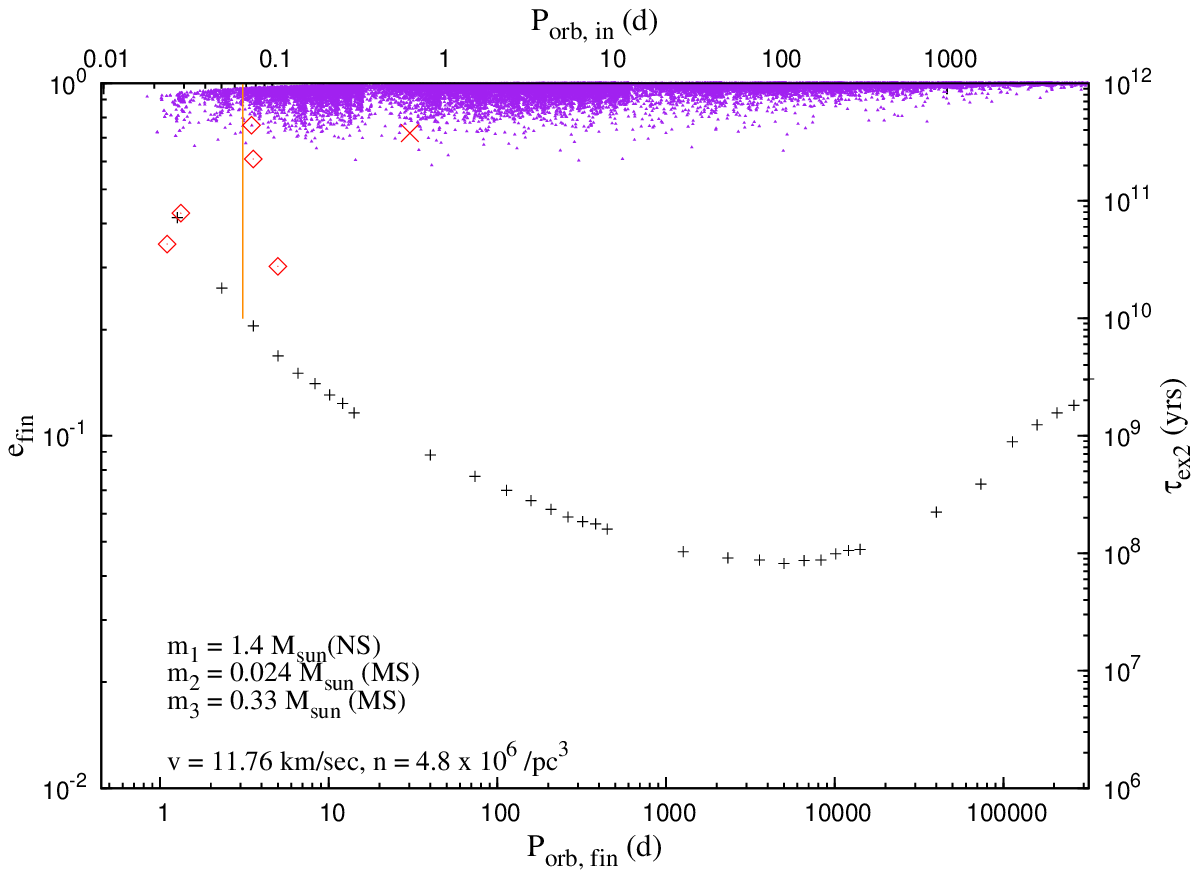}}
{\includegraphics[width=0.5\textwidth,
height=0.3\textheight]{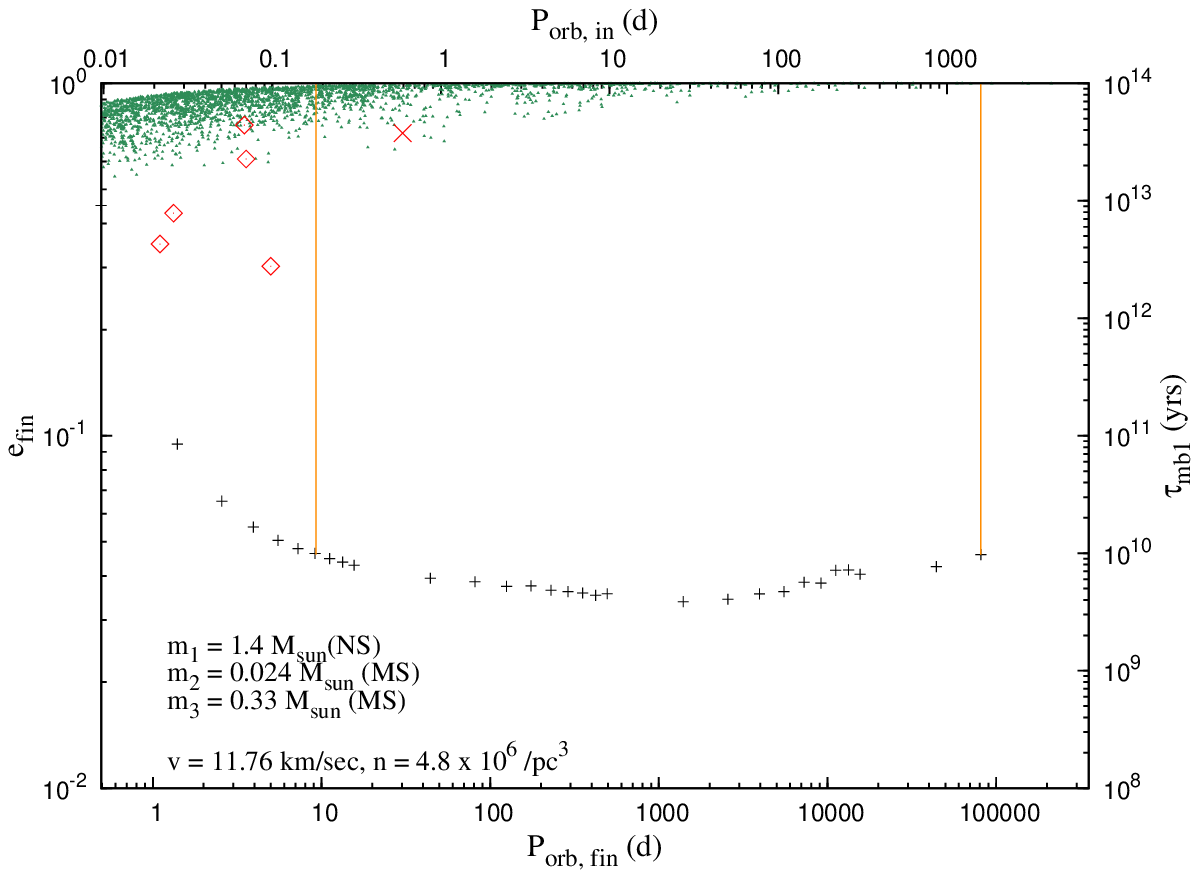}} \caption{\footnotesize{Time
scales (denoted by black `+') and final eccentricity distributions
(scatter-plot of points) with initial and final orbital periods
($\Delta~=~0$ in Eqn. \ref{eq:afin_gen}) for exchange (purple points) and merger (green points)
interactions with different stellar parameters. We plot $P_{orb,
in}$  along the top x-axis and $P_{orb, fin}$  along the bottom
x-axis. The left y-axis gives the final eccentricities while the
right y-axis gives the time scales of interactions. The vertical
orange lines form the boundaries of the allowed orbital period
regions where interaction time scales are less than $10^{10}$
yrs. The stellar parameters are as follows : $m_1=1.4~M_{\odot}$,
$m_3=0.33~M_{\odot}$ and $m_2$ varies from $0.024~M_{\odot}$ to
$0.16~M_{\odot}$ and then to $0.40~M_{\odot}$. }}
\label{fig:terzan_exch_merg_m3_l}
\end{figure}

\begin{figure}
{\includegraphics[width=0.5\textwidth,
height=0.3\textheight]{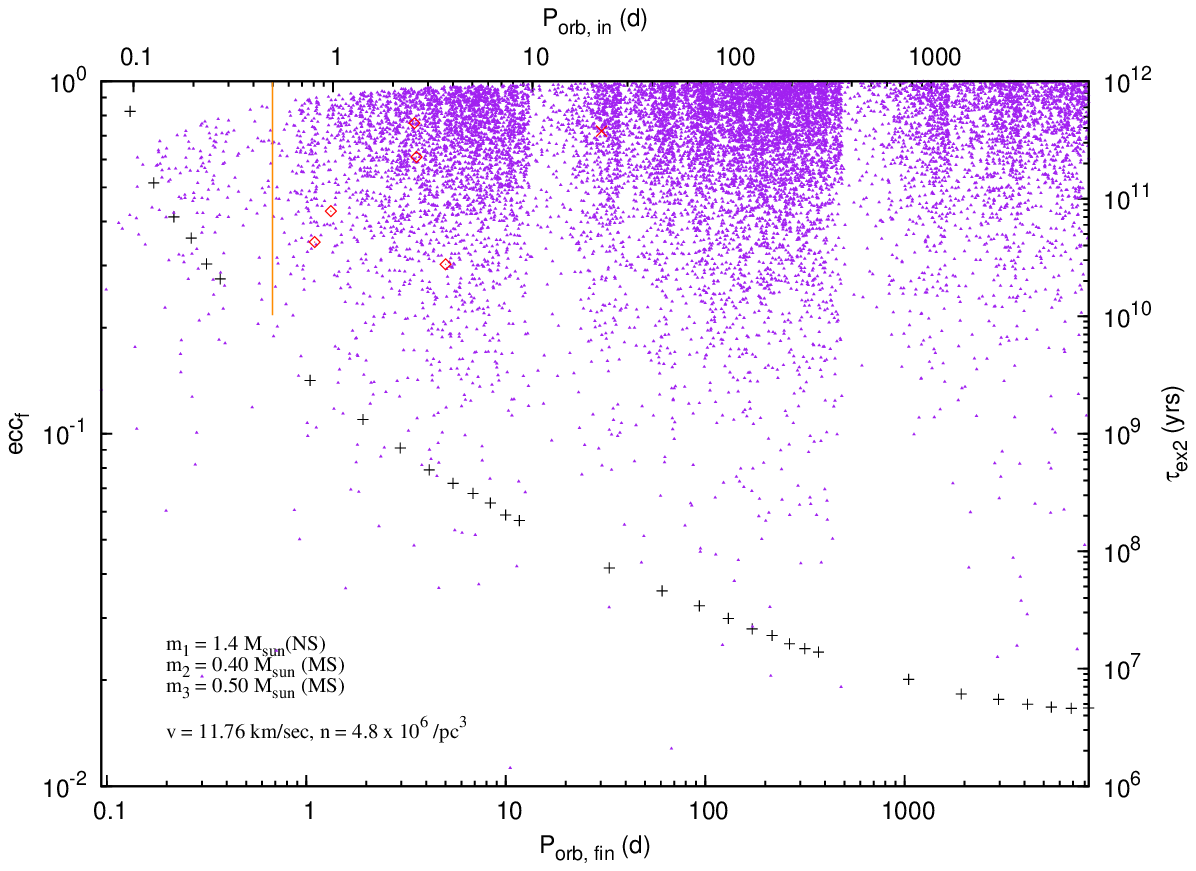}}
{\includegraphics[width=0.5\textwidth,
height=0.3\textheight]{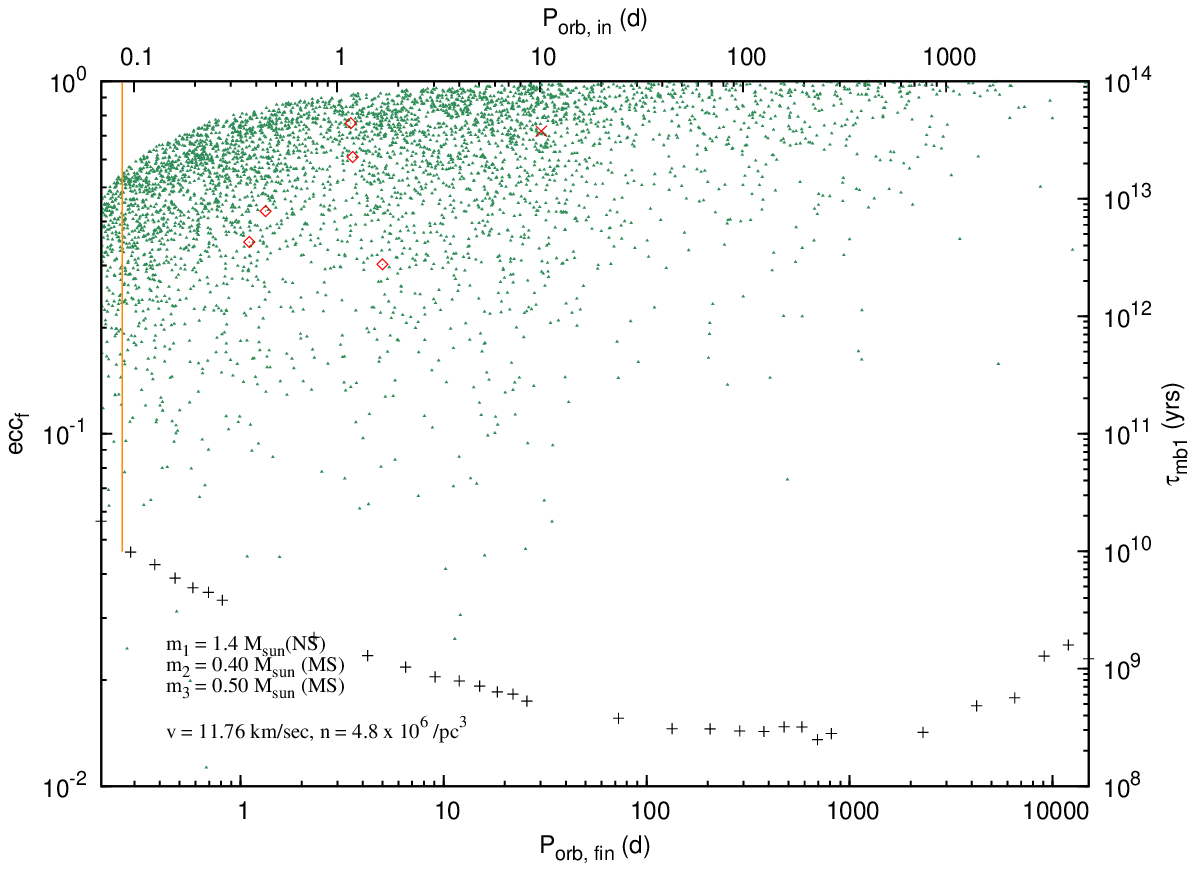}}
{\includegraphics[width=0.5\textwidth,
height=0.3\textheight]{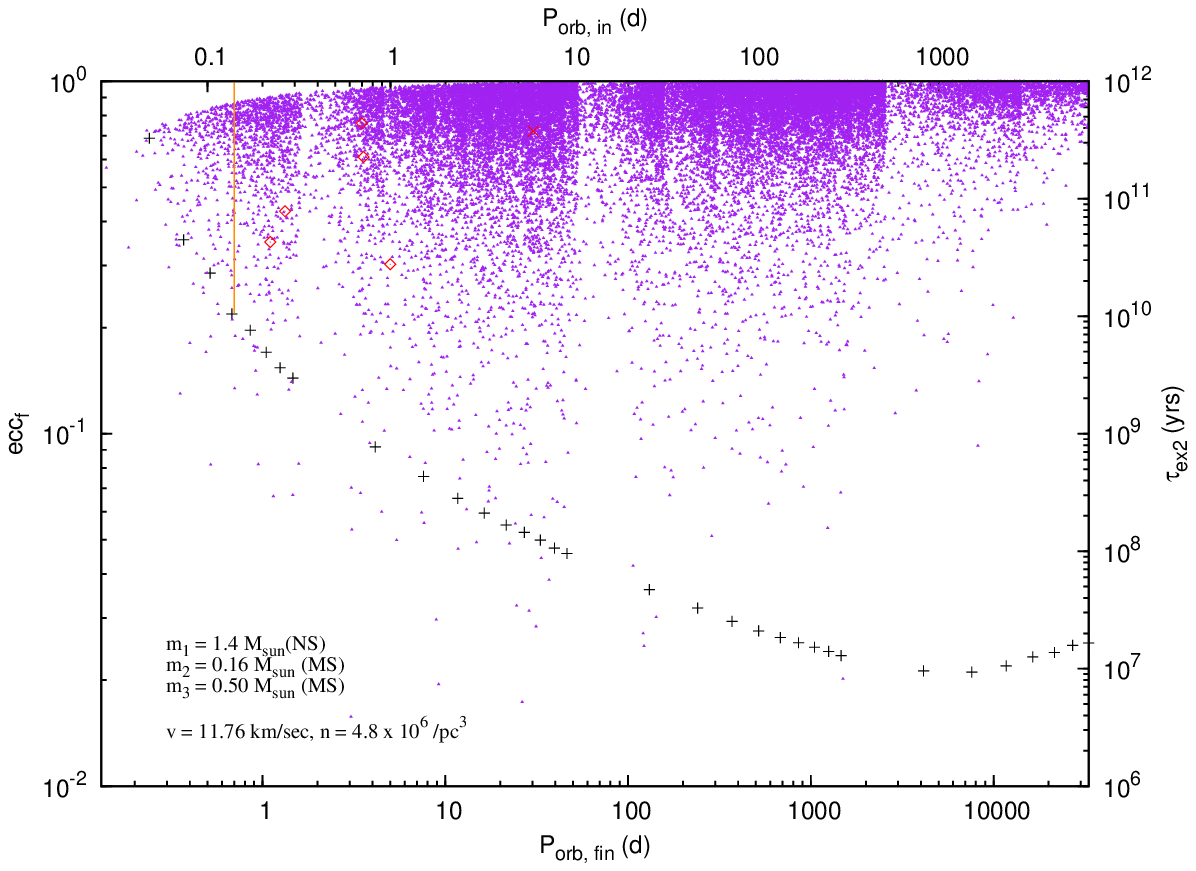}}
{\includegraphics[width=0.5\textwidth,
height=0.3\textheight]{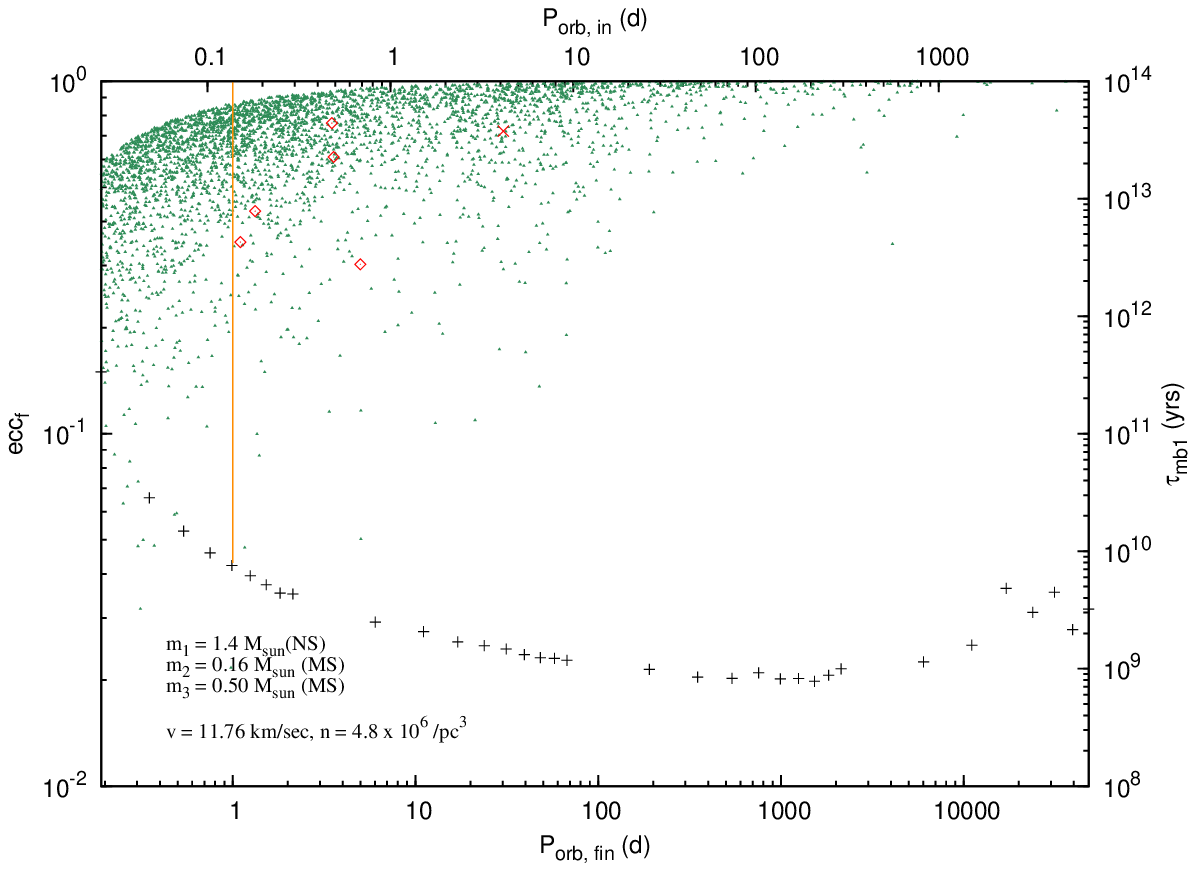}}
{\includegraphics[width=0.5\textwidth,
height=0.3\textheight]{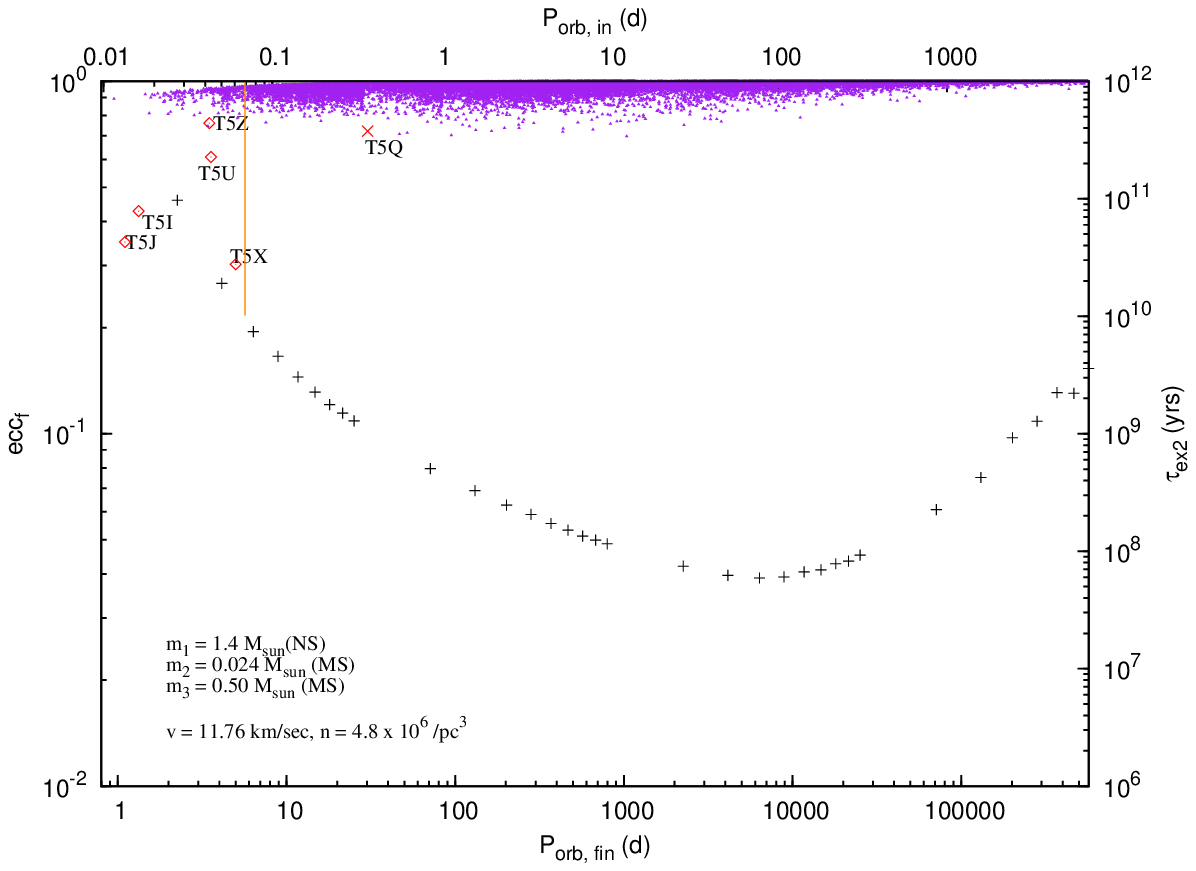}}
{\includegraphics[width=0.5\textwidth,
height=0.3\textheight]{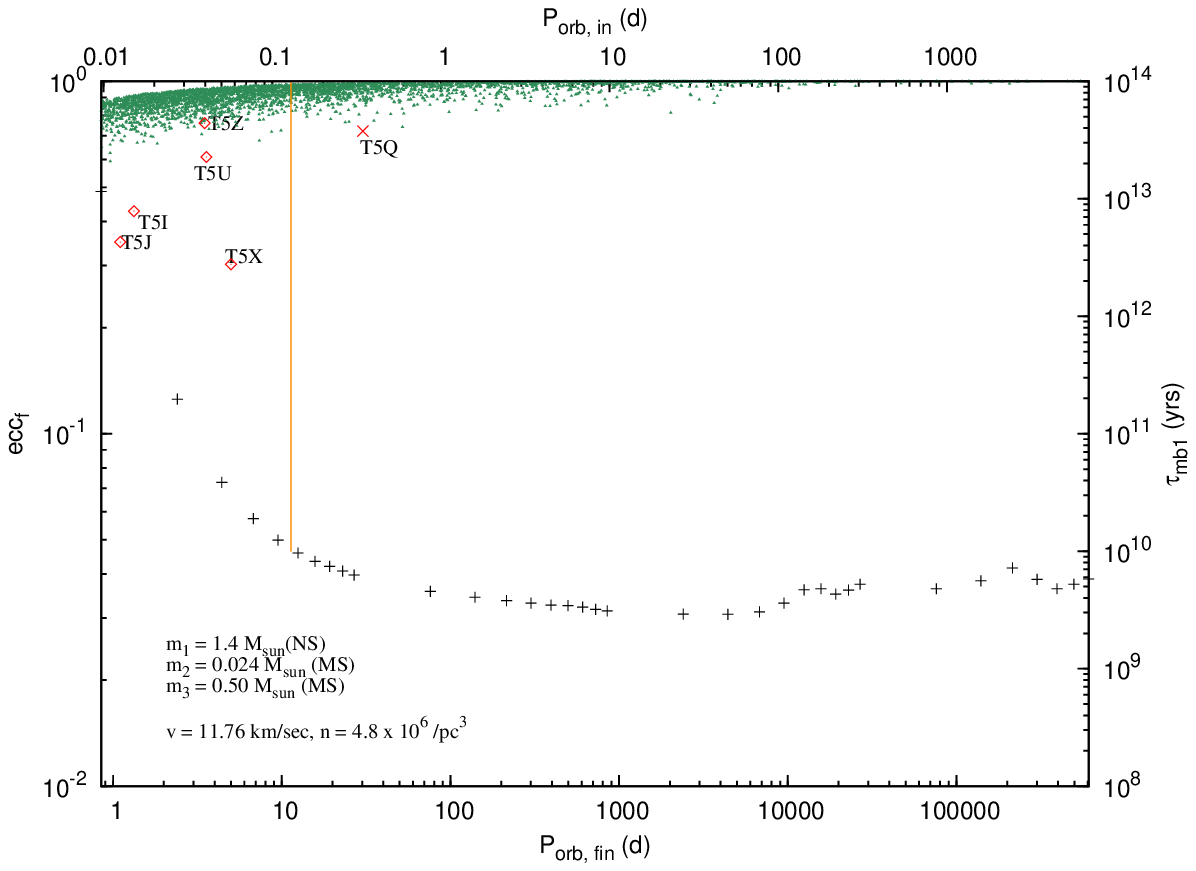}} \caption{\footnotesize{Plots similar to those in Fig \ref{fig:terzan_exch_merg_m3_l} with only difference that here $m_3=0.50~M_{\odot}$}.}
\label{fig:terzan_exch_merg_m3_h}
\end{figure}

In each panel of Figs. (\ref{fig:terzan_exch_merg_m3_l}) and (\ref{fig:terzan_exch_merg_m3_h}),
we plot $P_{orb, in}$ of the initial binary (comprising of stars $m_1$ and $m_2$)
along the top x-axis and $P_{orb, fin}$  along the bottom x-axis. The left y-axis gives the final
eccentricities while the right y-axis gives the time scales of the
interactions. Purple points in the left panels are for exchange interactions while the green points
in the right panels are for the merger interactions. Interaction time scales are plotted with black $+$s.
The vertical orange lines form the boundaries of the allowed orbital period regions
where interaction time scales are less than $10^{10}$ years. It is clear from the
scatter plots (Figs. \ref{fig:terzan_exch_merg_m3_l}, \ref{fig:terzan_exch_merg_m3_h}) that the final binaries will most probably have $e>0.1$ if they undergo either exchange or merger events.
The observed binary pulsars with $e>0.1$ in Ter 5 can be found in
Table \ref{tab:gcpsr_parms} where the companion masses ($m_{c}$) are for inclination
angle $i~=~60^{\circ}$. These binaries are also shown in all panels of  Figs. \ref{fig:terzan_exch_merg_m3_l} and \ref{fig:terzan_exch_merg_m3_h} (red in color, named in lower panels of Fig. \ref{fig:terzan_exch_merg_m3_h}). In Fig \ref{fig:terzan_exch_merg_m3_l}, $m_1=1.4~M_{\odot}$, $m_3=0.33~ M_{\odot}$ and we vary $m_2$ from $0.024~ M_{\odot}$ to $0.16~ M_{\odot}$ and $0.40~ M_{\odot}$.  We have also performed the similar kind of numerical simulations by taking $m_3 = 0.50~M_{\odot}$ and found that the outcomes of the simulations i.e. the appearance of the scatter plots (Fig \ref{fig:terzan_exch_merg_m3_h}) does not change much in comparison to the change with the variation in the value of $m_2$ (for fixed $m_3$). 

In Section \ref{sec:fly_by}, we have seen that PSR B1638+36B (in M13) has such a combination of $P_{orb} - e $ that it can not be explained by fly-by interactions. Moreover, the scatter plots (Figs. \ref{fig:terzan_exch_merg_m3_l}, \ref{fig:terzan_exch_merg_m3_h}) also reveal that its eccentricity can not be caused by exchange or mergers as these interactions imparts eccentricities larger than $0.01$ whereas the upper limit of the eccentricity of PSR B1638+36B is 0.001. One possibility is that  the true eccentricity of this system is significantly lower than the upper limit so that it falls in the $t_{fly} < 10^{10}$ yrs region. Another possibility is that it could have undergone eccentricity pumping of the inner orbit due to the presence of a third star in a wide outer orbit, i.e. in a hierarchical triplet like the PSR B1620-26 in M4 system \cite{sson93, ras95b, thor99, ford00}.

If an observed eccentric binary pulsar lies in the region where time scale for
a particular interaction is greater than $10^{10}$ years, then
that interaction can not be responsible for its eccentricity.
Moreover, an exchange interaction to be the origin of the eccentricity of
a particular pulsar, we should have $m_c \approxeq m_3$ and for merger $m_c \approxeq m_2+m_3$, where $m_c$ is the companion mass. Considering these two facts, what we can conclude about possible dynamic history about the eccentric pulsars in Ter 5 which are summarized below where ``$E$" means possible exchange and ``$M$" means possible merger, ``$N_E$" means exchange scenario satisfies the mass condition, but should be excluded on the basis of time-scale condition, ``$N_M$" means merger scenario satisfies the mass condition, but should be excluded on the basis of time-scale condition. Values of $m_c$s are given below the pulsar names. All masses are in the unit of $M_{\odot}$. It seems that $m_3=0.33~M_{\odot}$ is a better choice than $m_3=0.50~M_{\odot}$ as the earlier one can explain the origin of all six eccentric pulsars in Ter 5.

\vskip 0.5 cm

\begin{tabular}{|c||c|c|c|c|c|c||c|c|c|c|c|c|}
\cline{1-13}
& \multicolumn{6}{c||}{$m_1 = 1.4,~m_3 = 0.33$} &  \multicolumn{6}{c|}{$m_1= 1.4, ~m_3 = 0.50 $}\\ \cline{1-13}
& \multicolumn{6}{|c||}{Pulsars} &  \multicolumn{6}{|c|}{Pulsars}\\ \cline{1-13}
\multicolumn{1}{|c||} {$m_2$} & I & J & Q & U & X &  Z  & I & J & Q & U &X & Z \\ 
\multicolumn{1}{|c||} {} & .24 & .39 & .53 & .46 & .29 & .25 &   .24 &  .39 & .53 & .46 & .29 & .25 \\ \cline{1-13}
\multicolumn{1}{|c||}{0.024} & $N_E$ & $N_E$, $N_M$ &  &   &$ N_E, N_M$ &  E &  &  & E, M &  $N_E, N_M$ &  &  \\ \cline{1-13}
\multicolumn{1}{|c||} {0.16}& E & E, M & M &  M & E & E &   &  & E & E  &  & \\ \cline{1-13}
\multicolumn{1}{|c||}{0.40} & E & E &   &   & E & E &  &  & E &  E &  &  \\ \cline{1-13}
\end{tabular}
\vskip 0.5 cm

\clearpage

\clearpage

\clearpage

We should remember that these conditions on stellar masses are mainly indicative and need not to be satisfied very accurately and we allow a deviation of $\pm 0.1 ~ M_{\odot}$ from the exact mass conditions because: 1) a slightly different choice of stellar masses can give the same or similar output of the simulation and 2) the companion masses are not known exactly in most cases; the masses are obtained from the mass function in terms of $m_c \, sin \, i$ and with the assumption that the orbital inclination angle is $60^{\circ}$ (Table \ref{tab:gcpsr_parms}).

\subsection{Ionization}
\label{sec:ion}

\begin{figure}[h]
{\includegraphics[width=0.32\textwidth,
height=0.2\textheight]{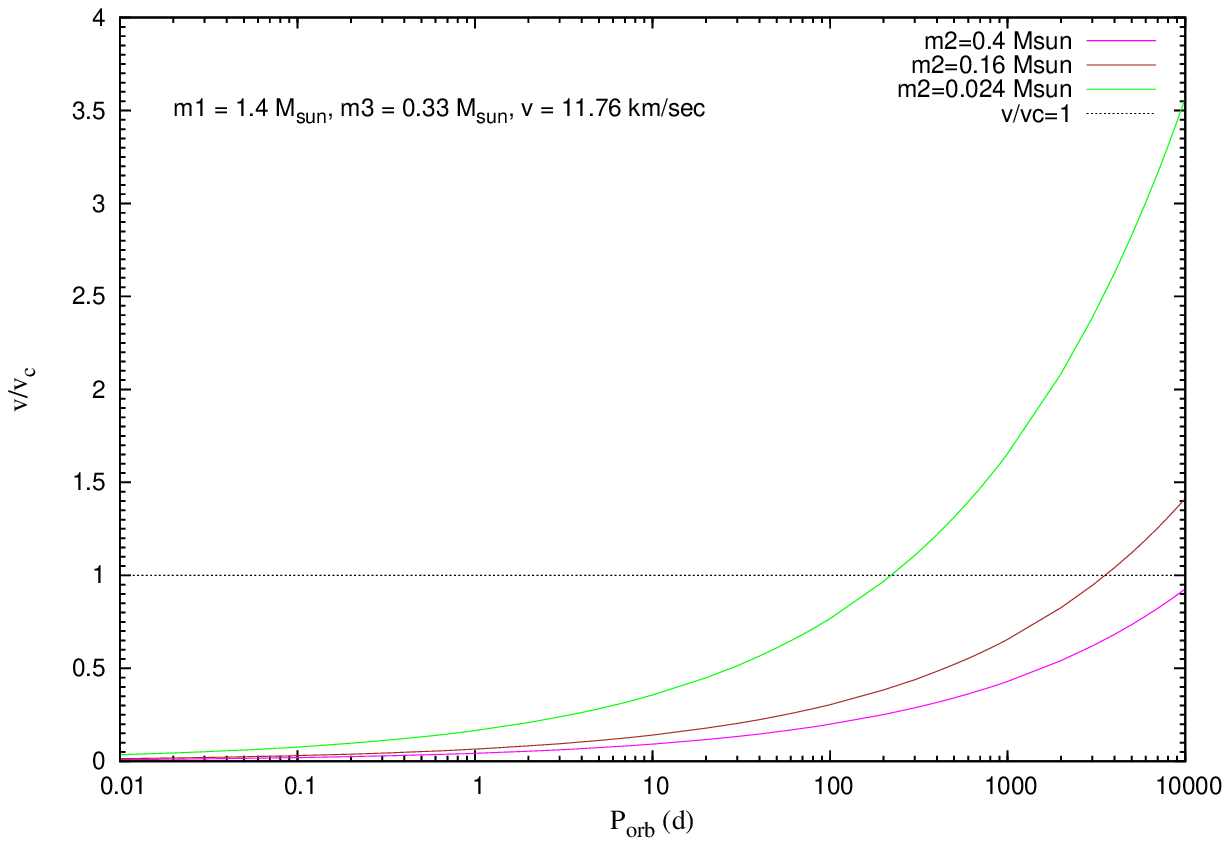}}
{\includegraphics[width=0.32\textwidth,
height=0.2\textheight]{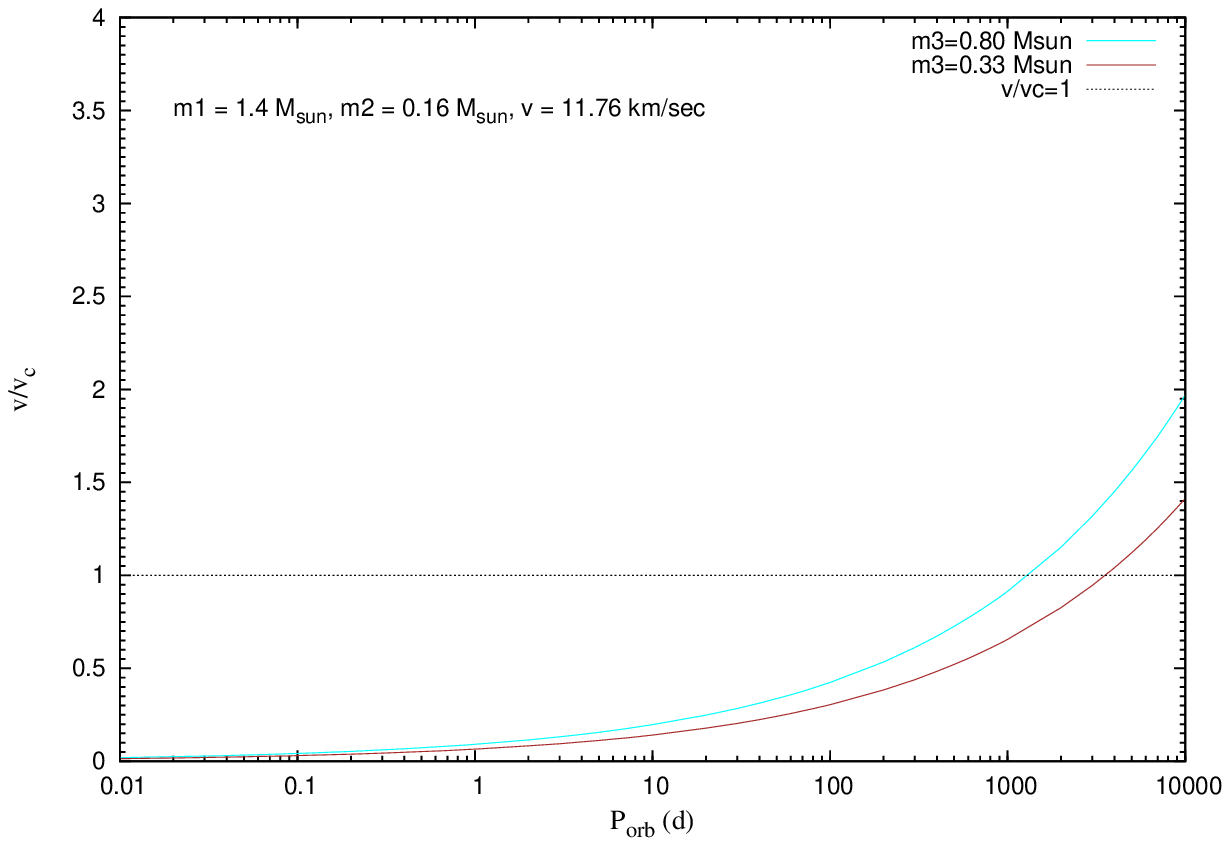}}
{\includegraphics[width=0.32\textwidth,
height=0.2\textheight]{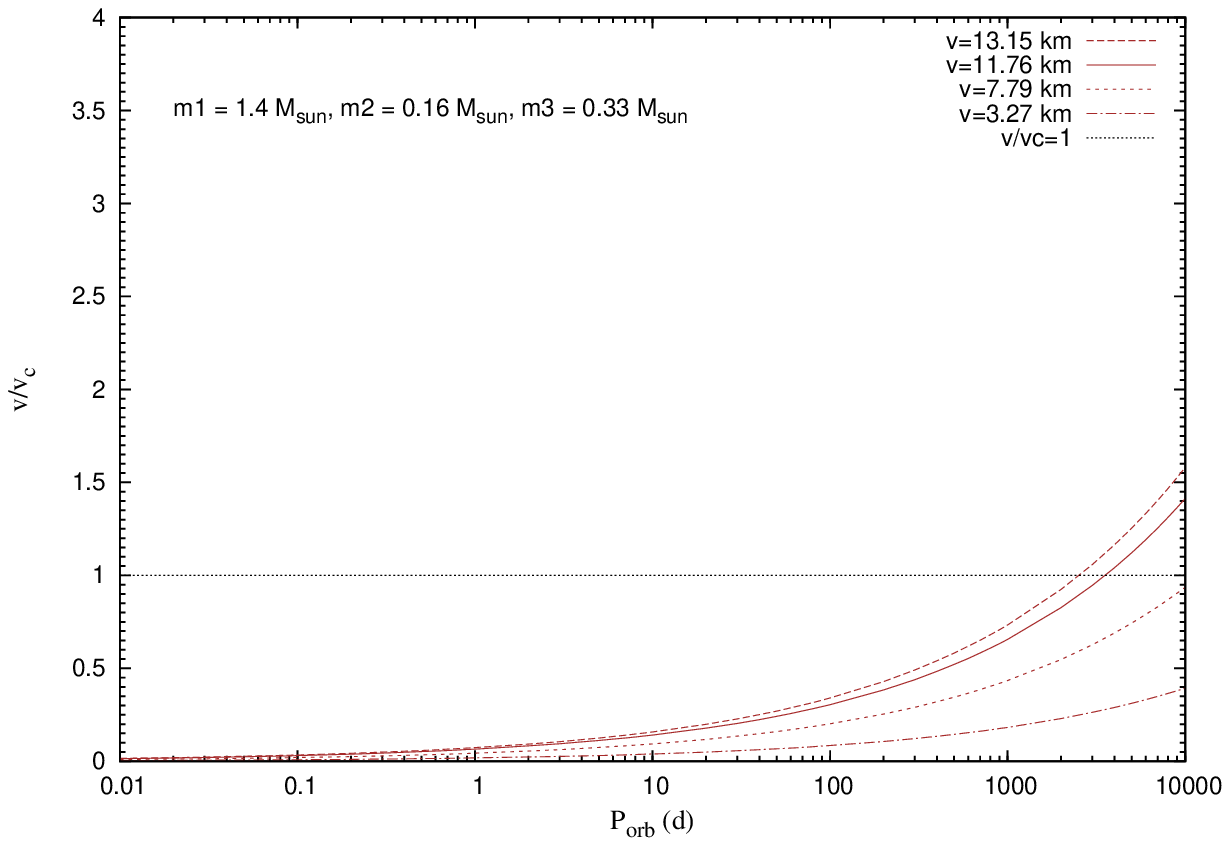}}
\caption{\footnotesize{Variation of $v/v_c$ with $P_{orb, in}$ for different parameters. The left panel shows the variation of $v/v_c$ with $P_{orb, in}$ for different values of $m_2$, the middle panel shows the variation of $v/v_c$ with $P_{orb, in}$ for different values of $m_3$ and the right panel shows the variation of $v/v_c$ with $P_{orb, in}$ for different values of $v$.}}
\label{fig:vbyvc}
\end{figure}

The condition for ionization to occur is $v/v_c > 1$ (where $v_c$ is the
critical velocity giving three unbound stars at zero energy
defined in Eq. \ref{eq:vc}). Fig. \ref{fig:vbyvc} shows plots of $v/v_c$ with $P_{orb}$.
It is clear that ionization starts only at high values of
$P_{orb, in}$; and a lower value of $m_2$ or a higher value of
$m_3$ or a higher value of $v$ facilitate ionization for obvious reasons of available
kinetic and binding energies. 

STARLAB runs agree with the expectation above.  As an example, keeping
$m_1=~1.4~M_{\odot}$ and $m_3=~0.33~M_{\odot}$ fixed, the minimum
value of $P_{orb, in}$ (at which ionization starts) increases from $\sim 200$ days to $\sim 3000$
days when we increase $m_2$ from $0.024~M_{\odot}$ to
$0.16~M_{\odot}$. For $m_2 = 0.40~M_{\odot}$, ionization occurs when $P_{orb} > 10000$.

There are three disk binaries with $P_{orb} > 1000$ days such as (i) PSR
B0820+20 with $P_{orb}=1232.40400$ days, $e=0.0118689$, $m_c=0.23~M_{\odot}$, characteristic age $t_c = 1.31 \times 10^{8}$ years (ii) PSR B1259-63 with $P_{orb}=1236.72404$ days, $e=0.8698869$,  $m_c = 4.14~M_{\odot}$, $t_c = 13.32 \times 10^{5}$ years and (iii) PSR J1638-4725 with $P_{orb}=1940.9$ days, $e=0.9550000$,  $m_c = 8.078~M_{\odot}$, $t_c = 2.53 \times 10^{6}$ years. There are no GC pulsar with $P_{orb} > 1000$ days. As GCs are very old systems, primordial binaries having one member as massive as the companions of PSR B1259-63 or PSR J1638-4725 would not exist still today - the massive companions would have evolved by now and depending upon the nature of its evolution, a low mass WD as a binary companion may be the end product. For significantly low values of the mass of the WD, the resultant binary would have been ionized as have seen from Fig \ref{fig:vbyvc} that if $m_2 \lesssim 0.16 ~M_{\odot}$, ionization can be effective for $P_{orb} \sim 1000$ days. But in case of a comparatively massive WD companion, the binary will survive ionization. Although such long orbital period binaries can not originate through tidal capture process, there is a possibility of formation of such wide binaries through exchange or merger interactions (see Figs. \ref{fig:terzan_exch_merg_m3_l} and \ref{fig:terzan_exch_merg_m3_h}). Moreover, there are 12 pulsars in the galactic disk having $ P_{orb}$ in the range of 100 $-$ 1000 days, one of them has $m_c = 15.82 ~M_{\odot}$, one has $m_c = 0.78 ~M_{\odot}$ and the other 10 have $m_c$ in the range of $0.14 - 0.38~ M_{\odot}$; all of these have characteristic ages $10^{9} - 10^{10}$ years. But there are only 3 pulsars in GCs having $ P_{orb}$ in the range of 100 $-$ 1000 days, having $m_c$s in the range of $0.21 - 0.35~ M_{\odot}$. As our study has shown that ionization to be operative at $P_{orb} \leq 1000 ~{\rm days} $, $m_2 $ should be less than $0.07~ M_{\odot}$ (which is very unlikely), it seems to be surprising that the number of binary pulsars having $ P_{orb}$ in the range of 100 $-$ 1000 days are significantly less than the number of these objects in the disk and there is no binaries with $P_{orb} > 1000$ days in GCs at all. It is possible that such long period binaries in GCs have been misidentified as isolated pulsars as many globular clusters pulsars ($e.g.$ those in 47 Tuc) are very weak, usually weaker than the detection threshold and one can see them very rarely only when their intensity is boosted by scintillation (M. Kramer, private communication). These occasions may be too rare and/or too short to identify them as binaries, because the larger the value of $P_{orb}$, the less is the effect of the binary motion in the observed spin period and only observations over a longer time span can reveal the binary nature.

\subsection{Comments on Zero Eccentricity Pulsars}
\label{sec:zero}

The nearly  circular binaries in GCs (group III) have $P_{orb}$ in the range of $0.106-4.0$ days. 
These systems lie in the $e-P_{orb}$ plane in such a region that stellar encounters are favorable.
But it is clear that they have not encountered any kind of eccentricity imparting stellar encounters $e.g.$ fly-by or exchange or merger. Dynamical formation of ultra-compact binaries involving
intermediate mass main sequence stars in the early life of the GCs
can be the origin of these pulsars \cite{cam05}. These companions must
have been massive enough (beyond the present day cluster turn-off
mass of $0.8 \; M_{\odot}$) so that the initial mass transfer
became dynamically unstable, leading to common envelope evolution
and subsequent orbital decay and circularization. Alternately,
the present day red giant and NS collisions may lead to a prompt
disruption of the red giant envelope when the system ends up as
an eccentric NS-WD binary \cite{ras91}. These binaries could decay
to the group III pulsars by gravitational radiation if they had
short $P_{orb} \leq 0.2 \; \rm d$. In any one of the scenarios mentioned above, the resultant binaries have to be young enough not to face any stellar encounter.  In this connection, it will be interesting to determine the ages of these binaries by performing optical and/or infrared observations of their WD companions to determine WD ages from their temperatures. Secondly, these binary pulsars may in fact lie outside cluster cores with a higher $v_{10}/n_4$ so that the timescales for stellar interactions are actually $>~10^{10}$ years (even though in projection many of them appear to be inside the cluster cores). 

Here we wish to mention that the positions of the binary pulsars in GCs with respect to the GC center
in the projected image is given in Table \ref{tab:gcpsr_parms}.
There are a number of binary pulsars for which position offsets are not known. Also the binaries which appear to be inside the cluster core in the projected image can actually lie outside the core.
The observed values of spin period derivatives ($\dot{P_{s}}$) of
the GC pulsars can give some indication of the pulsar position
and its environment since negative or positive $\dot{P_{s}}$ indicate positions in the back half or in
the front half of the cluster respectively \cite{phi92}; however, the
actual positions can not be determined with the knowledge of $\dot{P_{s}}$ only.
Dispersion measure (DM) can also be a tool to measure pulsar positions with respect to
cluster center if DM variations were dominated by intra-cluster gas \cite{ransom07}. However, it has been shown that for Ter 5, DM is mainly due to the interstellar medium \cite{ransom07}, so DM values alone
can not help to determine pulsar positions with respect to the center
for this cluster.


\begin{table}[h]
\caption{Globular cluster parameters taken from Webbink (1985) \cite{web85}.}
\begin{center}
\begin{tabular}{cccccc}
\hline
group &GC & d (from sun) & $v_{10}$  &$n_{4}$ & $v_{10}/n_{4}$  \\
& &  (kpc) & ($10~ {\rm km /sec}$)  & ($10^4~\rm{pc^{-3}}$)  & ($10^{-3}~{\rm km~s^{-1}~pc^{3}}$) \\ \hline
1&Ter 5 & 10.3 & 1.18  & 479.77 & 0.0024 \\ \hline
2&NGC 6440 & 8.4 & 1.30   & 102.10 & 0.013 \\
&M 30 & 8.0 & 0.52  & 36.39 & 0.014 \\
&NGC 1851  &12.1 &  0.98  &60.40  & 0.016  \\
&M 62 & 6.9 & 0.52  & 32.36 & 0.016 \\
&M 15 & 10.3  & 0.86  & 36.65  &  0.023 \\ \hline
3&NGC 6441 & 11.2 & 1.44  & 31.70 & 0.046 \\
&NGC 6544 & 2.7 & 0.59  & 9.75 & 0.060 \\
&47 Tuc & 4.5 & 1.32  & 31.99 & 0.062 \\ \hline
4&M 28 & 5.6 & 1.06 &  9.59 & 0.110 \\
&NGC 6342   & 8.6 & 0.45  &3.49 & 0.130 \\
&NGC 6752   & 4.0 & 0.78   & 5.97 & 0.131 \\
&NGC 6760   & 7.4 & 0.58  & 4.17 & 0.138 \\
&NGC 6539   & 8.4 & 0.45   & 3.09 & 0.146 \\
&NGC 6397   & 2.3 & 0.48  & 3.16 & 0.152 \\ \hline
5&M 4 &  2.2 & 0.51  &   1.63 & 0.32 \\
&M 5 & 7.5 & 0.84  &   2.57 & 0.328  \\
&M 3 & 10.4 &0.82  & 1.44 & 0.568  \\
&M 22 & 3.2 & 0.83 &  1.96 & 0.423 \\ \hline
6&M 71 & 4.0 & 0.33 &  0.35 & 0.934 \\
&M 13 & 30.4 & 0.78 &  0.72 & 1.082 \\
&NGC 6749   & 7.9 &  0.39    & 0.23 & 1.713\\
&M 53 & 17.8 & 0.65  & 0.30 & 2.167 \\
\hline
\end{tabular}
\end{center}
\label{tb:gc_v_n}
\end{table}

\clearpage

\begin{table*}
\caption{Parameters for 73 binary pulsars with known orbital
solutions in 23 globular clusters. Pulsars marked with $^{*}$ are not used in the analysis. Pulsar offsets from the cluster cores are in the units of the core radius $r_c$ (taken
from Freire's webpage at http://www.naic.edu/$\sim$pfreire/GCpsr.html and S. Ransom's
webpage at http://www.cv.nrao.edu/$\sim$sransom/). In Freire's webpage, eccentricities of 32 pulsars are listed as zero, which we have replaced by an arbitrarily small value 0.0000003 (see text for details). Last column shows the median values of companion masses $i.e.$ for inclination angle $i=60$ degree (again from Freire's webpage) taking pulsar masses as $1.35~ M_{\odot}$.  For some pulsars, better mass measurements are available from the measurement of rates of periastron advance or from the spectral analysis of the optical counterpart. These pulsars are listed here : (i) 47Tuc H has $m_p = 1.44~M_{\odot},~ m_c = 0.17~M_{\odot}$ \cite{fre03} (ii) NGC 1851A has $m_p = 1.35~M_{\odot},~ m_c = 1.103~M_{\odot}$ \cite{fre07} (iii) M5B has $m_p = 2.08~M_{\odot},~ m_c = 0.21~M_{\odot}$ \cite{fre08c} (iv) Ter 5 I has $m_p = 1.87~M_{\odot},~ m_c = 0.30~M_{\odot}$ \cite{ran05} (v) Ter 5 J has $m_p = 1.73~M_{\odot},~ m_c = 0.47~M_{\odot}$  \cite{ran05} (vi) NGC 6440B has $m_p = 2.74~M_{\odot},~ m_c = 0.18~M_{\odot}$ \cite{fre08b} (vii) NGC 6441A has $m_p = 1.26~M_{\odot},~ m_c = 0.67~M_{\odot}$ \cite{fre08b} (viii) PSR J1911-5958A has $m_p = 1.34~M_{\odot},~ m_c = 0.175~M_{\odot}$ (using spectral analysis of the optical counterpart \cite{bas06}) (ix) M15 C has $m_p = 1.358~M_{\odot},~ m_c = 1.354~M_{\odot}$ \cite{jac06}.} \footnotesize
\label{tab:gcpsr_parms}
\begin{center}
\begin{tabular}{llllllllll}
\hline \hline \multicolumn{1}{l}{No}          &
\multicolumn{1}{l}{GC}             &
\multicolumn{1}{l}{Pulsar}        &
\multicolumn{1}{l}{offset}             &
\multicolumn{1}{l}{$P_{s}$}           &
\multicolumn{1}{l}{$\dot{P_{s}}$}            &
\multicolumn{1}{l}{DM} &
\multicolumn{1}{l}{$P_{orb}$}             &
\multicolumn{1}{l}{$e$}             &
\multicolumn{1}{l}{$m_c$}             \\
\multicolumn{1}{l}{}            & \multicolumn{1}{l}{}    &
\multicolumn{1}{l}{}         & \multicolumn{1}{l}{(in} &
\multicolumn{1}{l}{}             &
\multicolumn{1}{l}{$(10^{-20}$}          &
\multicolumn{1}{l}{} & \multicolumn{1}{l}{} &
\multicolumn{1}{l}{} & \multicolumn{1}{l}{} \\
\multicolumn{1}{l}{}            & \multicolumn{1}{l}{}    &
\multicolumn{1}{l}{}         &
\multicolumn{1}{l}{$r_c$)}             &
\multicolumn{1}{l}{(ms)}             & \multicolumn{1}{l}{${\rm
sec/sec})$}         & \multicolumn{1}{l}{${(\rm cm^{-3} pc})$} &
\multicolumn{1}{l}{(d)} & \multicolumn{1}{l}{} &
 \multicolumn{1}{l}{($M_\odot$)}
 \vspace{1mm}  \\ \hline \hline
1 & 47Tuc & J0024-7205E &  1.477 &   3.536  & 9.851 & 24.23 &  2.25684 &     0.0003152    &    0.18  \\
2 & 47Tuc & J0024-7204H &  1.75  &   3.210  & -0.183    & 24.36 &  2.35770 &   0.070560     &  0.19     \\
3 & 47Tuc &J0024-7204I &  0.659 &   3.485  &  -4.587      &  24.42 &  0.22979 &   $<0.0004$        &  0.015   \\
4 & 47Tuc &J0023-7203J (e) &  2.273 &   2.100  & -0.979     & 24.58 & 0.12066 &      $<0.00004$  &       0.024   \\
5 &  47Tuc &J0024-7204O &  0.136 &   2.643  & 3.035       & 24.36 &  0.13597  &    $<0.00016$    &    0.025   \\
6 &  47~Tuc &J0024-7204P  &  ?     &  3.643  & ? & 24.30 & 0.1472   &    0.0000003 &     0.02   \\
7 &  47~Tuc &  J0024-7204Q &  2.227 &   4.033 & 3.402    &  24.29&  1.18908 &   0.000085    &    0.21   \\
8 &  47Tuc & J0024-7204R (e) &  ?     &  3.480   & ? & 24.40 & 0.0662    &  0.0000003 &     0.030  \\
9  &  47Tuc & J0024-7204S &  0.432  &  2.830  & -12.054  & 24.35 &  1.20172  &  0.000394   &    0.10  \\
10 &  47Tuc & J0024-7204T &  0.773  &  7.588  & 29.37    &  24.39 &  1.12618  &  0.00040    &    0.20  \\
11  & 47Tuc & J0024-7203U &  2.136  &  4.343  & 9.523    & 24.33 &  0.42911  &  0.000149   &    0.14  \\
12 &  47Tuc & J0024-7204V  (e?) &   ?    &  4.810    & ? & 24.10 & 0.227   & 0.0000003    &   0.35  \\
13  &  47Tuc & J0024-7204W (e) & 0.182  &  2.352  & ? & 24.30 &  0.1330  &  0.0000003  &    0.14    \\
14  & 47Tuc &  J0024-7204Y  &   ?    &  2.197  & ? & 24.20 & 0.52194  &     0.0000003   &    0.16  \\
\hline
15 & NGC1851 & J0514-4002A  & 1.333  &  4.990  & 0.117    & 52.15 & 18.78518 &   0.8879773    &    1.10 \\
\hline
16 & M53 & B1310+18    &     ?   &   33.163    & ? & 24.00 & 255.8    &     0.01  &  0.35  \\
\hline
17  & M3 &    J1342+2822B &  0.254   &  2.389   & 1.858   & 26.15 &  1.41735  &   0.0000003   &    0.21     \\
18 & M3  &   J1342+2822D &  0.418   &  5.443  & ? & 26.34 &
128.752 & 0.0753      &   0.21   \\
\hline
\hline
\end{tabular}
\end{center}
\end{table*}

\addtocounter{table}{-1}
\begin{table*}
\caption{(continued).}
\vspace{-2.0mm}
\footnotesize
\begin{center}
\begin{tabular}{llllllllll}
\hline \hline \multicolumn{1}{l}{No}          &
\multicolumn{1}{l}{GC}             &
\multicolumn{1}{l}{Pulsar}        &
\multicolumn{1}{l}{offset}             &
\multicolumn{1}{l}{$P_{s}$}           &
\multicolumn{1}{l}{$\dot{P_{s}}$}            &
\multicolumn{1}{l}{DM} &
\multicolumn{1}{l}{$P_{orb}$}             &
\multicolumn{1}{l}{$e$}             &
\multicolumn{1}{l}{$m_c$}             \\
\multicolumn{1}{l}{}            & \multicolumn{1}{l}{}    &
\multicolumn{1}{l}{}         & \multicolumn{1}{l}{(in} &
\multicolumn{1}{l}{}             &
\multicolumn{1}{l}{$(10^{-20}$}         &
\multicolumn{1}{l}{} & \multicolumn{1}{l}{} &
\multicolumn{1}{l}{} &
\multicolumn{1}{l}{}\\
\multicolumn{1}{l}{}            & \multicolumn{1}{l}{}    &
\multicolumn{1}{l}{}         &
\multicolumn{1}{l}{$r_c$)}             &
\multicolumn{1}{l}{(ms)}             & \multicolumn{1}{l}{${\rm
sec/sec})$}         & \multicolumn{1}{l}{${(\rm cm^{-3} pc})$} &
\multicolumn{1}{l}{(d)} & \multicolumn{1}{l}{} &
\multicolumn{1}{l}{($M_\odot$)}
\vspace{1mm}  \\
\hline  \hline
19 & M5 &     B1516+02B  &   0.545  &   7.947  & -0.331      & 29.45 & 6.85845  &   0.13784    &   0.13   \\
20 & M5 &   J1518+0204C (e)  &  ?   &    2.484   & ? & 29.30 & 0.087     &  0.0000003    &   0.038   \\
21 & M5 &  J1518+0204D  &   ?   &    2.988  &  ?  & 29.30 & 1.22
& 0.0000003     &   0.20
  \\
22 & M5 &   J1518+0204E  &   ?   &    3.182   & ? & 29.30  & 1.10  &  0.0000003   &    0.15  \\
\hline
23 & M4 & B1620-26   &   0.924  &  11.076  & -5.469      & 62.86 & 191.44281  &  0.02531545     &   0.33   \\
\hline
24 & M13 & B1639+36B     &  ?  &   3.528  & ? & 29.50 & 1.25911  &  $<0.001$   &   0.19   \\
25  & M13 &  J1641+3627D  &   ?   &     3.118   & ? & 30.60 & 0.591     & 0.0000003    &   0.18    \\
26  & M13 &   J1641+3627E (e?) &    ?  &      2.487  & ? & 30.30 &  0.117   &   0.0000003     &   0.02  \\ \hline
27 & M62 &    J1701-3006A  & 1.778  &    5.241 & -13.196       & 115.03  & 3.80595 & 0.000004    &    0.23  \\
28 & M62 &  J1701-3006B (e) & 0.155    &  3.594  & -34.978       & 113.44 & 0.14455  & $<0.00007$       &  0.14   \\
29 & M62 &   J1701-3006C &  0.972 &     3.806 &-3.189       & 114.56 & 0.21500 & $<0.00006$     &   0.08   \\
30  & M62 &  J1701-3006D  &   ?      &  3.418    & ? & 114.31 & 1.12     & 0.0000003     &   0.14  \\
31  & M62 &   J1701-3006E (e) &   ?     &   3.234   & ? & 113.78  & 0.16   &    0.0000003    &   0.035   \\
32  & M62 &   J1701-3006F  &   ?    &    2.295  & ? & 113.36 & 0.20 &   0.0000003     &   0.02  \\
\hline
33 & NGC6342  & B1718-19 (e)  &  46.000 &  1004.04   & $1.59 \times 10^{5}$  & 71.00 & 0.25827 &  $<0.005$     &   0.13    \\ \hline
34 & NGC6397 & J1740-5340 (e) &  18.340   &   3.650  & 16.8    & 71.80 & 1.35406 &   $<0.0001$    &     0.22    \\
\hline
35 & Ter5  &  J1748-2446A (e) &  2.778  &  11.5632 & -3.400     & 242.10 & 0.075646 &  0.0000003    & 0.10    \\
36 & Ter5  &  J1748-2446E   & out(1.6)  & 2.19780 & ?     & 236.84 & 60.06   &    0.02    &        0.25    \\
37 & Ter5  &  J1748-2446I   &  ?    &     9.57019  & ?    & 238.73 & 1.328 &     0.428   &   0.24   \\
38 & Ter5  &  J1748-2446J   &  ?    &    80.3379  & ?    & 234.35 & 1.102 &     0.350  &        0.39   \\
39 & Ter5  &  J1748-2446M  & in(0.48) &  3.56957 & ?    & 238.65 & 0.4431 & 0.0000003        & 0.16    \\
40 & Ter5 &   J1748-2446N  & in(0.39) &  8.66690 & ? &    238.47 & 0.3855 & 0.000045   &    0.56       \\
41 & Ter5 &   J1748-2446O (e) &  in(0.45) &  1.67663 & ? & 236.38 & 0.2595  &   0.0000003    &   0.04  \\
42 & Ter5  &  J1748-2446P (e) &  in(0.74)  & 1.72862 & ? & 238.79 & 0.3626  &   0.0000003    &   0.44   \\
43 & Ter5  &  J1748-2446Q  & out(1.45) &  2.812  & ? & 234.50 & 30.295   &  0.722   &   0.53      \\
44 & Ter5 &   J1748-2446U   &   ?     &   3.289   & ? & 235.46 & 3.57   &     0.61 &         0.46    \\
45 & Ter5  &  J1748-2446V &  in(0.90)  & 2.07251  & ? & 239.11  & 0.5036  &  0.0000003    &   0.14        \\
46 & Ter5  &  J1748-2446W &  in(0.42)  & 4.20518 &  ? & 239.14 & 4.877 &  0.015   &    0.34     \\
47 & Ter5   &  J1748-2446X    &  ?    &    2.99926  & ? & 240.03 & 4.99850 &  0.3024     &    0.29        \\
48 & Ter5 &   J1748-2446Y  &  in(0.55)  & 2.04816 &  ? & 239.11 & 1.16443  & 0.00002     &    0.16       \\
49 & Ter5 &   J1748-2446Z   &   ?   &     2.46259  & ? & 238.85 & 3.48807     & 0.7608     &   0.25      \\
50 &Ter5  &  J1748-2446ad (e) &   ?   &     1.39595  & ? & 235.60 & 1.09443     &  0.0000003      &  0.16     \\
51 & Ter5 &   J1748-2446ae &  in(0.42)   & 3.65859  & ? & 238.75
& 0.17073 &  0.0000003       & 0.019 \\  \hline
52 & NGC6440 & J1748-2021B  & 0.530   &   16.760 & -32.913 & 220.92 & 20.550 &     0.570     &   0.090  \\
53  & NGC6440  & J1748-2021D (e) &  4.230    &   13.496 &  58.678           & 224.98 &  0.286 & 0.0000003       & 0.14  \\
54  & NGC6440  &  J1748-2021F   & 0.690    &   3.794 & 31.240     & 224.10 &  9.83397  &  0.0531    &   0.35 \\
\hline
55 & NGC6441 & J1750-37A   &  1.910   &  111.609  & 566.1 &     233.82 &17.3   &     0.71   &   0.70    \\
56 & NGC6441 & J1750-3703B &  3.000   &    6.074    & 1.92 &      234.39 & 3.61   & 0.0000003      &   0.19   \\ \hline
 57 & NGC6539 & B1802-07   &   0.463   &   23.1009  & 47.0 &     186.38 &  2.61676  &   0.21206     &    0.35    \\ \hline
58 & NGC6544 & B1802-07    &   ?     &     3.05945  & ? & 134.0 & 0.071092   & 0.0000003       & 0.010   \\ \hline
\hline
\end{tabular}
\end{center}
\end{table*}

\addtocounter{table}{-1}
\begin{table*}
\caption{(continued).}
\vspace{-2.0mm}
\footnotesize
\begin{center}
\begin{tabular}{llllllllll}
\hline \hline \multicolumn{1}{l}{No}          &
\multicolumn{1}{l}{GC}             &
\multicolumn{1}{l}{Pulsar}        &
\multicolumn{1}{l}{offset}             &
\multicolumn{1}{l}{$P_{s}$}           &
\multicolumn{1}{l}{$\dot{P_{s}}$}            &
\multicolumn{1}{l}{DM} &
\multicolumn{1}{l}{$P_{orb}$}             &
\multicolumn{1}{l}{$e$}             &
\multicolumn{1}{l}{$m_c$}             \\
\multicolumn{1}{l}{}            & \multicolumn{1}{l}{}    &
\multicolumn{1}{l}{}         & \multicolumn{1}{l}{(in} &
\multicolumn{1}{l}{}             &
\multicolumn{1}{l}{$(10^{-20}$}         &
\multicolumn{1}{l}{} & \multicolumn{1}{l}{} &
\multicolumn{1}{l}{} &
\multicolumn{1}{l}{}\\
\multicolumn{1}{l}{}            & \multicolumn{1}{l}{}    &
\multicolumn{1}{l}{}         &
\multicolumn{1}{l}{$r_c$)}             &
\multicolumn{1}{l}{(ms)}             & \multicolumn{1}{l}{${\rm
sec/sec})$}         & \multicolumn{1}{l}{${(\rm cm^{-3} pc})$} &
\multicolumn{1}{l}{(d)} & \multicolumn{1}{l}{} &
\multicolumn{1}{l}{($M_\odot$)}
\vspace{1mm}  \\
\hline  \hline
59 & M28  &   J1824-2452C  &  ?   &     4.159    & ? & 120.70 & 8.078    &  0.847    &      0.30     \\
60 & M28  &   J1824-2452D &   ?    &   79.832  & ? & 119.50 & 30.404 &   0.776  &      0.45     \\
61 & M28  &   J1824-2452G  &  ?   &     5.909   & ? & 119.40 & 0.1046 &  0.0000003     &   0.011    \\
62 & M28  &   J1824-2452H (e) &  ?   &    4.629   & ? & 121.50 & 0.435     & 0.0000003     &  0.20     \\
63 & M28  &   J1824-2452I (e) & ?    &    3.93185 & ? & 119.00 & 0.45941 &  0.0000003 &     0.20     \\
64 & M28  &   J1824-2452J   &  ?   &    4.039    & ? & 119.20 & 0.0974    &  0.0000003    &    0.015    \\
65 & M28  &   J1824-2452K  &  ?    &    4.46105 & ? & 119.80 & 3.91034  &  0.001524      &    0.16     \\
66 & M28   &  J1824-2452L   & ?    &    4.10011  & ? & 119.00 &0.22571  &   0.0000003     &  0.022    \\ \hline
67 & M22   &  J1836-2354A  &  ?   &       3.35434  & ? & 89.10 & 0.20276   &  0.0000003       &   0.020    \\ \hline
68  & NGC6749  & J1905+0154A   &  0.662    &    3.193  & ? &  193.69 &   0.81255  &  0.0000003     &    0.090    \\ \hline
69  & NGC6752  &  J1911-5958A   & 37.588    &   3.26619 & 0.307 &       33.68 &  0.83711  &  $<0.00001$      &   0.22     \\ \hline
70  & NGC6760  & J1911+0102A   &  1.273   &    3.61852 & -0.658  & 202.68 &  0.140996  &   $<0.00013$       &   0.020      \\ \hline
71  & M71   &   J1953+1846A (e)   &  ?      &    4.888   & ? & 117.00 & 0.1766  &  0.0000003     &    0.032     \\ \hline
72  & M15    &  B2127+11C    &  13.486   &   30.5293   & 499.1      & 67.13 & 0.33528   &    0.681386     &    1.13     \\ \hline
73  & M30    &  J2140-2310A (e)  &  1.117    &  11.0193   & -5.181      & 25.06 & 0.17399   &    $<0.00012$       &   0.11     \\ 
$74^{*}$  & M30    & ${ \rm J2140-2310B}^{*}$   &  ?   &  13.0   &   ?    & 25.09 &  $> 0.8$  &    $> 0.52$       &   $> 0.02$     \\
\hline
\hline
\end{tabular}
\end{center}
\end{table*}

\clearpage

\begin{table*}
\caption{Binary pulsars in the galactic disk compiled from the online ATNF pulsar catalog http://www.atnf.csiro.au/research/pulsar/psrcat/. Pulsars marked with $^{*}$ are not used in the analysis.  Last column shows the median values of companion masses $i.e.$ for inclination angle $i=60$ degree taking pulsar masses as $1.35~ M_{\odot}$.}
 \footnotesize
\label{tab:diskpsr_parms}
\begin{center}
\begin{tabular}{llcccccc} \hline \hline
No. & Pulsar & $P_s$ & $\dot{P_s}$ & DM & $P_{orb }$ & e & $m_c$ \\ 
 &  & ms & ${ \rm 10^{-20}~s/s }$  & ${\rm cm^{-3}~pc}$ & d &  & $M_{\odot}$ \\   \hline
1  &   J0034-0534  &      1.87718    &       0.4959      &      13.76     &       1.58928     &      1.4E-5      &       0.165   \\
2  &   J0218+4232  &      2.32309   &        7.7389    &       61.25    &         2.02885    &       2.22E-5    &        0.196   \\
3  &   J0407+1607  &      25.70174        &     7.9          &    35.65      &      669.0704      &         0.0009368    &      0.222    \\
4  &   J0437-4715   &     5.75745   &      5.72937   &      2.64     &       5.74105     &      1.9180E-5    &      0.164   \\
5$^{*}$ &    J0610-2100$^{*}$    &    3.86132       &    1.235   &      60.67    &         0.28602    &       ?          &       0.025   \\
6  &   J0613-0200    &    3.06184     &     0.9591     &       38.78      &      1.19851     &     5.5E-6       &      0.150   \\
7  &   J0621+1002   &     28.85386        &    4.732     &       36.60      &       8.31868        &    0.0024574   &      0.527   \\
8  &   B0655+64      &   195.67094        &     68.53        &     8.77         &     1.02867     &     0.0000075      &    0.796    \\
9  &   J0737-3039A   &    22.699378    &      175.993    &     48.92     &         0.10225     &    0.0877775   &       1.559   \\
10 &    J0737-3039B  &     2773.46077          &     $8.92 \times 10^{4}$        &    48.92       &       0.10225     &    0.0877775      &    1.739    \\
11 &   J0751+1807    &    3.47877   &         0.77860    &      30.25      &       0.26314    &    7.1E-7    &        0.150   \\
12 &   B0820+02      &   864.87280         &     $1.04550\times 10^{4}$    &     23.73       &    1232.404         &       0.0118689   &       0.226     \\
13 &   J0900-3144      &  11.10965    &       4.912    &   75.70    &        18.73764      &      1.03E-5    &        0.423    \\
14$^{*}$  &  J1012+5307$^{*}$     &   5.25575   &        1.71342   &       9.02    &         0.60467    &      ?         &       0.125    \\
15   & J1022+1001     &   16.45293     &      4.334    &        10.25      &       7.80513   &      9.729E-5    &       0.853   \\
17  &  J1045-4509      &  7.47422       &     1.755     &       58.17      &       4.08353    &       2.41E-5    &        0.186   \\
18  &  J1125-6014      &  2.63038       &     0.401      &      52.95        &      8.75260         &   7.9E-7      &       0.328    \\
19   &  J1141-6545     &   393.89783        &      $4.294683\times 10^{5}$   &    116.08   &          0.19765      &     0.171884   &        1.213   \\
20  &  J1157-5112      &  43.58923         &     14.3     &        39.67       &       3.50739       &     0.000402     &      1.456    \\
21  &  J1216-6410      &  3.53937       &      0.162   &         47.40     &         4.03673    &        6.8E-6      &       0.182    \\
22   &  J1232-6501     &   88.28191           &    81.0       &      239.4        &      1.86327      &       0.00011    &        0.166     \\
23  &  B1257+12       &   6.21853     &      11.43341      &   10.16     &       25.262    &            0.0        &        0.000   \\
24  &  B1259-63        &  47.76251          &    $2.276484\times 10^{4}$    &   146.72     &     1236.72404      &         0.8698869    &      4.140   \\
25  &  J1420-5625     &   34.11713         &     6.8       &       64.56        &     40.29452        &      0.003500     &      0.438      \\
26   & J1435-6100      &  9.34797        &     2.45    &       113.7       &       1.354885     &      1.05E-5    &        1.079     \\
27   &  J1439-5501     &   28.63489       &      14.18  &         14.56    &          2.11794    &       4.99E-5     &       1.376     \\
28  &  J1454-5846     &   45.24877         &     81.7       &     115.95      &      12.42306       &      0.001898      &     1.047     \\
29  &  J1455-3330     &   7.98720       &      2.428      &      13.57      &      76.17457     &        0.0001700    &      0.297     \\
30  &  J1518+4904    &    40.93499     &       2.7190   &        11.61       &     8.63400    &      0.2494845    &     0.993    \\
31  &  J1528-3146     &   60.82223          &   24.9         &    18.16       &       3.18034      &     0.000213   &        1.155     \\
32  &  B1534+12       &   37.90444     &      242.26281    &  11.61       &     0.42074    &    0.2736767       &   1.624   \\
33  &  J1600-3053      &  3.59793        &    0.94959     &     52.32       &      14.34846    &      0.0001737     &   0.240    \\
34  &  J1603-7202     &   14.84195          &   1.574      &      38.05     &          6.30863     &      9.28E-6     &       0.338    \\
35$^{*}$  &  J1618-39$^{*}$        &  11.98731          &          ?  &               117.5       &      22.8          &          ?          &       0.202     \\
36   &  J1638-4725    &    763.9335           &         $4.8\times 10^{4}$    &        552.1   &        1940.9      &             0.955    &          8.078    \\
37  &  J1640+2224    &    3.16331   &        0.28276    &     18.43   &        175.46066     &       0.0007973   &     0.290    \\
38  &  J1643-1224     &   4.62164         &   1.849      &     62.41    &       147.01739      &       0.0005058   &      0.139    \\
39  &   J1709+2313   &     4.63120       &     0.363    &   25.35    &        22.71189    &        0.0000187   &       0.319    \\
40 &   J1711-4322     &   102.61829           &    2666.0     &      191.5    &        922.4708      &          0.002376     &      0.237    \\
41 &   J1713+0747    &    4.57014   &        0.85289   &        15.99      &      67.82513     &     0.0000749   &    0.324    \\
42 &   J1732-5049     &   5.31255         &    1.38         &    56.84      &        5.26300      &      9.8E-6     &        0.209    \\
43  &  J1738+0333    &    5.85009      &      2.409     &       33.78      &        0.35479   &       4.0E-6    &         0.104    \\
44  &  J1740-3052      &  570.30958            &    $2.54969\times 10^{6}$     &    740.9     &       231.02965      &         0.5788720  &       15.820    \\
45$^{*}$  &  J1741+1354$^{*}$     &   3.74715               &    ?   &                24.0        &      16.335           &       ?      &          0.281    \\
46  &  J1744-3922      &  172.44436          &     155.0     &       148.1        &      0.19141      &       0.000      &        0.097    \\
47  &  J1745-0952     &   19.37630        &      9.5       &       64.47      &        4.94345       &      1.8E-5       &      0.126    \\
48  &  J1751-2857     &   3.91487     &       1.126   &        42.81   &         110.74646      &       0.0001283     &     0.226    \\
49  &  J1753-2240     &   95.13781           &    7.9   &      158.6      &       13.63757       &      0.303582      &     0.582    \\
50  &  J1756-2251     &   28.46159         &     101.71   &      121.18   &          0.31963     &       0.180567   &        1.353    \\
51  &  J1757-5322     &   8.86996       &      2.63      &       30.82      &        0.45331    &      4.0E-6     &        0.667    \\
\hline
\end{tabular}
\end{center}
\end{table*}

\addtocounter{table}{-1}
\begin{table*}
\caption{(continued).}
\vspace{-2.0mm}
\footnotesize
\begin{center}
\begin{tabular}{llcccccc} \hline \hline
No. & Pulsar & $P_s$ & $\dot{P_{s}}$ & DM & $P_{orb }$ & e & $m_c$ \\ 
 &  & ms & ${ \rm 10^{-20}~s/s }$  & ${\rm cm^{-3}~pc}$ & d &  & $M_{\odot}$ \\   \hline
52  &  J1802-2124      &  12.64759        &     7.2    &       149.6      &        0.69889     &      3.2E-6    &         0.982    \\
53  &  B1800-27       &   334.41543            &    1711.3      &    165.5       &     406.781        &         0.000507     &      0.168     \\
54  &  J1804-2717    &    9.34303      &       4.089   &         24.67    &         11.12871     &        3.5E-5    &         0.235     \\
55  & J1810-2005     &   32.82224         &     15.1     &      240.2       &      15.01202        &      2.5E-5       &      0.329     \\
56  &  J1811-1736    &    104.18195          &     90.1     &       476.0        &       18.77917     &       0.828011    &       1.041     \\
57  &  J1822-0848    &    834.83927            &    $1.35\times 10^{4}$      &      186.3        &    286.8303         &       0.058962      &     0.383     \\
58  &    B1820-11     &     279.82870           &     $1.37862\times 10^{5}$   &      428.59    &       357.76199     &          0.794608     &      0.779    \\
59  &   J1829+2456   &      41.00982     &         5.25   &           13.8      &         1.17603     &      0.1391412   &       1.569     \\
60  &  B1831-00       &   520.95431            &    1053.0     &      88.65        &      1.81110     &       0.004472    &       0.073     \\
61  &  J1841+0130   &     29.77277        &       817.0      &      125.88    &        10.47163      &         8.2E-5     &        0.111     \\
62  &  J1853+1303    &    4.09180    &        0.885     &        30.57    &       115.65379      &       0.0000236 9   &      0.281    \\
63  &  B1855+09      &    5.36210      &      1.771     &       13.29     &       12.32717     &       2.09192E-5    &     0.284     \\
64  &  J1903+0327    &    2.14991 5     &      1.879    &       297.54    &        95.17412     &         0.4366784 11  &      1.084    \\
65  &  J1904+0412    &    71.09490           &    11.0          &   185.9       &      14.93426        &       2.2E-4     &        0.258     \\
66  &  J1906+0746    &    144.07193             &   $2.0280\times 10^{6}$       &    217.78      &       0.16599      &      0.085303     &      0.976     \\
67  &  J1909-3744     &   2.947108    &      1.40241   &       10.39      &       1.53345     &    1.302E-7      &     0.229     \\
68  &  J1910+1256    &    4.98358       &     0.977        &    38.06     &       58.46674        &    0.0002302 2   &      0.224     \\
69  &  J1911-1114     &   3.62574        &    1.416       &     30.97     &        2.71656     &       1.9E-5      &       0.140     \\
70  &  B1913+16      &    59.03000        &    862.713   &     168.77     &        0.32300   &      0.6171338   &       1.056    \\
71  &  J1918-0642     &   7.64587        &     2.4       &       26.60     &       10.91318       &      2.2E-5     &        0.278     \\
72  &  J1933-6211     &   3.54343        &     0.37       &      11.50     &        12.81941    &        1.2E-6      &       0.382    \\
73  &  B1953+29      &    6.13317       &      2.9739    &      104.58     &      117.34910     &         0.0003303   &       0.208    \\
74  &  B1957+20       &   1.60740    &       1.68515    &      29.12        &     0.38197    &      0.00000      &      0.025    \\
75  &  J2016+1948     &   64.94039            &       ?     &              34.0     &          635.039      &        0.00128    &        0.342    \\
76  &  J2019+2425     &   3.93452     &      0.70237     &      17.20      &       76.51163       &     0.0001111       &   0.364    \\
77  &  J2033+17        &  5.94895           &       1.1      &        25.2     &         56.31         &        0.00013    &        0.219    \\
78  &  J2051-0827     &   4.50864     &       1.2737    &       20.74       &      0.09911     &     0.0000        &     0.031    \\
79  &  J2129-5721     &   3.72635         &    2.981      &      31.86      &        6.62549       &      1.91E-5      &      0.154    \\
80  &  J2145-0750     &   16.05242      &     2.9757    &        9.00      &        6.83890     &    1.928E-5    &       0.503    \\
81  &  J2229+2643    &    2.97782      &      0.146      &       23.02      &       93.01589      &       0.0002556    &      0.142    \\
82  &  B2303+46       &   1066.37107           &    $5.6909 \times 10^{4}$     &      62.06      &       12.33954       &     0.658369       &    1.431    \\
83  &  J2317+1439    &    3.44525      &      0.242     &        21.91    &         2.45933     &      0.0000005      &    0.201    \\ \hline
\hline
\end{tabular}
\end{center}
\end{table*}

\clearpage

\section{Eccentricity Pumping in Hierarchical Triple Systems}
\label{sec:triple}

In the previous section, we have discussed how stellar interactions in globular clusters lead to binary millisecond pulsars in eccentric orbits. In the absence of such stellar interactions $i.e.$ in the galactic disk, millisecond pulsars are expected to be in circular orbits as during the accretion phase (recycling process), tidal coupling leads to orbit circularization. On the other hand, if the millisecond pulsar is the part of a hierarchical triple, consisting of inner and outer binaries, then as a result of Kozai oscillation, the eccentricity of the inner binary can be large. 

To understand this mechanism, let us assume that a star of mass $m_0$ forms the inner binary with its companion of mass $m_1$. Another star of mass $m_2$ forms an outer binary with the center of mass of the inner binary. One can consider the inner and outer binaries to be isolated binaries of masses $m_0$, $m_1$ and $m_0+m_1$, $m_2$ respectively. Let us denote the eccentricities, semi-major axes, inclinations with respect to the sky plane and the arguments of the periastrons (with respect to their lines of nodes) of the inner and outer binaries by $e_1,e_2, a_1, a_2, i_1, i_2, \omega_1 $ and $\omega_2$, respectively and the mutual inclination angle between the two orbits $i_m=i_1+i_2$. Then the magnitudes of the angular momenta ${\cal G}_1 $ and ${\cal G}_2$  of the inner and outer binaries will be as follow :
\begin{subequations} 
\label{g12_Eq}
\begin{align}
{\cal G}_1 &= \eta_i\, \biggl \{ G\, M_i^3\, a_1\, (1-e_1^2) \biggr \}^{1/2} 
\,,\\
{\cal G}_2 &=  m_2 \biggl \{ 
\frac{ G\, M_i^2}{ ( M_i + m_2) } \, a_2\, (1-e_2^2) 
\biggr \}^{1/2} 
\,,
\end{align}
\end{subequations} where G is the gravitational constant, $M_i = (m_0 + m_1)$ is the total mass of the inner binary and $\eta_i$ is the symmetric mass ratio of the inner binary, given by $\eta_i = m_0\, m_1 /M_i^2$. The temporal evolution of the inner binary can be probed with the help of secular perturbation theory, applicable to Newtonian hierarchical triples containing point masses, while including the dominant  quadrupolar order  interactions between the two orbits. In other words, the dynamical equations invoked are accurate to order $\left ( a_1/a_2 \right )^2$, where $\left ( a_1/a_2 \right )$ is the small parameter in the perturbative expansion. The relevant equations providing secular temporal evolution for the eccentricity and the argument of periastron of the inner binary including the dominant order general relativistic effect that causes the periastron of an isolated compact binary to advance are \cite{BLS02}: 
\begin{subequations} 
\label{etgt_Eq}
\begin{align}
\label{dgdt}
\dot{\omega}_1 & = 
\frac{ 6\,C_2}{{\cal G}_1} \biggl \{ 4\theta^2+(5\cos2 \omega_1-1)(1-e_1^2-
                 \theta^2) \biggr \} 
\nonumber\\
 &
+\frac { 6\, C_2\, \theta }{  {\cal G}_2} \biggl \{ 2+e_1^2(3-5\cos2 \omega_1) \biggr \} 
\nonumber\\
 &
+{3\over c^2\, a_1\, (1-e_1^2)}\,
\biggl [ \frac {G\, M_i }{ a_1} \biggr ]^{3/2}
\,,\\
\label{dedtht}
\dot{e}_1 & =\frac{ 30\, C_2}{ {\cal G}_1} \, {e_1(1-e_1^2)}\, (1-\theta^2)\, \sin 2 \omega_1
\,,
\end{align}
\end{subequations}
where $ \theta = \cos i_m $. The quantity $C_2$ is
\begin{equation} 
\label{c2_def}
C_2 =  \frac{ G\, M_i\, \eta_i} {16\, a_2} \, \frac{ m_2}{ (1-e_2^2)^{3/2} } \, 
\left({a_1\over a_2} \right)^2
\,,\\
\end{equation} Note that the Newtonian contributions to Eqns.~(\ref{etgt_Eq}) originate from certain `doubly averaged' Hamiltonian, which is derivable from the usual Hamiltonian for a hierarchical triple at the quadrupolar interaction order \cite{FKR00}.  The `doubly averaged' Hamiltonian, suitable for describing secular (long-term) temporal evolution of a hierarchical triple, is independent of the mean anomalies of the inner and outer orbits. This implies that their respective conjugate momenta and hence the semi-major axes, $a_1$ and $a_2$, are constants of motion. Here changes in $a_1$ and $e_1$ due to gravitational wave radiation have been neglected. First two terms in Eqn. (\ref{dgdt}) give the rate of change of periastron due to the hierarchical triple configuration ($\dot{\omega}_{1, ht}$). The third term gives the rate of change of periastron due to the general relativistic effect ($\dot{\omega}_{1, gr}$) up to 1PN order only and neglecting the spin-orbit coupling (see Eqn. \ref{eq:per_adv}).

It turns out that if the mutual inclination angle is fairly high and in a certain window, the orbital eccentricities experience periodic oscillations over time-scales that are extremely large compared to the respective orbital periods. The above effect, usually referred to as the Kozai resonance, arises due to the tidal torquing between the two orbits \cite{K62}. The Kozai resonance can force initially tiny eccentricity of the inner binary to oscillate through a maximum value, given by ${e_1}^{\rm max} \simeq \left( 1 - \frac{5}{3}\, \cos^2 i_{m,0} \right)^{1/2} $ where $i_{m,0}$ is the initial value for the mutual inclination angle. Due to the obvious restriction, namely  $| \cos i_{m,0} <  (3/5)^{1/2} |$, $i_{m,0}$ is required to lie in the range $39^{\circ} - 141^{\circ}$. An approximate expression for the period of the Kozai oscillation reads  \cite{MS79} :
\begin{equation}
\label{t_Kozai}
 \tau_{\rm Kozai}  = P_{orb, 1} \, \frac{ M_i}{m_2}\, \left ( \frac{a_2}{a_1} \right )^3\, ( 1 - e_2^2 )^{3/2} \,,
\end{equation}
where $P_{orb, 1}$ is the orbital period of the inner binary. 

The general relativistic periastron advance of the inner binary, in principle, can interfere with the Kozai resonance and even terminate the eccentricity oscillations. This is because the extra contribution $\dot{\omega}_{1, gr}$ to $\dot{\omega}_{1}$ can indirectly affect the evolution of $e_1$. The following useful criterion can be used to infer the possibility for this not to happen \cite{BLS02}:
\begin{equation}
\label{alpha_GR}
 \left ( \frac{ a_2}{a_1} \right ) < \biggl [ \frac{3}{4}\, \frac{ m_2 }{M_i} \, \frac {\tilde a_1} { \tilde M_i } \,
\left ( \frac{ 1 - e_1^2 } { 1 -e_2^2}  \right )^{3/2}  \biggr ]^{1/3} 
\,,
\end{equation}
\newline
where $ \tilde M_i $ is $M_i/M_{\odot}, \tilde a_1 = a_1/L_{\odot} $ and $ L_{\odot} = 1.476625$ km. 

Another constraint for $ a_2 / a_1 $ can be obtained by invoking an empirical relation needed for the Newtonian coplanar prograde orbits in hierarchical triple configurations \cite{MA_01} :
\begin{equation}
\label{alpha_MA}
 \left ( \frac{ a_2}{a_1} \right ) >  \frac{ 2.8} { 1 -e_2} \, \biggl [  \biggl ( 1 + \frac{ m_2}{M_i} \biggr )\, 
\frac{ ( 1 + e_2) }{ ( 1-e_2)^{1/2} } \biggr ] ^{2/5}
\,.
\end{equation}
We treat the above inequality to be rather conservative \cite{BLS02} as the inclined orbits, relevant for our investigation, are expected to more stable than the coplanar triples of Eq.~(\ref{alpha_MA}).

\subsection{PSR J$1903+0327$ : Open Question }

Recently a millisecond pulsar ($P_s \simeq 2.15$ ms) PSR J$1903+0327$ in a highly eccentric orbit around a solar mass companion in the galactic plane has been discovered \cite{cha08}. The discoverers almost discarded the  ``born-fast" scenario for this pulsar and strongly favored a binary origin for it. So they had to explore unconventional formation scenarios. They favored the idea that this pulsar is a member of the inner binary of a triple system and Kozai resonance is responsible for its high eccentricity. They have also speculated a globular cluster origin for the system, involving the ejection of the system from a globular cluster or the disruption of the cluster itself. 

Accurate radio timing analysis of the pulsar (using `DDGR' timing model \cite{TW}) reveals that the binary has the following properties - the orbital period is $95.1741176(2)$ days, eccentricity is $0.436678411(12)$, pulsar mass is $1.74(4)~ M_{\odot}$, companion mass is $1.05(15)~ M_{\odot}$, the longitude of periastron is $141.65779^{\circ}$, the advance of periastron is $2.46(2) \times 10^{-4}~{\rm deg~ yr^{-1}}$  \cite{cha08}. Here  the numbers in the parenthesis are twice the 1-$\sigma$ uncertainties in the least significant digit(s). The companion mass was determined using the advance of periastron and the shapiro shape (``$s$") parameter, assuming the advance of periastron to be only due to the general relativistic effect $i.e.$ the third term in Eqn. (\ref{dgdt}). Infrared observations with Gemini North telescope yielded a possible main-sequence companion star for the pulsar. 

The hierarchical triplet hypothesis for this system is that the pulsar and the companion (either a neutron star or white dwarf as the pulsar is recycled - implying past accretion from the red-giant phase of its companion) forms the inner binary and the main-sequence star is in a much wider orbit forming the outer binary. This means that the white-dwarf in the inner binary contributes in the timing analysis and the main-sequence star in the outer binary has been observed in infrared observation. So, in Eqns. (\ref{g12_Eq}), (\ref{dgdt}), (\ref{dedtht}), (\ref{c2_def}), (\ref{t_Kozai}), (\ref{alpha_GR}) and (\ref{alpha_MA}) one can use $m_0 =1.74 ~ M_{\odot}$, $m_1 = 1.05~ M_{\odot}$, $e_1 = 0.436678$, $\omega_1=141.658^{\circ}$, ${\dot {\omega}}_1 = 2.46(2) \times 10^{-4}~{\rm deg~ yr^{-1}}$. $a_1$ can be calculated from the orbital period ($\sim 95.174$ days) using Kepler's law. But the problem here is the companion mass was found to be $1.05~ M_{\odot}$ with an assumption that $\dot{\omega}_{1}$ is only due to general relativistic effect (third term in Eqn. \ref{dgdt}, $\dot{\omega}_{1, gr}$) on the other hand, for a hierarchical triple system all the terms in Eqn. (\ref{dgdt}) would contribute. So if this system is the part of a hierarchical triple, the value of the companion mass of the inner binary is likely to be different from the above mentioned value.

\begin{figure}[h]
{\includegraphics[width=0.9\textwidth,
height=0.45\textheight]{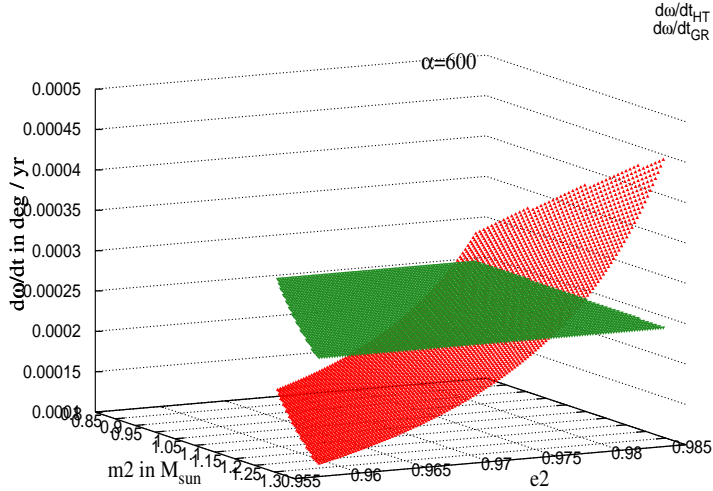}} 
\caption{\footnotesize{Variation of $\dot{\omega}_{ht}$ and  $\dot{\omega}_{gr}$ with $m_2$ and $e_2$ for  $\alpha = 600$. Red points are for $\dot{\omega}_{ht}$ and green points are for $\dot{\omega}_{gr}$. $\dot{\omega}_{gr}$ is independent of $m_2$ and $e_2$ (see the third term of Eqn. \ref{dgdt}) and matches with the observed value \cite{cha08}. Points are plotted only for such combinations of $m_2$ and $e_2$ that inequalities (\ref{alpha_GR}) and (\ref{alpha_MA}) are satisfied.}} \label{fig:gdot_htgr}
\end{figure}

So the timing-analysis of this system was continued. The latest value of the companion mass measured using only the Shapiro shape (``$s$") and range parameter (``$r$") is $1.03 \pm 0.04~M_{\odot}$. This matches with the earlier estimate using $s$ and $ \dot{\omega}_{1}$ supporting the assumption that the $\dot{\omega}_{1}$ is purely relativistic \cite{fre09}. However, because of the lower precision in the measurement of $r$, it is impossible to exclude small contributions to $\dot{\omega}_{1}$ at present. Using Eqn. \ref{dgdt}, we found that for preferable combination of $\alpha = a_2/a_1,~m_2,~ e_2$ the contribution form the first two terms in Eqn. \ref{dgdt} can be negligible in comparison to the third term. As an example,  in Fig. \ref{fig:gdot_htgr} we plot the variation of $\dot{\omega}_{1, ht}$ and $\dot{\omega}_{1, gr}$ with $m_2$ and $e_2$ for a fixed $\alpha = 600$ (for such combinations of $m_2$, $e_2$ and $\alpha$ that both the inequalities (\ref{alpha_GR}) and (\ref{alpha_MA}) are satisfied); red points are for $\dot{\omega}_{1, ht}$ and green points are for $\dot{\omega}_{1, gr}$. $\dot{\omega}_{1, gr}$ is independent of $m_2$ and $e_2$ (see the third term of Eqn. \ref{dgdt}) and matches with the observed value \cite{cha08}. We find that  $\dot{\omega}_{1, ht}$ can be significantly less than $\dot{\omega}_{1, gr}$ for sufficiently small values of $e_2$. In that case the companion mass measured using $s$ and $r$ can match with the measurement done with $s$ and $ \dot{\omega}_{1}$ assuming $\dot{\omega}_{1} = \dot{\omega}_{1, gr}$. But the precision of $r$ is still improving significantly with continued timing and it might be possible in future to confirm whether there is even any small deviation of $ \dot{\omega}_{1}$ from $\dot{\omega}_{1, gr}$.

In principle, $\dot{e}_{1}$ measurement can also help to test the validity of hierarchical triple scenario. For a hierarchical triple system, the observed $\dot{e}_{1}$ can be the combined effect of three terms like $\dot{e}_{1, obs}=\dot{e}_{1, ht}+\dot{e}_{1, gw}+\dot{e}_{1, geo}$, where $\dot{e}_{1, ht}$ is the rate of change of eccentricity because of Kozai oscillation in a hierarchical triple system given by Eqn. (\ref{dedtht}),  $\dot{e}_{1, gw}$ is the rate of change of eccentricity because of gravitational wave radiation given by Eqn. (\ref{eq:dedtgr}) and $\dot{e}_{1, geo}$ is due to geodetic precession of the pulsar spin axis (see Eqns. 8.58, 8.82 and 8.91 of \cite{lk05}). It is clear from Eqn. (\ref{eq:dedtgr}) that $\dot{e}_{1, gw}$ is always negative and we found it to be $-1.45975808 \times 10^{-24}~{\rm s^{-1}}$ for the inner binary as mentioned above. Eqn. (\ref{dedtht}) shows that $\dot{e}_{1, ht}$ can be positive or negative depending on whether ${\rm sin} \, 2 \omega_1$ is positive or negative and for $\omega_1 = 141.65779^{\circ}$, ${ \rm sin} \, 2 \omega_1 = -0.97312 $ making $\dot{e}_{1, ht}$ negative. So the magnitude of  $\dot{e}_{1, ht}$ should be such that $\left( -1.45975808 \times 10^{-24}~{\rm s^{-1}}+\dot{e}_{1, ht}+\dot{e}_{1, geo} \right) = \dot{e}_{1, obs}$. The problem here is that although we can calculate $\dot{e}_{1, ht}$ using Eqn. (\ref{dedtht}) \cite{gop09} using the parameters of the inner binary as mentioned earlier and varying $i_{m}$, $a_2$, $e_2$ and $m_2$ over realistic ranges satisfying conditions (\ref{alpha_GR}) and (\ref{alpha_MA}), we can not calculate $\dot{e}_{1, geo}$ as the orientation of the spin axis of the pulsar with respect to the total orbital angular momentum is not known, which is needed to calculate $\dot{e}_{1, geo}$ (see Eqns. 8.58, 8.82 and 8.91 of \cite{lk05}). Moreover, the presently determined value of $\dot{e}_{1}$ from timing analysis is $ \left( 8 \pm 15 \right) \times 10^{-17}~{\rm s^{-1}}$ which is merely the measurement uncertainty (unpublished, provided by Paulo Freire through private communication). So this approach can not help us to test the hierarchical triple hypothesis at present.

On the other hand, the observed value of the spin period derivative ${\dot P}_{s}$ can be used to test the hierarchical triple scenario. A rough estimate for ${\dot P}_{s}$ can be obtained by treating the inner binary as a single object of mass $m_0+m_1$ which forms a wide binary with stellar mass companion of mass $m_2$ \cite{JR97}. Further, we let the projected semi-major axis of this new binary to be $600$ times that of the inner binary and choose a value for $e_2$ consistent with inequalities (\ref{alpha_GR}) and (\ref{alpha_MA}). The associated line of sight acceleration leads to an estimate of ${\dot P}_{s} $ of the order of $ 10^{-17} {\rm s \, s^{-1}}$ \cite{{lk05}} and this is much higher than the reported value ${\dot P}_{s} \sim  10^{-20} {\rm s  \, s^{-1} }$ \cite{cha08}. This is fact seems to be a clear indication that this system is not the part of a hierarchical triple.

There is an alternative hierarchical triple scenario \cite{cha08, heuvel08}, such that the original inner companion (WD) of the pulsar has been evaporated by now and we see only the outer binary at present with $m_0 \simeq 1.74~M_{\odot}$, $m_2 \simeq 1.051~M_{\odot}$, $\omega_2 = 141.65^{\circ}$, $e_2 \simeq 0.436678$, $\dot{\omega}_2 = 2.46(2) \times 10^{-4}~{\rm deg~ yr^{-1}}$ and $P_{orb,~2} = 95. 174$ days. This scenario is difficult to test as we don't have any information about the parameters of the extinguished inner binary like $m_1$, $e_1$, $\omega_1$ and $a_1$. In this case, the agreement between the measurement of the present day companion mass ($m_2$) using $s$ and $\dot{\omega}_2$ with the same using $s$ and $r$ is expected as $\dot{\omega}_2$ is only due to general relativistic effect (third term in Eqn. \ref{dgdt}) because there is only one binary at present.

\begin{figure}[h]
{\includegraphics[width=0.9\textwidth,
height=0.45\textheight]{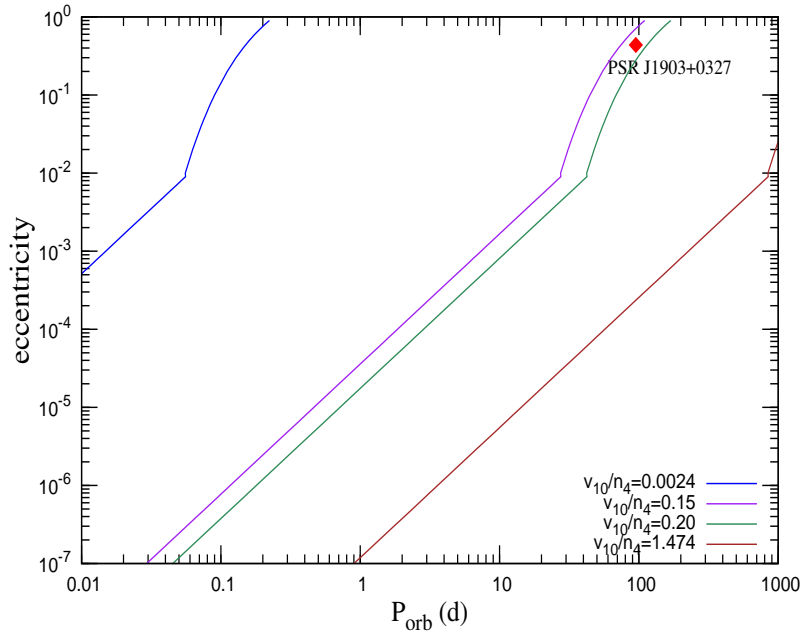}} 
\caption{\footnotesize{Fly-by isochrones for $ {\rm t_{fly} } = 10^{9}$ years in the eccentricity - orbital period plane for different values of the GC parameter $v_{10}/n_{4}$. The region to the lower right of each of these lines have shorter timescales. This indicates that PSR J1903$+$0327 could have originated only in a rather special GC environment with characteristics belonging to the upper-left portion of the isochrone with $v_{10}/n_{4} = 0.15$ (the unit of $v_{10}/n_{4}$ is $10^{-3}~{\rm km~s^{-1}~pc^{3}}$ as mentioned earlier). }} \label{fig:kozai_gc}
\end{figure}

Let us now comment on the possible globular cluster origin for the system, motivated by the fact that in globular clusters, stellar interactions can impart eccentricity to millisecond pulsar binaries and indeed $ \sim20 \%$ of the millisecond binaries in globular clusters have eccentricity $>0.1$. Using the formalism discussed in the earlier section for fly-by interaction (see \ref{eq:tfly2}), we find that the PSR J1903+0327 like systems could have been produced in globular cluster systems up to a mean  $v_{10}/n_4 = 0.15$ but not in any globular cluster with a higher value of this parameter. Here we consider maximum allowed timescale for stellar interaction to be $10^{9}$ years as the characteristic age of the pulsar is $1.8 \times 10^{9}$ years which is usually an upper limit of the true age of a radio pulsar \cite{cha08}. Thus the putative origin of PSR J1903+0327 inside a globular cluster could have taken place only in a restricted set of globular clusters. This further restricts the environment for production of such pulsars and therefore the overall probability for origin of this pulsar in a globular cluster must be smaller than that estimated by the dicoveres \cite{cha08}. Here it is noteworthy to mention that the other types of eccentricity inducing interactions $e.g.$ exchanges and mergers also scale with the parameter $v_{10}/n_4$ - so these interactions also can lead to  PSR J1903+0327 like systems only for a restricted set of globular clusters. So the eccentricity of this system still remains to be a puzzle to us. 

\section{Summary}
\label{sec:summary}

In this article we discuss how orbital parameters of binary radio pulsars reveal the history of their formation and evolution including dynamic interactions with other objects, the physics of the matter at extreme densities as inside the pulsar as well as the properties of the pulsar magnetosphere. Firstly we try to understand the importance of studying DNSs. If at least one of the NS of a DNS is a radio pulsar, we can determine the moment of inertia of the pulsar through accurate radio timing analysis by measuring the extra advancement of the periastron of the orbit due to the spin-orbit coupling over the standard post-Newtonian one. Then by matching this observed value of the moment of inertia with the theoretically predicted value, one can constrain the dense matter EoS. One necessary condition for this to be done is that the pulsar must be rapidly spinning and its companion NS must be much slower; the double pulsar system PSR J0737-3039(A/B) seems to be the most potential candidate for this purpose till date as we know that PSR J0737-3039A has $P_{s} = 22.7 $ ms whereas PSR J0737-3039B has $P_{s} = 2.77 $ s. Moreover the masses of both the NSs are well determined. We also checked how good the system really is for this purpose by studying the effect of various parameters on the spin-orbit coupling term as a larger value of this term will facilitate the moment of inertia measurement. We found that this term has larger effect if the following conditions are satisfied - (i) short spin period  (ii) short orbital period, (iii) high eccentricity.  Unfortunately, PSR J0737-3039 lacks conditions (iii) and even (i) to some extent.  So it will be nice if we can find better systems satisfying all the above conditions. Globular clusters are potential sources for favorable DNS systems in the above sense as DNSs can be formed by exchange collisions due to the high stellar density and the abundance of millisecond pulsars in globular clusters. On the contrary, only one DNS has been found in a GC so far which might be a result of selection effects as GCs are usually at larger distances in comparison to the disk pulsars. The resulting binaries are more probable to be highly eccentric with a millisecond pulsar facilitating  moment of inertia estimation. We have also discussed how the study of the eclipse of the radio beam of  PSR J0737-3039A by the magnetosphere of  PSR J0737-3039B can be used as a tool to understand the magnetospheric properties of a pulsar in a better extent. Some suggestive works about the formation of PSR J0737-3039A/B have been mentioned too.

We explain the presently observed orbital eccentricity and
period data of GC binary pulsars by stellar interaction scenarios
of fly-bys, exchanges and mergers with field stars. Binaries with $e >
0.1$ are most probably the result of exchange or merger events
whereas binaries with $0.01 > e > 0.00001 $ are products of
fly-by of single stars. A number of wide orbit intermediate eccentricity pulsars seen in the galactic
disk are absent in the GC sample because they have been kicked
up to relatively high eccentricities by passing stars in the
dense stellar environments in GCs. In some GCs such as Ter 5, the
stellar densities are so high, and the velocity dispersion so
modest that the interaction timescale for exchange and fly-by
interactions is relatively short. In such GCs a typical binary
system may undergo multiple interactions. If the original binary
contains a spun-up millisecond pulsar in a relatively ``soft" binary,
then the exchange interaction may even produce a single millisecond pulsar in the
cluster. This may explain the higher incidence of isolated millisecond pulsars
in GCs compared to that in the galactic disk \cite{cam05}; in GCs 52 out of 59 isolated pulsars are millisecond pulsars whereas in the disk 23 out of 1565 isolated pulsars are millisecond pulsars. In addition, exchange interactions, as we have seen, can
lead to highly eccentric orbits and the system can be ejected
from the cluster core. If the last encounter took place not too
long ago, the system can be at a relatively large offset from the
cluster core, albeit being still spatially co-located with the GC. We have found that the eccentricity of PSR B1638+36B (in M13) is difficult to explain by any type of interactions - fly-by, exchange or merger. So we propose that either its true eccentricity is  lower than the presently known upper limit or it is a member of a hierarchical triple. We have also considered the effects of collision induced
ionization on the present day distribution of orbital parameters
of radio pulsars in GCs. In the galactic disk we find 12 pulsars with $ 100 < P_{orb} < 1000 $ days (although few of them have more massive companions than the pulsar binaries in the
GCs), while there are only 3 pulsars in this range of orbital period. Although it is tempting to speculate that these systems are missing from the present day GCs because they have been ionized in the past, we find that the ionization probability becomes substantial in this orbital period range only for very small values of companion masses. Observational selection effects can be the cause of this discrepancy between the number of large orbital period pulsars in the disk in comparison to that in GCs. Many galactic disk binary pulsars are seen in the region of $e-P_{orb}$ plane predicted by Phinney due to the
fluctuation dissipation of convective eddies and the resultant
orbital eccentricities that are induced. These pulsars are
missing from the GC sample. There is no reason not to expect
these systems to form in the GCs (although due to the lower
metallicity of the stellar companions in GCs, they are expected
to be in slightly different region in the $e-P_{orb}$ plane). This can be
explained by the substantial probability of them being knocked
out of their original phase space due to stellar interaction in GCs. Indeed there are some wide binaries in the present day GCs with moderately high eccentricities (e.g. $0.01< e <0.1$ and
$60 < P_{orb} < 256 \; \rm d$) which could have arisen out of
fly-bys or exchanges from progenitor binaries with ``relic"
eccentricities $e \sim 10^{-4}$ (merger in unlikely because of moderately low values of $m_c$s). The circularity of group III pulsars in GCs are thought to be due to either their special formation channels and young age or due to their locations at the outskirts of the GCs.

PSR J1903$+$0327 is the only one eccentric millisecond pulsar in the galactic disk discovered so far. Its millisecond spin-period suggests the spin-up scenario through accretion in the past. As such recycled pulsars are expected to be in circular orbits due to tidal-coupling induced circularization, this system has created lots of interest among pulsar researchers. As it is well known that in globular clusters, stellar interactions can impart eccentricity to an initially circular orbit, the possibility of past GC association of this system have been studied and almost rejected. Moreover, the possibility of this system being a member of a hierarchical triple system also seems to be less unlikely. So this system remains an open question for us.

We hope in the future with more advanced technologies like SKA, many more new interesting types of pulsars will be discovered and that the exciting and challenging situation will ultimately lead to a better understanding of the physics of pulsars.

\section{Acknowledgments}

This article uses some of the results of the collaborative works with Alak Ray and Achamveedu Gopakumar whom the author wishes to thank. The author also thanks Michael Kramer and Paulo Freire for discussions; Bhaswati Bhattacharyya and Ranjeev Misra for comments on the manuscript.

{}

\end{document}